\documentclass[11pt,a4paper]{article}
\usepackage{amscd,amsmath,amssymb,amsfonts,xspace,mathrsfs,mathtools,amsthm}
\usepackage{xr-hyper}
\usepackage{jheppub}
\usepackage[dvipsnames]{xcolor}
\usepackage[utf8]{inputenc}
\usepackage{bm}
\usepackage{bbm}
\usepackage{graphicx}
\usepackage[color,matrix,arrow]{xy}
\usepackage{stmaryrd}
\SetSymbolFont{stmry}{bold}{U}{stmry}{m}{n}
\usepackage[shortlabels]{enumitem}
\usepackage{array,amsmath,booktabs}
\usepackage{slashed}
\usepackage{tensor}    
\usepackage{caption}
\usepackage{subcaption}   
\usepackage{enumitem} 
\usepackage[textsize=tiny]{todonotes}
\usepackage{diagbox}
\usepackage{pdflscape}
\usepackage{silence}
\usepackage{placeins}
\usepackage{booktabs, cellspace, hhline}
\newcommand{\RN}[1]{%
  \textup{\uppercase\expandafter{\romannumeral#1}}%
}
\newcommand{\MD}[1]{}
\newcommand{\JM}[1]{}

\newcommand{\SW}{\text{SW}}
\newcommand{\UV}{\text{UV}}

\newcommand{\vdim}{\text{vd}}

\newcommand{\bea}{\begin{equation} \begin{aligned}}
\newcommand{\eea}{\end{aligned} \end{equation}}
\newcommand{\be}{\begin{equation}} 
\newcommand{\ee}{\end{equation}}  
\newcommand{\bes}{\begin{equation*}}
\newcommand{\ees}{\end{equation*}}
\newcommand{\sgn}{\mathrm{sgn}}

\newcommand{\CE}{\mathcal{E}}  
\newcommand{\CF}{\mathcal{F}}

\newcommand{\CL}{\mathcal{L}} 
\newcommand{\CM}{\mathcal{M}}  
\newcommand{\CN}{\mathcal{N}}
\newcommand{\CO}{\mathcal{O}} 
\newcommand{\CP}{\mathcal{P}}

\newcommand{\CR}{\mathcal{R}}
\newcommand{\CS}{\mathcal{S}}
\newcommand{\CT}{\mathcal{T}}

\newcommand{\BR}{\mathbb{R}}

\newcommand{\BC}{\mathbb{C}}

\newcommand{\eps}{\epsilon}

\newcommand{\bfk}{{\boldsymbol k}}

\newcommand{\bfmu}{{\boldsymbol \mu}}

\renewcommand{\epsilon}{\varepsilon}

\title{Equivariant Localization for $\CN=2$ Theories with Hypermultiplets
}

\abstract{Equivariant localization in the $\Omega$-background is a powerful technique for the evaluation of the partition function of $\CN=2$ supersymmetric field theory on a toric four-manifold. For the complex projective plane $\mathbb{CP}^2$, we apply equivariant localization to supersymmetric Yang-Mills theories with gauge groups $SU(2)$ and $SO(3)$ with $N_f$ massive hypermultiplets in the fundamental representation, and the $\CN=2^*$ theory with the hypermultiplet in the adjoint representation. We evaluate equivariant correlation functions, which provide an equivariant extension of intersection numbers of moduli spaces of instantons on $\mathbb{CP}^2$, such as Euler numbers and Segre invariants. In the non-equivariant limit, we compare our results with the evaluation using low energy field theory and integration over the $u$-plane. Using this comparison, we identify a specific overall factor, which captures the difference between a partition function for gauge group $U(2)$ and $SU(2)$ or $SO(3)$ on a compact four-manifold. We also discuss our results in the context of $S$-duality of the $N_f=4$ and $\CN=2^*$ theory, which includes the  triality automorphism group of the flavor group for $N_f=4$.
\\
\\
\\
}

\author{Matthias Dennemann, Jan Manschot\\
\vspace{5pt}
\it School of Mathematics, Trinity College, Dublin 2, Ireland\\
 \it Hamilton Mathematical Institute, Trinity College, Dublin 2, Ireland
\vspace{20pt} 
}

\preprint{}
\begin{document}
\maketitle
\baselineskip=18pt

%%%%%%%%%%%%%%%%%%%%%%%%%%%%%%%%

\section{Introduction}
 
Correlation functions of topologically twisted quantum field theories with $\CN=2$ supersymmetry are of much interest for the study of non-perturbative phenomena, as well as for the  intersection theory of moduli spaces of instantons and monopoles \cite{Witten:1988ze, Donaldson90, Moore:1997pc, LoNeSha, Witten:1994cg, Vafa:1994tf, Ne}.

Over the last years, $\CN=2$ supersymmetric theories with hypermultiplets receive increasing interest
\cite{LoNeSha, Ne, Dedushenko:2017tdw, Manschot:2021qqe, Aspman:2022sfj, Aspman:2023ate, Manschot:2023rdh}, in part because their moduli space of $Q$-fixed equations include monopole solutions besides the anti-self-dual instantons. The hypermultiplet masses are equivariant parameters for the flavour groups acting on the hypermultiplets, and the path integrals reduce to integrals involving equivariant Chern classes of the index bundle with respect to the action of these groups on the moduli spaces \cite{LoNeSha}. 
 For fundamental hypermultiplets, these invariants are known as Segre invariants in the algebraic geometric context \cite{Marian:2015, Gottsche:2020ass, Gottsche:2021ihz}, while for the adjoint hypermultiplet these are suitably defined Euler numbers \cite{Vafa:1994tf, Tanaka:2017jom, Manschot:2021qqe}. 
 
 Among the theories with hypermultiplets are two deformations of superconformal theories, the  theory with $N_f=4$ hypermultiplets in the fundamental representation and the $\CN=2^*$ theory with a single hypermultiplet in the adjoint representation. Electric-magnetic duality implies modularity for the partition functions of these theories \cite{Vafa:1994tf, Labastida:1998sk, Manschot:2021qqe}. For $N_f=4$, electric-magnetic duality intertwines in an interesting way with the $S_3$ outer automorphism group of the $SO(8)$ flavor group. A crucial part of our analysis are the background fluxes for the flavor group. In the absence of these fluxes, the scalars of the hypermultiplet recombine to sections of the spinor bundle by the topological twist \cite{Hyun:1995mb, Labastida:1995zj}. Therefore non-trivial background fluxes are required if the theory is formulated on a non-spin four-manifold \cite{Moore:1997pc, Manschot:2021qqe, Aspman:2022sfj}, while the infinite set of background fluxes provide a rich setting to study correlation functions. These background fluxes form a ``generalized Spin$^c$ structure" introduced in \cite{Moore:2024vsd}. For the canonical background flux induced by a complex structure, the $\CN=2^*$ partition function equals the partition function of the Vafa-Witten twist of the $\CN=4$ theory \cite{Vafa:1994tf, Manschot:2021qqe}.

Various approaches have been developed in physics and mathematics to evaluate correlation functions, such as the $u$-plane integral \cite{Moore:1997pc, LoNeSha, Manschot:2021qqe, Aspman:2022sfj, Aspman:2023ate}, equivariant localization \cite{Ne, Moore:1997dj, Moore:1998et, Lossev:1997bz, Pestun:2007rz, Bawane:2014uka, Pestun:2016zxk} and algebraic-geometric techniques \cite{Mochizuki2009, Gottsche:2010ig, Tanaka:2017jom, Gottsche:2020ass}. This paper evaluates topological correlation functions for theories with hypermultiplets for the complex projective plane $\mathbb{CP}^2$, which is an example of a smooth, compact four-manifold without boundary. We use the toric action on $\mathbb{CP}^2$, to evaluate the partition functions using equivariant localization in the $\Omega$-background associated to the $U(1)\times U(1)$ symmetry \cite{Ne}. The $\Omega$-background is characterized by two equivariant parameters $\eps_1$ and $\eps_2$, which were introduced in Refs \cite{Moore:1997dj, Moore:1998et, Lossev:1997bz}, and used in Ref. \cite{Moore:1998et} to regularize integrals over moduli spaces of instantons on $\mathbb{R}^4$. Correlation functions can be evaluated using toric localization \cite{Nekrasov:2003vi} as a contour integral for the Coulomb branch parameter $a$ on various toric manifolds such as the complex projective plane $\mathbb{CP}^2$ and $\mathbb{F}_0=S^2\times S^2$. The correlation functions  depend on these equivariant parameters. This approach has by now been applied to a variety of theories in four-dimensions \cite{Gottsche:2006tn,
     Bershtein:2015xfa, Bonelli:2017, Bonelli:2020xps} and five dimensions \cite{Nakajima:2005fg, Gottsche:2006bm, Kim:2025fpz, Khlaif:2026rnb}. See \cite{Pestun:2016zxk} for a review.

This paper applies the ultraviolet approach of \cite{Bershtein:2015xfa, Bonelli:2017, Bonelli:2020xps} to theories with hypermultiplets in the fundamental and adjoint representation. We study in detail the inclusion of background fluxes in the equivariant setup, as an $\eps_i$-dependent shift of the masses $M_j$. The quantization conditions for these fluxes are manifest in the equivariant approach as a requirement for the perturbative part to reduce to a product with a finite number of terms. We obtain explicit results for $N_f\leq 4$ and $\CN=2^*$, which allows us to study mass decoupling limits, reflection symmetry of the $\epsilon_i$, and the action of electric-magnetic duality for $N_f=4$ and $\CN=2^*$. 
 
 An interesting aspect of our study is the identification of a specific function $D^{(N_f)}$ \eqref{eq:DefDNf} as the ratio of partition functions of a theory with gauge group $U(2)$ or $SO(3)$ for a compact four-manifold.  The equivariant partition function $Z^{(N_f)}_{\mu}$ \eqref{eq:NekPartFctnGeneral} is a contour integral of a product of Nekrasov partition functions for $\mathbb{C}^2$. This partition function is naturally the partition function for gauge group $U(2)$ with a fixed first Chern class or 't Hooft flux. The $SU(2)$ and $SO(3)$ partition functions follow by dividing out a specific factor as is familiar from the AGT conjecture \cite{Alday:2009aq} and recursion relation for the $\mathbb{C}^2$ partition function. For compact toric manifolds, in particular $\mathbb{CP}^2$, we determine that the $SO(3)$ partition function $H^{(N_f)}_{\mu}$ follows by dividing from the $U(2)$ partition function $Z^{(N_f)}_{\mu}$ the factor $D^{(N_f)}$ defined in \eqref{eq:DefDNf}. Together with the appropriate action on the background flux for $N_f=4$ \eqref{eq:actionTBj}, we find that the $N_f=4$ partition function $H^{(N_f)}_\mu$ is invariant under the triality group. 
 
 As mentioned before, we have restricted our analysis to the four-manifold $\mathbb{CP}^2$. With $b_2=1$, its topology is relatively simple and the above aspects can be analysed without the difficulties of wall-crossing effects. We leave analysis of other four-manifolds with $b_2^+=1$ and $b_2^+>1$ for the future \cite{FMM:future}.

One motivation of our work is to compare with results for topological correlators obtained using the low energy approach by evaluating the $u$-plane integral \cite{Moore:1997pc, LoNeSha, Manschot:2021qqe, Aspman:2022sfj, Aspman:2023ate}. The evaluation of the integral is non-trivial due to the intricate Coulomb branch geometry, especially if the masses are chosen generically \cite{Aspman:2021vhs}. Explicit results are achieved in \cite{Aspman:2023ate}. 
While this approach is qualitatively different from the infrared approach, we find agreement of $H^{(N_f)}_\mu$ in the non-equivariant limit with results for topological correlators of Ref. \cite{Manschot:2021qqe} and \cite{Aspman:2023ate}. 
In particular, we find precise agreement for the path integrals for $N_f\leq 3$. When observables are inserted in the path integral, we find slightly different results for some instances of $N_f\leq 3$, which we attribute to ambiguities in the low energy effective expressions for the observables. 
 
For the superconformal theories, we moreover find agreement for
$N_f=4$ with the evaluation of the $u$-plane by Malmendier and Ono
\cite{Malmendier:2008db}; for $\CN=2^*$ with $w_2(P)=0,1$ and
background fluxes $B=\pm 1$; and for $w_2(P)=1$ with the canonical
Spin$^c$ structure $B=\pm 3$ \cite{Vafa:1994tf, Manschot:2021qqe}. For
the remaining case where  results are available, $w_2(P)=0$ with
$B=\pm 3$, the application of the equivariant residue calculation does
appear to correctly reproduce terms for positive instanton number
$k\in \mathbb{N}_+$, however the subtle constant term of the
non-equivariant result \cite{Vafa:1994tf, Manschot:2021qqe} is not reproduced.\footnote{During the completion of this project, we learned of \cite{Khlaif:2026rnb} on $\CN=1^*$ theory on $X\times S^1$, which includes similar findings to ours for $\CN=2^*$ for $X=\mathbb{CP}^2$ with background flux equal to the canonical Spin$^c$ structure.} It appears that there are further contributions for vanishing instanton number in this case, which we leave for further investigation.

The outline of this paper is as follows. Section \ref{sec:topN2YM} reviews relevant aspects of topologically twisted theories in the non-equivariant setting. Section \ref{sec:EqLocC2} reviews equivariant localization on $\mathbb{C}^2$ in the presence of hypermultiplets, while Section \ref{sec:EqLocP2} discusses the equivariant partition function for $\mathbb{CP}^2$. Section \ref{sec:ExResults} presents explicit results and discusses physical aspects. Section \ref{sec:compwuplane} concludes with a detailed comparison with non-equivariant results in the literature.

\subsection*{Acknowledgments}
We thank Elias Furrer, Osama Khlaif, Greg Moore, Sergey Mozgovoy,
Boris Pioline and Alessandro Tanzini for useful discussions and correspondence. The work of MD is supported by a Hamilton Scholarship.

\section{\texorpdfstring{$\mathcal{N}=2$}{N=2} Yang-Mills theories and their topological twists}
\label{sec:topN2YM}
 We review a few relevant aspects about $\CN=2$ supersymmetric
 Yang-Mills theories and their topological twists.

\subsection{Classical aspects}
\label{sec:classasp}
Throughout the gauge group is either $U(2)$, $SU(2)$ or $SO(3)$. The field content consists of the vector multiplet $(A_\mu,\phi,\Psi^I_\alpha,\bar \Psi^I_{\dot \alpha})$, $I=1,2$ and hypermultiplets $(q,\tilde q,\lambda_\alpha,\bar \lambda_{\dot \alpha},\chi_\alpha,\bar \chi_{\dot \alpha})$.
We distinguish six theories for $SU(2)$ or $SO(3)$: the four asymptotically free theories with
$0\leq N_f\leq 3$ hypermultiplets in the fundamental representation; and the two deformations of superconformal theories:
the theory with $N_f=4$ fundamental hypermultiplets, and the $\CN=2^*$ theory with a single adjoint hypermultiplet. Where convenient, we will use $N_f=*$ as a shorthand for the $\CN=2^*$ theory. 

Continuous parameters of the Yang-Mills theories are the hypermultiplet masses $m_j$, $j=1,\dots, N_f$, the complexified $\UV$-coupling $\tau_\UV$, $q_\UV=e^{2\pi i \tau_\UV}$ and the $\UV$-scale $\Lambda_\UV$. The product of $q_\UV$ and $\Lambda_\UV$ defines the dynamical scale $\Lambda$ of the theories for $N_f\leq 3$,
\be
\label{eq:dynscale}
\Lambda^{4-N_f}=\Lambda_\UV^{4-N_f}q_\UV.
\ee
We will add the subscript $N_f$, ie $\Lambda=\Lambda_{N_f}$, in case we want to stress that it is the scale of the theory with $N_f$ hypermultiplets. Since the superconformal theories, $N_f=4$ and $\CN=2^*$ do not give rise to a separate dynamical scale, we use $\Lambda$ as short-hand for $\Lambda_\UV$.

To use uniform notation for the six theories, we introduce the variable $q_{N_f}:=q$ as
\be
\label{eq:defq}
q=\left\{ \begin{array}{rr} q_\UV,&\qquad N_f=4,\,\, \CN=2^*, \\ 
 \Lambda^{(4-N_f)}, & \qquad N_f=0,\dots,3.
 \end{array} \right.
\ee
For $N_f=4$, $q_\UV$ is often parametrized in terms of a second coupling $\tau_{\SW}$ through \cite{Grimm:2007tm, Poghossian:2009mk}
\begin{equation}
\label{eq:qUVqSW}
    q_{\rm UV} = \frac{\vartheta_2(2\tau_{\rm SW})^4}{\vartheta_3(2\tau_{\rm SW})^4}, 
\end{equation}
with $\vartheta_j$ the classical Jacobi theta functions,
\begin{equation}
    \vartheta_2(\tau) = \sum_{n = -\infty}^\infty q^{(n + 1/2)^2/2}, \qquad \vartheta_3 (\tau) = \sum_{n = -\infty}^\infty q^{n^2/2},\qquad q=e^{2\pi i\tau}.
\end{equation}
The coupling $\tau_{\SW}$ is the complex structure of the $N_f=4$ Seiberg-Witten curve in the massless limit. From Eq. (\ref{eq:qUVqSW}), one can derive a $q_{\rm UV}$-series for $q_{\SW}$,
\begin{equation}
    16\,q_{\SW} = q_\UV + \frac{1}{2}\,q_\UV^2 + \frac{21}{64}\,q_\UV^3 + \frac{31}{128}\,q_\UV^4 + \CO(q_\UV^5).
    \label{eq:qSW_expand}
\end{equation}

  Throughout, we let $A_\mu$ be the connection of an $SO(3)$ gauge principal bundle $P$, with field strength $F_{\mu\nu}$. A central role will be played by the instanton or anti-self-dual solutions,
\be
\label{eq:ASD}
F^+=0.
\ee
The instanton number $k$ reads
\be
\label{eq:instnum}
k=-\frac{1}{8\pi^2}\int_X {\rm Tr}_{\bf 2}[F\wedge F],
\ee
where $X$ is the spacetime manifold.
The Donaldson-Uhlenbeck-Yau theorem states that such solutions on a compact, simply connected K\"ahler surface are in 1-to-1 correspondence with semi-stable vector bundles \cite{Donaldson1987, Uhlenbeck:1986ntc}. This connection provides an important technique for the evaluation of partition functions.
We note that the set of principal $SO(3)$ bundles contains the set of
$SU(2)$ bundles. The $SO(3)$ bundles with a vanishing second
Stiefel-Whitney class $w_2(P)\in H^2(X, \mathbb{Z}_2)$ can be lifted
to  a principal $SU(2)$ bundle, while $w_2(P)\neq 0$ provides the
obstruction to such a lift. If $w_2(P)$ is not an anti-self-dual two-form, the relevant analogue of \eqref{eq:ASD} are the Hermitean Yang-Mills equations.

Let $E$ be the rank 2 bundle associated to $P$ for the 2-dimensional, spinorial representation of $SO(3)$. If $w_2(P)\neq 0$, this bundle does not exist globally. There is no obstruction to lift an $SO(3)$ bundle to a principal $U(2)$ bundle $\tilde E$
by multiplying $E$ by the square root of a line bundle $\CL$ with
$c_1(\CL)=w_2(P) \mod H^2(X,2\mathbb{Z})$. The Chern classes of the
$U(2)$ bundle $\tilde E$ are $c_j(\tilde E)$ with $c_1(\tilde E)=w_2(P)\mod
H^2(X,2\mathbb{Z})$.  In terms of the Chern classes, the instanton number (\ref{eq:instnum}) reads
\be
\label{eq:kchern}
k=\int_X c_2(\tilde E)-\frac{1}{4}c_1(\tilde E)^2.
\ee

\subsection{Coulomb branches}
\label{sec:Coulbranch}
The Coulomb branch of the theory is the phase where the $SU(2)$ gauge symmetry is spontaneously broken to $U(1)$ by a vacuum expectation value of the vectormultiplet scalar $\phi$, which can be brought to the form,
\be
\phi=\left( \begin{array}{cc} a & 0 \\ 0 & -a \end{array}\right).
\ee
For $X=\BR^4$, the classical order parameter on the Coulomb branch is the variable $u$,
\be
u=\frac{1}{16\pi^2} \langle {\rm Tr}[\phi^2] \rangle_{\mathbb{R}^4}.
\ee
The non-perturbative prepotential fully specifies the low energy effective action on the Coulomb branch. In the notation of Ref. \cite{Aspman:2021vhs} with masses $(m_1,\dots,m_{N_f})$, for $N_f \leq 3$ the leading terms of the prepotential read
\be
\label{eq:prepotential}
\begin{split}
&F(a,m_j)=\frac{2i}{\pi}a^2 \left( \log(4a/\Lambda_{N_f}) - \frac{3}{2} \right)-\frac{1}{2}\sum_{j=1}^{N_f}\left( n_j\frac{m_j}{\sqrt{2}}a+ C(a^2+m_j^2/2) \right)\\
&\quad -\frac{i}{4\pi}\sum_{j=1}^{N_f} \left(a+\frac{m_j}{\sqrt{2}}\right)^2 \log\left((a+\tfrac{m_j}{\sqrt{2}})/\Lambda_{N_f} \right)+ \left(a-\frac{m_j}{\sqrt{2}}\right)^2 \log\left((a-\tfrac{m_j}{\sqrt{2}})/\Lambda_{N_f} \right)\\
&\quad + O(q),
\end{split}
\ee 
 where the integers $n_j$ are known as winding
 numbers. Single-valuedness of the $u$-plane integral requires
 $n_j=-1\mod 4$ \cite{Aspman:2021vhs}. The coefficient
 $C=\frac{1}{2}+\frac{i}{2\pi}\log(2)-\frac{3i}{2\pi}$ of the
 classical terms is fixed by considering the decoupling limit $m_j\to
 \infty$ to $N_f-1$ hypermultiplets. The $N_f=4$ and $\CN=2^*$ theory are superconformal, yet their perturbative partition function depends on $\Lambda$.
 The $O(q)$ in
 Eq. \eqref{eq:prepotential} represents the subleading polynomial terms in
 $\Lambda^{4-N_f}$ and $q_\UV$, which are the non-perturbative instanton
 contributions.

 The $\CN=2^*$ prepotential reads
 \be
 \begin{split}
&F(a,m)=\frac{1}{2}\tau_\UV a^2-\frac{1}{8\pi i } \left[ 2(2a)^2
  \log(2a/\Lambda)-(2a+m)^2\log((2a+m)/\Lambda)\right.\\
  &\qquad \left.-(2a-m)^2\log((2a-m)/\Lambda)-m^2\log(m/\Lambda)+\frac{9}{2}
  m^2 \right]+O(q).
\end{split}
 \ee
  
 The full non-perturbative expansion is determined by the Seiberg-Witten solution \cite{Seiberg:1994rs, Seiberg:1994aj} in terms of a family of  elliptic curves depending on the vacuum expectation value $u$. For general $N_f\leq 3$, $u$ can be obtained from the prepotential $F$ as
\be
u=\frac{4\pi i}{4-N_f}\Lambda_{N_f} \frac{\partial F(a,m_j)}{\partial \Lambda_{N_f}}+\frac{1}{4-N_f} \sum_j m_j^2,
\ee
while for $N_f=4$ and
\be
u=2\frac{\partial F(a,m)}{\partial \tau_{\UV}},
\ee
for $\CN=2^*$,
\be
u=2\frac{\partial F(a,m)}{\partial \tau_{\UV}}-\frac{m^2}{12}E_2(\tau_\UV),
\ee
with $E_2$ the $SL(2,\mathbb{Z})$ Eisenstein series of weight 2.

The $N_f=4$ theory is of special interest due to its electric-magnetic duality group involving triality, ie the action on the hypermultiplets of the outer automorphism group $S_3$ of the $SO(8)$ flavor symmetry combined with the $SL_2(\mathbb{Z})$ action on the coupling constant $\tau_\SW$ \cite{Seiberg:1994aj, Aspman:2021evt}.
The symmetry is generated by two generators $\CS$ and $\CT$. Their action on $\tau_\SW$ and the masses $m_j$ is,
\be
\label{eq:STtriality}
\begin{split}
&\CS:\qquad 2\tau_\SW \mapsto -\frac{1}{2\tau_\SW},\qquad \left( \begin{array}{c} m_1 \\ m_2 \\ m_3 \\m_4  \end{array} \right)\mapsto \frac{1}{2}\left(\begin{array}{c} m_1 +m_2+m_3+m_4 \\ m_1+m_2-m_3-m_4 \\ m_1-m_2+m_3-m_4 \\m_1-m_2-m_3+m_4  \end{array} \right),\\
\vspace{-.3cm} &\\
&\CT: \qquad 2\tau_{\SW}\mapsto 2\tau_{\SW}+1,\qquad \left(\begin{array}{c} m_1 \\ m_2 \\ m_3 \\m_4  \end{array} \right) \mapsto \left( \begin{array}{c} m_1 \\ m_2 \\ m_3 \\-m_4  \end{array} \right).
\end{split}
\ee
By \eqref{eq:qUVqSW}, $\CS$ and $\CT$ act on $q_\UV$ as
\be
\begin{split}
\label{eq:STtrialityqUV}
&\CS:\qquad q_\UV \to 1-q_\UV,\\
&\CT:\qquad q_\UV\to -\frac{q_\UV}{1-q_\UV}.
\end{split}
\ee

Finally, the partition functions of $\CN=2^*$ theory transform as a vector-valued modular form for the standard $SL_2(\mathbb{Z})$ action on $\tau_\UV$ \cite{Vafa:1994tf, Labastida:1998sk, Manschot:2021qqe}.

\subsection{Topological twisting}
\label{sec:toptwist}
We apply the topological twist to formulate the above theory on a smooth, compact four-manifold $X$. The action of the twist on the supersymmetry generators results in a scalar supercharge $Q$ \cite{Witten:1988ze}. Observables of the topological theory are elements of the $Q$-cohomology. We let $L$ be the lattice $H^2(X,\mathbb{Z})$ modulo torsion. A lift of an element of $H^2(X,\mathbb{Z})$ to $L$ is denoted by an overline. 
We set
\be 
2\bfmu=\bar w_2(P),
\ee 
with $\bar w_2(P)\in L$ the lift of the second Stiefel-Whitney class $w_2(P)\in H^2(X,\mathbb{Z}_2)$ of the gauge principal bundle $P$. 

Furthermore for $1\leq N_f\leq 4$, $\CL_j$ are the flavor line bundles with Chern class $c_1(\CL_j)$. We will also use $\bfk_j$ to denote the background flux, related to the lift $\bar c_1$ of $c_1$ by 
\be
\label{eq:defbfk}
2\bfk_j=\bar c_1(\CL_j).
\ee
The topological twist reorganizes the hypermultiplet scalars into chiral spinors. The spinor of hypermultiplet $j$ is a section of
\be
S^+\otimes E \otimes \CL_j^{1/2},
\label{eq:Spinc_bundle}
\ee
with $ S^+$ the chiral spin bundle and $E$ the rank 2 gauge bundle introduced above Eq. \eqref{eq:kchern}. The product bundle \eqref{eq:Spinc_bundle} must be globally well-defined for the proper formulation of the hypermultiplet fields. Therefore the Chern classes $c_1(\CL_j)$ need to satisfy \cite{Aspman:2022sfj}
\be
\label{eq:quantcond}
c_1(\CL_j) \in w_2(X)+w_2(P)\mod H^2(X,2\mathbb{Z}).
\ee
For the $\CN = 2^\ast$ theory with an adjoint hypermultiplet, the quantization condition for $c_1(\CL)$ is in this case \cite{Manschot:2021qqe}
\be
c_1(\CL)=w_2(X) \mod H^2(X,2\mathbb{Z}),
\label{eq:quantcond_2*}
\ee
independently of $w_2(P)$. Partition functions for canonical background flux $c_1(\CL)=K_X$ are equivalent to the Vafa-Witten twist of $\CN=4$ Yang-Mills \cite{Vafa:1994tf}, while other choices of background flux are known as $L$-twisted Vafa-Witten theory in the mathematics literature \cite[Section 6]{Tanaka:2017jom}. 

Once a consistent formulation of the theory is chosen, the partition function will localize to the solution space $\CM_k^{Q,N_f}$ of the $Q$-fixed equations, called the \textit{non-Abelian monopole equations} \cite{ Hyun:1995hz, Hyun:1995mb, Labastida:1995gp,  Labastida:1995zj}\footnote{A closely related set of equations, the so-called “SO(3) monopole equations,” was studied in the mathematical literature \cite{pidstrigach1995localisation, Bradlow:1996alg, Teleman:1996, okonek1996recent, Feehan:1997gj}.}: 
\begin{equation}
        \label{eq:nonab_mono_eqs}
        \begin{split}
        &\left( F_{\dot{\alpha} \dot{\beta}}^a \right)^+ + \frac{i}{2} \sum_{j=1}^{N_f} \bar{M}^j_{(\dot{\alpha}} T^a M^j_{\dot{\beta})} = 0, \\
        &\slashed{D} M^j = 0,
        \end{split}
\end{equation}
with $T^a$ the generators of $\mathfrak{su}(2)$ in the 2-dimensional fundamental representation, $M^j_{\dot \alpha}$ the
hypermultiplet spinors, $j = 1, \dots, N_f$, and $\slashed{D}_j$ the Dirac operator for sections $M^j_{\dot \alpha}$ of the bundle \eqref{eq:Spinc_bundle}. For $\CN=2^*$, $T^a$ is a generator in the adjoint representation.

The virtual dimension of the moduli space of non-Abelian monopoles with gauge group $SU(2)$ and $N_f$ monopole fields is
\be
{\rm vdim}_\BC(\CM^{Q,N_f}_{k})=(4-N_f)k+\frac{1}{4}\left( -3\chi -(3+N_f)\sigma+\sum_{j=1}^{N_f} c_1(\CL_j)^2\right).
\label{eq:vdim_def}
\ee
The non-Abelian monopole equations for the adjoint hypermultiplet of the $\CN=2^*$ theory have a similar form \cite{Labastida:1998sk, Manschot:2021qqe}. The corresponding dimension of $\CM^{Q,*}_{k}$ is
\be
{\rm vdim}_\BC(\CM^{Q,*}_{k})=3\frac{c_1(\CL)^2-2\chi-3\sigma}{8}.
\ee

The correlators of the relevant topological observables $\CO$ for $N_f$ are a $U(1)^{N_f}$ equivariant integral over $\CM^{Q, N_f}_{k}$. This integral can be localized to different fixed point components of $U(1)^{N_f}$, ie the instanton component with vanishing monopole fields, $M^j_{\dot \alpha}=0$, and monopole components with $M^j_{\dot{\alpha}} \neq 0$ \cite{pidstrigach1995localisation, Hyun:1995hz}. This paper focuses on the toric manifold $\mathbb{CP}^2$, and the equivariant partition functions all result from integrals over the instanton component $\CM_k$ . 

The hypermultiplets do give rise to non-trivial characteristic classes in the integral over $\CM_k$. Each hypermultiplet gives rise to an index bundle $W_{k,j}$ of the Dirac operator $\slashed{D}_j$, $j=1,\dots, N_f$, over the moduli space of instantons $\CM_{k}$. A fibre of $W_{k,j}$ can be thought of as the solution space for the $j$-th hypermultiplet spinor Dirac equation in the background of the basepoint instanton. Note that anti-self-dual instantons $F=-*F$ form a subset of the solution space of the non-Abelian monopole equations. The rank of the index bundle is
\be
{\rm rk}(W_{k,j})=-k+\frac{1}{4}(c_1(\CL_j)^2-\sigma),
\ee 
which is an integer given the condition (\ref{eq:quantcond}). A correlation function of $Q$-closed observables then reads
\cite{LoNeSha, Aspman:2022sfj, Malmendier:2008db}
\begin{equation}
    \label{eq:topSQCD_correlator}
    \langle \CO_1 \dots \CO_p \rangle = \sum_k \Lambda_{N_f}^{{\rm vdim} \left( \CM^{Q, N_f}_{k}\right)} \int_{\CM_{k}} {\rm Eu}(-W_k)\,\mu_{\rm D} (\CO_1) \wedge \dots \wedge \mu_{\rm D}(\CO_p),
\end{equation}
with ${\rm Eu}(-W_k)$ the Euler class of (minus) the total index bundle and $\mu_{\rm D}$ is the Donaldson $\mu$-map. 

If we specialize the four-manifold $X$ to an algebraic surface $S$, we can state this more formally in algebraic geometry as an integral over the moduli space $M_k$ of semi-stable sheaves. To this end, we let $\alpha$ be a K-theory class on $S$ whose Chern classes are related to $\bfk_j$ (\ref{eq:defbfk}) as \cite{Furrer:2026byd},
\be
\begin{split}
&r(\alpha)=N_f, \\
&c_1(\alpha)=\frac{N_f K_S+2\sum_j \bfk_j}{2}, \\
&c_2(\alpha)=\frac{N_f(N_f-1)}{8}K_S^2+\frac{N_f-1}{2} B(K_S,\sum_j\bfk_j)+\sum_{i<j} \bfk_i\bfk_j, 
\end{split}
\ee
then without observables
\be
\int_{M_k} c(\alpha_M),
\ee
with $\alpha_M$ a K-theory class on $M_k$, whose Chern character ${\rm ch}(\alpha_M)$ is defined by \cite{Gottsche:2020ass}
\be
{\rm ch}(\alpha_M)={\rm ch}(-\pi_{M!}(\pi^*_S \alpha \cdot E \cdot {\rm det}(E)^{-1/2})).
\ee

For the $N_f=4$ theory, the action of the triality group (\ref{eq:STtriality}) on the background classes $\bfk_j$ \eqref{eq:defbfk} follows the action on the hypermultiplets and is therefore identical to the action on the masses. We thus have
\be
\label{eq:actionTBj}
\begin{split}
&\CS:\qquad \left( \begin{array}{c} \bfk_1 \\ \bfk_2 \\ \bfk_3 \\ \bfk_4  \end{array} \right)\mapsto \frac{1}{2}\left(\begin{array}{c} \bfk_1 +\bfk_2+\bfk_3+\bfk_4 \\ \bfk_1+\bfk_2-\bfk_3-\bfk_4 \\ \bfk_1-\bfk_2+\bfk_3-\bfk_4 \\ \bfk_1-\bfk_2-\bfk_3+\bfk_4  \end{array} \right),\\
&\CT: \qquad \left(\begin{array}{r} \bfk_1 \\ \bfk_2 \\ \bfk_3 \\\bfk_4  \end{array} \right) \mapsto \left( \begin{array}{c} \bfk_1 \\ \bfk_2 \\ \bfk_3 \\-\bfk_4  \end{array} \right).
\end{split}
\ee

\section{Equivariant localization on \texorpdfstring{$\mathbb{C}^2$}{C\texttwosuperior}}
\label{sec:EqLocC2}
We continue with a review of equivariant localization in the $\Omega$-background for evaluating the equivariant partition function $   Z_{\BC^2}(a, \epsilon_i)$ on $\mathbb{C}^2$ \cite{Moore:1997dj, Moore:1998et, Lossev:1997bz, Ne}. The $\Omega$-background involves the toric $U(1)^2$ action on  $\BC^2$ with equivariant parameters $\epsilon_1, \epsilon_2 \in \BR$.  The leading
term in the $(\eps_1, \eps_2)$-expansion  corresponds to the classical prepotential $F$ \cite{Ne,
  NekOk},
\be
F(a)= \lim_{\eps_i \to 0} \eps_1\eps_2 \log\left(     Z_{\BC^2}(a,\epsilon_i) \right).
\ee
Using this correspondence, non-equivariant observables can be derived
from the equivariant partition function \cite{Manschot:2019pog}. The equivariant partition function $Z_{\BC^2}$ is the partition function for gauge group $U(N)$ determined using the ADHM solutions. Fixing $N=2$, the $U(2)$ gauge group of $Z_{\BC^2}$ differs from the $SU(2)$ gauge group of Section \ref{sec:topN2YM} by a $U(1)$. In the $U(2)$ theory, $\phi$ contains two independent diagonal elements $a_1$ and $a_2$, which can be adapted classically to the $SU(2)$ theory by simply setting $a_1=-a_2$. We use the convention that $a=a_1=-a_2$.\footnote{Note that some papers, including Ref. \cite{Bonelli:2020xps}, use the alternative convention $a \coloneqq a_1 - a_2 \overset{SU(2)}{=} 2 a_1$.} Besides this identification of classical parameters, there is an extra factor to be divided out for $N_f=2,3,4$ \cite{Alday:2009aq, Poghossian:2009mk, Manschot:2019pog}. We comment on this in more detail in Section \ref{sec:InstPart}.

The partition function for the theory with $N_f$ hypermultiplets depends besides $\eps_i$ on the local coordinate $a$, the masses $M_j$ and the appropriate instanton counting parameter, either $q_\UV$ or the dynamical scale $\Lambda$ as in \eqref{eq:defq}.\footnote{The mass parameters $M_j$ in the equivariant setting, and $m_j$ in the non-equivariant, low energy setting differ by a numerical factor. We determine these factors in Section \ref{sec:compwuplane}. The mass parameter $M_j$ is not to be confused with the $j$-th hypermultiplet spinor $M^j_{\dot \alpha}$ of Section \ref{sec:toptwist}.}  It is given by the product of the perturbative partition function $Z^{(N_f)}_{\BC^2, \text{pert}}$ and the instanton partition function $Z^{(N_f)}_{\BC^2, \text{inst}}$,
\begin{equation}
    Z^{(N_f)}_{\BC^2}(a,\Lambda,q,   M_j, \eps_i) =  Z^{(N_f)}_{\BC^2,\text{pert}}(a,\Lambda,q,M_j, \eps_i)\,Z^{(N_f)}_{\BC^2,\text{inst}}(a,q, M_j,\eps_i).
    \label{eq:C2factorisation}
\end{equation}
In the following subsections, we review both $Z^{(N_f)}_{\BC^2,\text{pert}}$ and $Z^{(N_f)}_{\BC^2,\text{inst}}$ for theories with fundamental or adjoint hypermultiplets. 
To avoid overly heavy notation, we will sometimes suppress some of the arguments of the partition function.

\subsection{The perturbative partition function}
The perturbative contribution reduces to a product of contributions from the vector and hypermultiplets \cite{NekOk, Nakajima:2003uh, Manschot:2019pog}, 
\begin{equation}
   \label{eq:Zpert_C2}
   \begin{split}
    Z^{(N_f)}_{\BC^2,{\rm pert}}(a,\Lambda,q, M_j, \eps_i) &= Z_{\BC^2,{\rm pert, vec}}(a,\Lambda,\eps_i)\,Z^{(N_f)}_{\BC^2,{\rm pert, hyp}} (a, \Lambda,M_j, \eps_i)\\
    &\quad \times \left\{ \begin{array}{rr} 1,&\qquad N_f=1,2,3, \\ (q_\UV)^{-\frac{a^2}{\epsilon_1\epsilon_2}}, & \qquad N_f=4,\,\, \CN=2^*,\end{array}\right.
    \end{split}
\end{equation}    
where
\begin{equation}
Z^{(N_f)}_{\BC^2,{\rm pert, hyp}} (a, \Lambda,M_j, \eps_i)=\left\{\begin{array}{rr}
    \prod_{j=1}^{N_f} Z_{\BC^2,{\rm pert, fund}} (a, \Lambda,M_j,\eps_i), & \quad N_f\leq 4,\\
    Z_{\BC^2,{\rm pert, adj}} (a, \Lambda,M,\eps_i), & \quad \CN=2^*,
    \end{array}\right. 
\end{equation}
with\footnote{\label{note:diffpert}We follow the definition for the perturbative part of the vector multiplet as in \cite{Nakajima:2003uh}\cite{Bershtein:2015xfa}, which differs slightly from \cite[Eq. (3.3)]{Bonelli:2020xps} by a shift in the argument of $\gamma_{\epsilon_1,\epsilon_2}$.}
\be
\begin{split}
    Z_{\BC^2,{\rm pert, vec}}(a,\Lambda,\varepsilon_i) &= \exp \left[ -\gamma_{\epsilon_1, \epsilon_2} ( 2a ; \Lambda) - \gamma_{\epsilon_1, \epsilon_2} (-2a ; \Lambda) \right], \label{eq:Z1vecfund} \\
    Z_{\BC^2,{\rm pert, fund}}(a,\Lambda,M,\varepsilon_i) &= \exp \left[ \gamma_{\epsilon_1, \epsilon_2} (a + M-\varepsilon_+; \Lambda) + \gamma_{\epsilon_1, \epsilon_2} (-a + M-\varepsilon_+; \Lambda)\right], \\
    Z_{\BC^2,{\rm pert, adj}}(a,\Lambda,M,\varepsilon_i)&=\exp \left[ \gamma_{\epsilon_1, \epsilon_2} (2a + M-\varepsilon_+; \Lambda) + \gamma_{\epsilon_1, \epsilon_2} (-2a + M-\varepsilon_+; \Lambda)\right. \\
    &\qquad \quad  \left. + \,\gamma_{\epsilon_1, \epsilon_2} (M-\varepsilon_+; \Lambda)\right].
\end{split}
\ee
$\gamma_{\epsilon_1, \epsilon_2} (x; \Lambda)$ is defined in Eq. (\ref{eq:Defgammaeps}) and reviewed in Appendix \ref{App:gamma}, and where we have used
\begin{equation}
    \epsilon_+ = \frac{\epsilon_1 + \epsilon_2}{2}.
\end{equation}
 Eq. \eqref{eq:gammaepsExp} demonstrates that the leading terms of $\gamma_{\epsilon_1, \epsilon_2}$ reproduces the logarithmic terms in the prepotential $F(a,m_j)$ (\ref{eq:prepotential}).
\par
In \eqref{eq:Zpert_C2}, the factor $(q_\UV)^{-\frac{a^2}{4\epsilon_1\epsilon_2}}$ for $N_f = 4$, $\CN=2^\ast$ amounts to the ``classical part'' of the equivariant partition function. For $N_f = 1,2,3$, we have implicitly absorbed this term into a redefinition of the dynamical scale $\Lambda$ using \eqref{eq:gammaepsLambdashift}, while we have used $\Lambda$ to denote $\Lambda_\UV$ for the $N_f = 4$ and $\CN=2^\ast$ theory.

\subsection{The instanton part}
\label{sec:InstPart}
The $U(1)^2$ action on $\BC^2$  lifts to an action on the moduli space of instantons on $\mathbb{C}^2$. The fixed point locus $\mathfrak{F}^{\text{point}}$ of this action corresponds to pointlike instantons located at the origin of $\mathbb{C}^2$. 
For gauge group $SU(N)$, the fixed points are labelled by a set $\vec Y = \{ Y_1, Y_2,\dots, Y_N \}$ of $N$ Young diagrams $Y_j$ \cite{Ne}\cite{Bruzzo_2003}. For each Young diagram $Y=(\lambda_1\geq \lambda_2 \geq \dots \geq 0)$, we denote the dual Young diagram by $Y^T=(\lambda_1'\geq \lambda_2' \geq \dots \geq 0)$ with total number of boxes $|Y|=\sum_\ell \lambda_\ell$. The total number of boxes in $\vec Y$ is $|\vec Y|=\sum_{j=1}^N |Y_j|$, and corresponds to the instanton number. We will henceforth restrict to $N=2$.

We denote the equivariant instanton partition function on $\mathbb{C}^2$ for $N_f$ hypermultiplets by $Z^{(N_f)}_{\rm inst}(a,q,M_j,\epsilon_i)$. It is a sum over the doublet $\vec Y$ of Young diagrams, 
\begin{equation}
    \label{eq:Zinst_iterative}
    Z^{(N_f)}_{\BC^2, \text{inst}}(a, q,M_j, \epsilon_i) = \sum_Y   Z^{(N_f)}_{\vec Y}(a,\Lambda_\UV,M_j,\epsilon_i)\,q_\UV^{|\vec Y|},
\end{equation}
where the summand $Z^{(N_f)}_{\vec Y}$ factorizes as a product over the different multiplets of the theory,
\begin{equation}
Z^{(N_f)}_{\vec Y}(a,\Lambda_\UV,M_j,\epsilon_i)=Z_{\vec Y, {\rm vec}}(a,\Lambda_\UV,\epsilon_i)\prod_{j=1}^{N_f} Z_{\vec Y, {\rm hyp}}(a,\Lambda_\UV,M_j,\epsilon_i).
\end{equation}
Depending on the theory, $Z_{\vec Y, \text{hyp}}$ is the factor due to $N_f$ fundamental hypermultiplets, or adjoint hypermultiplet factor. The dependence on $\Lambda_\UV$ and $q_\UV$ on the right hand side of \eqref{eq:Zinst_iterative} reduces to dependence on $q$ \eqref{eq:defq}.

The contribution of the vector multiplet can be calculated explicitly as \cite{Moore:1997dj, Moore:1998et, Lossev:1997bz, Ne, Poghossian:2009mk, Bruzzo_2003, FlumPogh02AnAlgorithm} 
\be
    \label{eq:Zinst_iterative_vec}
    Z_{\vec Y, {\rm vec}}(a,\Lambda_\UV,\epsilon_i) = \prod_{\alpha, \beta = 1}^2 \prod_{s\in Y_\alpha} \frac{\Lambda_\UV^2}{E_{\alpha, \beta}(a_\alpha, a_\beta,s)(\epsilon_1 + \epsilon_2 - E_{\alpha, \beta}(a_\alpha, a_\beta,s))}, 
\ee    
where for a given pair $\vec Y$,  $E_{\alpha,\beta}: \mathbb{C}^2 \times \mathbb{N}^2\to \mathbb{C}$ is the function\footnote{We use expressions consistent with \cite{Huang:2011qx, FlumPogh02AnAlgorithm}. We expect that there is a misprint in \cite[Eq. (3.3)]{Poghossian:2009mk} in the subscripts of $Y$. This is irrelevant for $N_f=0$, but does affect results for $N_f>0$. That is to say, we find that the recursion formula (\ref{eq:InstRecRel_Nf1}) below does not hold with \cite[Eq. (3.3)]{Poghossian:2009mk}.
},
\be 
\label{eq:DefE}
    E_{\alpha, \beta}(x, y, s) = y - x - \epsilon_1 L_{Y_\alpha}(s) + \epsilon_2 (A_{Y_\beta}(s) + 1).
\ee
Here $A_{Y}(s)$ and $L_{Y}(s)$ denote the arm length, respectively leg length, of the box $s=(i,j)$,
\be
A_Y(s)=\lambda_i-j,\qquad L_Y(s)=\lambda_j'-i.
\ee
Note that for a box $s \in Y$ the arm and leg length are both positive. On the other hand if $\alpha\neq \beta$ for $E_{\alpha,\beta}$ (\ref{eq:DefE}), $l_{Y_\beta}(s)<0$   if the box $s\in Y_\alpha$ does not correspond to a box in $Y_\beta$. \par 
Further, note that $Z_{(Y_1, Y_2), {\rm vec}} (a,\Lambda,-\epsilon_i) = Z_{(Y_2, Y_1), {\rm vec}} (a,\Lambda,\epsilon_i)$. Hence, the partition function of the pure theory, $Z^{(0)}_{\rm inst}$, is invariant under $(\epsilon_1, \epsilon_2) \mapsto (-\epsilon_1, -\epsilon_2)$.

\subsubsection*{Fundamental hypermultiplet}
The contribution of the fundamental hypermultiplet is given by \cite{Ne, NekOk, Poghossian:2009mk}\footnote{Note that relative to \cite{Poghossian:2009mk}, we have shifted the masses as $m \to M + \frac{\epsilon_1 +  \epsilon_2}{2}$. Ref. \cite{Poghossian:2009mk} gives this as the expression for the anti-fundamental representation, and that the fundamental representation is given by the substitution $m\mapsto \epsilon_1+\epsilon_2-m$. This corresponds to $M\mapsto -M$ in our convention.
}
\begin{align}
\label{eq:Zinst_iterative_fund}
    Z_{\vec Y, {\rm fund}} (a, \Lambda_\UV,M, \epsilon_i) &= \prod_{\alpha=1}^2 \prod_{s_\alpha \in Y_\alpha} \frac{\chi_{\alpha, s_\alpha} + M}{\Lambda_\UV} \\
    \chi_{\alpha, s_\alpha} &= a_\alpha + (i_{s_\alpha} - \tfrac{1}{2}) \epsilon_1 + (j_{s_\alpha} -\tfrac{1}{2})\epsilon_2.
\end{align}
Here $i_{s_\alpha}, j_{s_\alpha}$ denote the column, respectively row of the box $s_\alpha$ in the respective Young diagram.

The overall power of $\Lambda_\UV$ can be factored out of the product over $s\in Y_\alpha$ and $Y_\beta$, which gives rise to the factor
\be
\label{eq:instpartC2_totalLambda}
\left(\Lambda_\UV^{4-N_f}\right)^{|\vec Y|},
\ee 
which combines with the factor $q_\UV^{|\vec Y|}$ to $q^{|\vec Y|}$. Indeed, $|\vec Y|$ corresponds to the instanton number. The expansion of $Z_{\BC^2,\text{inst}}$ in powers of $q$ can now be calculated explicitly. For example, for $N_f = 1$, we have to second order
\begin{equation}
\label{eq:ZNf1firstterms}
Z^{(1)}_{\BC^2,\text{inst}} (a,  q,M, \epsilon_i) = 1 + \frac{2\,M\,q}{\epsilon_1 \epsilon_2 ((\epsilon_1 + \epsilon_2)^2 - (2a)^2 )} + \dots.
\end{equation}

\subsubsection*{Adjoint hypermultiplet}
The contribution from the adjoint multiplet is given by \cite{Huang:2011qx, Bershtein:2015xfa},\footnote{\label{fnsignM}We note that the formula for $Z_{\vec Y, {\rm adj}}$ in \cite[]{Huang:2011qx} differs by the sign of $a_\beta-a_\alpha$ from that in \cite[Eq. (4.4)]{Bershtein:2015xfa}, which is important once combined with the perturbative part. We follow \cite{Huang:2011qx} with $E_{\alpha,\beta}(a_\alpha,a_\beta,s)=\eps_1+\eps_2-E^{Y_\alpha Y_\beta}_{i,j}(a_\alpha-a_\beta)$ in terms of $E^{Y_\alpha Y_\beta}_{i,j}$ \cite[(A.1)]{Huang:2011qx}. Moreover with this identification, \cite[Eq. (A.3)]{Huang:2011qx} has the opposite sign for the mass $M$ compared to our formula for $Z_{\vec Y, {\rm adj}}$. This sign is crucial since with the appropriate regularization of \eqref{eq:gammalog} for toric localization on $\mathbb{CP}^2$, the perturbative prepotential is not invariant under a change of sign of $M$. 
We find that the opposite sign of $M$ in $Z_{\vec Y, {\rm adj}}$ results in singularities for the non-equivariant limit, $\epsilon_i\to 0$, in the results for $\mathbb{CP}^2$. Also note that $\gamma(x)$ \cite[(A.4)]{Huang:2011qx} corresponds to $\gamma_{-\epsilon_1,-\epsilon_2}(x;1)$.}
\be
    \label{eq:Zinst_adj}
    \begin{split}
&Z_{\vec Y, {\rm adj}}(a,\Lambda_\UV,M,\epsilon_i) = \\ &\qquad \prod_{\alpha, \beta = 1}^2 \prod_{s\in Y_\alpha} \frac{E_{\alpha, \beta}(a_\alpha, a_\beta+M-\eps_+,s)(\epsilon_1 + \epsilon_2 - E_{\alpha, \beta}(a_\alpha+M-\eps_+, a_\beta ,s))}{\Lambda_\UV^2}.
\end{split}
\ee   

\subsubsection*{Recursion formulae for \texorpdfstring{$N_f = 1, 2, 3,4$}{Nf = 1,2,3,4} and \texorpdfstring{$\CN=2^*$}{N=2^*}}

Remarkably, there exist recursion formulae for $Z_{\rm inst}$ \cite{Poghossian:2009mk}, such that it is not necessary to determine $Z_{\rm inst}$ through the sum over partitions as discussed above.
These relations are originally derived using the connection to recursion relations for conformal blocks \cite{Zamolodchikov:1984eqp} in conformal field theory through the AGT correspondence \cite{Alday:2009aq}. Whereas the sum over partitions typically leads to large expressions with many spurious poles in $a$, the recursion formulae provide more compact expressions without spurious poles.
Since the pole structure of  $Z_{\mathbb{C}^2}$ is more manifest from the recursion relation, it simplifies the evaluation of the contour integral for the partition function on $\mathbb{CP}^2$. 

To state the recursion formulae, we start by introducing the set of functions $C^{(N_f)}, N_f=0,\dots,4,*$ by
\be
\label{eq:instrecrel_def_C}
\begin{split}
&C^{(4)}(a, q_{\rm UV},M_j,\epsilon_i) =  \left( \frac{q_{\rm UV}}{16\, q_{\rm SW}}\right)^{\frac{a^2}{\epsilon_1 \epsilon_2}} (1-q_\UV)^{\frac{1}{4\epsilon_1 \epsilon_2} (\epsilon_1 + \epsilon_2 + \sum_{j=1}^4 M_j)^2} \\
&\qquad \qquad \qquad \times   \vartheta_3(2\tau_{\SW})^{-\frac{(\epsilon_1 + \epsilon_2)^2}{\epsilon_1 \epsilon_2} + \frac{2}{\epsilon_1 \epsilon_2} \sum_{j=1}^4 M_j^2 },
\\
&C^{(3)}(\Lambda,M_j,\epsilon_i)= \exp\!\left(-\frac{ \Lambda}{2\epsilon_1 \epsilon_2}\left(  \frac{\Lambda}{32} + \epsilon_1 + \epsilon_2 + \sum_{j=1}^3 M_j\right) \right),\\
&C^{(2)}(\Lambda, \epsilon_i)= \exp\!\left( -\frac{\Lambda^2}{2\epsilon_1 \epsilon_2}\right),\\
&C^{(0)} = C^{(1)} =  1,\\
&C^{(*)}(q_\UV,M, \epsilon_i)=\left(\prod_{n=1}^\infty (1-q_{\UV}^n)\right)^{-\frac{2(M+(\epsilon_1-\epsilon_2)/2)(M-(\epsilon_1-\epsilon_2)/2)}{\epsilon_1\epsilon_2}},
\end{split}
\ee
where for $N_f=4$, $\tau_\UV$ and $\tau_\SW$ are related by \eqref{eq:qUVqSW}. 

The functions $C^{(N_f)}$ for $N_f<4$ follow from the decoupling limit from $C^{(4)}$. Note that only $C^{(4)}$ depends on the Coulomb parameter $a$. The factor $C^{(4)}$ occurs naturally in the recursion relation for conformal blocks in 2-dimensional CFT \cite{Zamolodchikov:1984eqp, Poghossian:2009mk}. 
We furthermore introduce the shorthand
\begin{align}
    \label{eq:apm}
       a_{\pm}^{m,n} &= \frac{m\epsilon_1 \pm n\epsilon_2}{2}.
\end{align} Then for later calculations, note the identity
\be
\label{eq:C4_factor_a_out}
C^{(4)}(a^{m,n}_+,  q_\UV, M_j)=\left(\frac{q_\UV}{16\,q_\SW} \right)^{mn}C^{(4)}(a^{m,n}_-, q_\UV,M_j ).
\ee
We now introduce the normalized partition functions,
\be
\label{eq:defH}
H^{(N_f)}_{\BC^2}(a,q,M_j,\epsilon_i)=\frac{Z_{\BC^2, {\rm inst}}^{(N_f)}(a,q,M_j,\epsilon_i)}{C^{(N_f)}}. 
\ee 
Based on our explicit results and comparison with results from the $u$-plane integral in Sections \ref{sec:ExResults} and \ref{sec:compwuplane}, we interpret $H^{(N_f)}_{\BC^2}$ as the partition function of $SO(3)$ or $SU(2)$ theory on $\mathbb{C}^2$. 
This is familiar from previous instances in the literature \cite{Alday:2009aq, Poghossian:2009mk, Manschot:2019pog}, although the precise factor is slightly different from 
the references. It is clear from the transformations (\ref{eq:STtriality}) and (\ref{eq:STtrialityqUV}) that $C^{(4)}$ is not invariant under the action of the triality group \eqref{eq:STtriality}. In Section \ref{sec:EqLocP2} and \ref{sec:resultsNf4}, we discuss this in more detail for the factor $D^{(4)}$ for compact four-manifolds.

We introduce the product for $N_f=1,2,3,4$,
\begin{align}
    \label{eq:InstRecRel_Nf1_Rmn}
    \CR^{(N_f)}_{m,n} (M_j,\epsilon_i) &= \frac{2{\prod'}_{r, s}  \prod_{j=1}^{N_f} \ \frac{1}{2}[2M_j - r\epsilon_1 - s\epsilon_2]}{{\prod''}_{k,l} \ k\epsilon_1 + l \epsilon_2},
 \end{align}
where the primed product $\prod'_{r,s}$ runs over the indices 
\be
\begin{split}
    r &= -m + 1, -m + 3, \dots, m -1,\\
    s &= -n + 1, -n + 3, \dots, n-1,
    \label{eq:instrec_primedset}
\end{split}
\ee
and the double-primed product $\prod''_{k,l}$ runs over 
\be
\begin{split}
    &k = -m+1, -m+2, \dots, m-1, m, \\
    &l = -n+1, -n+2, \dots, n-1, n, \\
    &(k,l) \notin \{ (0,0), (m,n) \}.
    \label{eq:instrec_doubleprimedset}
\end{split}
\ee
Moreover, for the $\CN=2^*$ we set 
\be
\label{eq:CR*}
\begin{split}
\CR^{(*)}_{m,n}(M,\epsilon_i)&=2\,\frac{\prod_{r=-m+1}^{m}\prod_{s=-n+1}^{n}M+(r - \frac{1}{2})\epsilon_1+(s - \frac{1}{2})\epsilon_2}{\prod''_{k,l} k\epsilon_1+l\epsilon_2}.
\end{split}
\ee
The recursion formula for $N_f\leq 4,*$ then takes the universal form \cite{Poghossian:2009mk, Zamolodchikov:1984eqp},
\be
    \label{eq:InstRecRel_Nf1}
    \begin{split}
&    H^{(N_f)}_{\BC^2}(a, q,M_j,\eps_i) \\
&\qquad = 1  - \frac{1}{4}\sum_{m,n = 1}^\infty \frac{X^{mn}}{a^2 - \left(a_+^{m,n} \right)^2}\,   \CR^{(N_f)}_{m,n}(M_j,\eps_i)\, H^{(N_f)}_{\BC^2}( a_-^{m,n}, M_j,q,\eps_i),
\end{split}
\ee
where
\be
X=\left\{\begin{array}{rr} 16\,q_{\SW},&\qquad N_f=4,\\
\Lambda^{4-N_f},& \qquad N_f=3,2,1,0,\\
q_\UV,&\qquad \CN=2^*.\end{array} \right.
\label{eq:defX}
\ee
Since $H^{(N_f)}_{\BC^2}=1+O(q)$ by \eqref{eq:defH}, this completely determines $H^{(N_f)}_{\BC^2}$. 

The recursion formulae demonstrate that the structure of the poles of $Z^{(N_f)}_{\BC^2, \text{inst}} (a,  q,M_j)$ is simpler than that of the individual terms of the combinatorial expression (\ref{eq:Zinst_iterative}). Namely, $Z^{(N_f)}_{\BC^2, \text{inst}}$ has single order poles at $a = \pm a_+^{m,n}$, for $m,n \geq 1$. The position of the poles is thus independent of the masses $M_j$.

\subsubsection*{Symmetries of \texorpdfstring{$Z^{(N_f)}_{\BC^2, \rm inst}$}{Zinst}}
The instanton partition functions $Z^{(N_f)}_{\BC^2, \rm inst}$, as well as $H^{(N_f)}_{\BC^2}$ satisfy various identities:
\begin{itemize}
\item Reflection symmetry of $a$:
\begin{equation}
    \label{eq:HC2_a_flip}
    H^{(N_f)}_{\BC^2} (-a,  q, M_j,\epsilon_i) = H^{(N_f)}_{\BC^2} (a,  q,M_j, \epsilon_i),
\end{equation}
and
\begin{equation}
    \label{eq:inst_a_flip}
    Z^{(N_f)}_{\BC^2,\rm inst} (-a,  q,M_j, \epsilon_i) = Z^{(N_f)}_{\BC^2, \rm inst} (a,  q,M_j, \epsilon_i),
\end{equation}
which is simply a consequence of this reflection being an element of the gauge group. 
\item Reflection symmetry for $\varepsilon_j$ for $N_f = 0,1,2, 3, 4, \ast$:
\begin{equation}
    \label{eq:HC2_epsilon_flip}
    H^{(N_f)}_{\BC^2}(a,  q, M_j,-\epsilon_i) = H^{(N_f)}_{\BC^2}(a,  q,M_j, \epsilon_i). 
\end{equation}
\end{itemize}
Note, that the $C^{(0)}, C^{(1)}, C^{(2)}, C^{(\ast)}$ and hence the respective $Z_{\BC^2,\rm inst}^{(N_f)}$ are invariant under $\epsilon_i \to - \epsilon_i$, while the $Z^{(3)}_{\BC^2, \rm inst}, Z^{(4)}_{\BC^2, \rm inst}$ are not. 

The above identities can be proven from the recursion formula \eqref{eq:InstRecRel_Nf1}. 
Eq. \eqref{eq:inst_a_flip} follows immediately from \eqref{eq:InstRecRel_Nf1} and \eqref{eq:instrecrel_def_C}, while \eqref{eq:HC2_epsilon_flip} follows from induction: Eq. \eqref{eq:InstRecRel_Nf1} demonstrates that
the function $H^{(N_f)}_{\BC^2}$ can be expanded as 
\begin{equation}
    H^{(N_f)}_{\BC^2} (a,  q, M_j,\epsilon_i)
    = \sum_{s=0}^\infty f_s (a, M_j, \epsilon_i)  \ X^{s},
\end{equation}
where $X=q$ for $N_f=0,1,2,3,*$, while for $N_f=4$, $q=q_\UV$ and $X$ are related by \eqref{eq:defX} and \eqref{eq:qSW_expand}.
Then by \eqref{eq:InstRecRel_Nf1}, we have $f_0 = 1$ and for $s \geq 1$ 
\begin{equation}
    \label{eq:recrel_fs}
    f_s(a, M_j, \epsilon_i) = - \frac{1}{4}\sum_{   m,n \geq 1, \ s' \geq 0 \atop s = mn + s' } \frac{\CR_{m,n}^{(N_f)}(M_j,\epsilon_i)}{ a^2 - \left(a_+^{m,n}\right)^2 } f_{s'} \left(a_-^{m,n}, M_j, \epsilon_i \right).
\end{equation}
First, note that the denominator on the RHS in \eqref{eq:recrel_fs} is invariant under $(\epsilon_1, \epsilon_2) \to -(\epsilon_1, \epsilon_2)$. The numerator
$\CR_{m,n}^{(N_f)}(M_j,\epsilon_i)$ for $N_f = 0,1,2,3,4$  \eqref{eq:InstRecRel_Nf1_Rmn} is also invariant because the index set $\Pi'$ in \eqref{eq:instrec_primedset} is invariant under $(r,s) \to - (r,s)$, and the product $\Pi''$ over \eqref{eq:instrec_doubleprimedset} has an even number of factors. Similarly, $\CR_{m,n}^{(*)}( M,\epsilon_i)$ \eqref{eq:CR*} is invariant.  
Since $0\leq s' < s$ in \eqref{eq:recrel_fs}, we can continue by induction. First, $f_{s}(-a^{m,n}_-,M_j,- \eps_i) = f_{s}(a^{m,n}_-, M_j,-\eps_i)$ for all $s' < s$ by Eq. \eqref{eq:HC2_a_flip}.
From \eqref{eq:recrel_fs}, we see that for any $N_f$, $f_s(a,M_j,\epsilon_i)$  are invariant under $(\epsilon_1, \epsilon_2) \to -(\epsilon_1, \epsilon_2)$ for $s=0,1$, and therefore by induction for all $s$.

\subsection{The abstruse duality}
\label{Chap:abstr_dual}
Ref. \cite{Bonelli:2020xps} proves a remarkable identity which is crucial in the analysis of the contour integral. Moreover, this relation is of independent interest due to the interplay between the perturbative and non-perturbative parts. This relation, known as the abstruse duality gives the behaviour of the (local) partition function around the poles $a_+^{m,n}$ and $a_-^{m,n}$ for the pure theory.
The abstruse identity extends to the 5-dimensional or K-theoretic theory  \cite{Kim:2025fpz}. 

We extend the abstruse duality here to $N_f=1,2,3,4,\ast$. For the ratio of instanton partition functions, we first deduce from the recursion relation (\ref{eq:InstRecRel_Nf1})  that 
\be
\begin{split}
& \lim_{a\to 0} a\,\frac{Z^{(N_f)}_{\BC^2,{\rm inst}}(a+a_+^{m,n},q,M_j,\eps_i)}{Z^{(N_f)}_{\BC^2,{\rm inst}}(a+a_-^{m,n},q,M_j,\eps_i)}= -\frac{1}{8\,a_+^{m,n}}\,q^{mn}\,\CR^{(N_f)}_{m,n}(M_j,\eps_i).
\end{split}
\ee
Here we used \eqref{eq:C4_factor_a_out} for the case $N_f = 4$.
For the perturbative contribution of the vector multiplet,
\be
\lim_{a\to 0}\,\frac{1}{a} \frac{Z_{\BC^2, \rm pert,\, vec}(a+a_+^{m,n},\Lambda, \epsilon_i)}{Z_{\BC^2, \rm pert,\, vec}(a+a_-^{m,n},\Lambda, \epsilon_i)}=-\frac{4a_+^{m,n}}{\Lambda^{4mn}} \,\sgn(\epsilon_1)\,{\textstyle \prod''_{k,l}}\, (k\epsilon_1+l\epsilon_2),
\ee
where $\prod''_{k,l}$ is defined in \eqref{eq:instrec_doubleprimedset}.
Next we note that for generic masses $M_j$, the limit $a\to 0$ of $Z_{\BC^2, \rm pert, fund}$ (\ref{eq:Z1vecfund}) is smooth. 
One finds for $N_f = 1, 2,3,4$ using \eqref{eq:gammalog} that for both $\epsilon_1 \epsilon_2>0$ and $<0$,
\begin{equation}
\frac{Z_{\BC^2,\rm pert,\, fund}(a_+^{m,n},\Lambda,M_j, \epsilon_i)}{Z_{\BC^2, \rm pert,\, fund}(a_-^{m,n},\Lambda, M_j,\epsilon_i)}= {\textstyle \prod_{j=1}^{N_f}}{\textstyle \prod'_{r,s}}\, \frac{2\Lambda}{2M_j-r\epsilon_1-s\epsilon_2},
\end{equation}
where $\prod'_{r,s}$ is defined in \eqref{eq:instrec_primedset}. For $N_f = \ast$, we have
\begin{equation}
    \frac{Z_{\BC^2,\rm pert,\, adj}(a_+^{m,n},\Lambda,M, \epsilon_i)}{Z_{\BC^2, \rm pert,\, adj}(a_-^{m,n},\Lambda,M, \epsilon_i)} = 
    \prod_{r = -m+1}^m \prod_{s = -n+1}^n \frac{2 \Lambda}{2M + (2r - 1)\eps_1 + (2s -1) \eps_2}.
\end{equation}
By \eqref{eq:Zpert_C2}, for $N_f = 4,\ast$ we get an additional factor of
\begin{equation}
    (q_\UV)^{-\frac{\left(a_+^{m,n}\right)^2}{\eps_1 \eps_2}+\frac{\left(a_-^{m,n}\right)^2}{\eps_1 \eps_2}} = (q_\UV)^{-mn}.
\end{equation}
Combining the perturbative and instanton contributions and substituting $\CR_{m,n}^{(N_f)}$ (\ref{eq:InstRecRel_Nf1_Rmn}), we arrive at\footnote{We note that this differs by a sign from \cite[Eq. (3.10)]{Bonelli:2020xps}, which we attribute to the different perturbative contribution $Z_{\rm pert}$. See footnote \ref{note:diffpert}.}
\be
\lim_{a\to 0}\,\frac{Z^{(N_f)}_{\mathbb{C}^2}(a+a_+^{m,n},\Lambda, q, M_j,\eps_i)}{Z^{(N_f)}_{\mathbb{C}^2}(a+a_-^{m,n},\Lambda, q,M_j,\eps_i)}=\sgn(\epsilon_1),
\ee
for each $N_f$. 

\section{Equivariant localization on \texorpdfstring{$\mathbb{CP}^2$}{P\texttwosuperior}}
\label{sec:EqLocP2}
We employ in this section the strategy of \cite{Nekrasov:2003vi, Bershtein:2015xfa, Bonelli:2017, Bonelli:2020xps, Kim:2025fpz} to evaluate the partition function of theories with hypermultiplets on $\mathbb{CP}^2$ using toric localization. This approach has been applied to a variety of theories and toric four-manifolds: 
 \begin{itemize}     
 \item pure SU(2) theory on $\mathbb{CP}^2$ \cite{Gottsche:2006tn,
     Bershtein:2015xfa, Bonelli:2017} and the Hirzebruch surfaces
   $\mathbb{F}_0$ and $\mathbb{F}_1$ \cite{Bonelli:2017, Bonelli:2020xps}. 
 \item contribution due to lowest instanton number for pure $SU(3)$ theory on \cite{Bonelli:2020xps}.
 \item the $SU(2)$, $\CN=2^*$ theory, the theory with adjoint
   hypermultiplet for $\mathbb{R}^4$ was studied for this theory in \cite[Chap 6]{NekOk}, and on $\mathbb{CP}^2$ with the VW twist \cite{Bershtein:2015xfa, Bonelli:2017}\footnote{See footnote \ref{fnsignM} spelling out differences between the equivariant partition functions on $\mathbb{C}^2$, which lead to quite different results for the equivariant partition functions on $\mathbb{CP}^2$.} See also \cite{Bonelli:2020xps} for a discussion on the non-holomorphic contribution mentioned above.
 \item the 5d $SU(2)$ on $\mathbb{CP}^2\times S^1$ theory giving K-theoretic invariants
   \cite{Nakajima:2005fg, Gottsche:2006bm, Kim:2025fpz}.
   \item the 5d $SU(2)$ on $\mathbb{CP}^2\times S^1$ theory with adjoint hypermultiplet giving the $\chi_y$-genus \cite{Khlaif:2026rnb}.
 \end{itemize}

The full $U(2)$ partition function on $\mathbb{CP}^2$ evaluates to a contour integral around $a=0$ of the product of $\chi(\mathbb{CP}^2) = 3$ partition functions $Z^{(N_f)}_{\BC^2}$ on $\BC^2$, one for each toric patch of $\mathbb{CP}^2$. The partition function $Z^{(N_f)}_{\mu} \coloneqq Z^{(N_f)}_{\mathbb{CP}^2,\mu}$ thus takes the form
\begin{equation}
\label{eq:NekPartFctnGeneral}
    Z^{(N_f)}_{\mu} ( \Lambda, q,M_j,b_j; \epsilon_i) = K \sum_{\substack{  p \\ {\rm semi-stable} }} \Theta_\mu (p) \ \frac{1}{2\pi i}  \oint_C da \prod_{l=1}^{3} Z^{(N_f)}_{\BC^2} ( a^{(l)}, \Lambda, q,M_j^{(l)}, \epsilon_i^{(l)}),
\end{equation}
where $C$ is a contour around $a = 0$, that does not include any poles of the integrand other than in $a = 0$, $Z_{\BC^2}$ is the partition function for $\BC^2$ as in \eqref{eq:C2factorisation}. Moreover, $K$ is a normalization factor. By comparison of our explicit results in Section \ref{sec:ExResults} for instanton number $k=\frac{3}{4}$ for which $\vdim(M_k)=0$ with mathematical results on the intersection numbers of instanton moduli spaces, we fix  $K$ for all theories as
\be
\label{eq:K}
K=\frac{2}{\Lambda}.
\ee

The sum over $p$ runs over triples which classify semi-stable toric bundles, $a^{(l)}, M_j^{(l)}, \epsilon_i^{(l)}$ are the parameters of the theory on patch $l$ explained in more detail below, and $C$ is a contour around the singularity $a=0$, but not surrounding any of the other singularities of the integrand. The function $\Theta_\mu$ is a $p$-dependent weight explained below. 

Ref. \cite{Bonelli:2020xps} argues that the path integral can be evaluated alternatively by a contour integral around all singularities giving rise to a sum over all triples $p\in \mathbb{Z}^3$. Intriguingly, it is argued that only the triples corresponding to semi-stable toric bundles contribute. See also Ref. \cite{Ruggeri:2025yqt}.

The equivariant topological correlators should satisfy various consistency conditions:
\begin{enumerate}
\item In the decoupling limit $M\to \infty$, the equivariant correlation function should reduce to the equivariant correlation function of the $N_f=0$ theory \cite{Bershtein:2015xfa},
\item Since $\mathbb{CP}^2$ is a compact four-manifold, the non-equivariant limit $\eps_1,\eps_2\to 0$ of the correlation functions should be finite.
\end{enumerate}

\subsection{Toric aspects of \texorpdfstring{$\mathbb{CP}^2$}{P\texttwosuperior}}
The $U(2)$-partition function on $\mathbb{CP}^2$ will then be given as in Eq. \eqref{eq:NekPartFctnGeneral}. We will now provide further details on all the ingredients.

First, note that the cohomology of $\mathbb{CP}^2$ is torsion free. Denote $b_2$ the second Betti number, and $b_2^+$ the number of positive eigenvalues of the intersection form of $\mathbb{CP}^2$. Then we have $b_2 = b_2^+ = 1$.

The complex projective plane $\mathbb{CP}^2$ is canonically covered by three patches, labelled by $l=1,\dots,3$. The toric action acts as $[z_0 : z_1  : z_2] \mapsto [z_0 : e^{i\epsilon_1}z_1  : e^{i\epsilon_2}z_2]$. Define the equivariant parameters
\begin{center}
\renewcommand{\arraystretch}{1.5}
\begin{tabular}{|c|r|r|}
\hline
$l$ & $\varepsilon_1^{(l)}$ & $\varepsilon_2^{(l)}$ \\
\hline 
1 & $\varepsilon_1$ & $\varepsilon_2$  \\
2 & $\varepsilon_2-\varepsilon_1$ & $-\varepsilon_1$ \\
3 & $-\varepsilon_2$ & $\varepsilon_1-\varepsilon_2$  \\
\hline
\end{tabular}
\label{tab:epsilon_l}
\end{center}
Then in local coordinates in the $l$-th open patch, the action is given by $(x^{(l)}, y^{(l)}) \mapsto (e^{i\epsilon^{(l)}_1} x^{(l)}, e^{i\epsilon^{(l)}_2} y^{(l)})$ \cite{Bershtein:2015xfa}.

This toric action on $\mathbb{CP}^2$ lifts to an action on vector bundles, and the moduli space of semi-stable vector bundles $M_k$. Semi-stable bundles of rank 2 on $\mathbb{CP}^2$ are special in the sense that the fixed point locus of the toric action on $M_k$ is zero-dimensional, while for higher rank and other toric surfaces the fixed loci are typically higher dimensional. The fixed points for rank 2 bundles are characterized by a 3-tuple of integers $p_l\in \mathbb{N},l=1,2,3$, which satisfy triangle inequalities \cite{BertinElencwajg1982, Klyachko1991}
\be
\label{eq:semistabconds}
p_l+p_{l+1} \geq p_{l+2},\qquad p_1+p_2+p_3=2\mu \mod 2,
\ee
with $\mu=w_2(E)/2$.  Each integer $p_l$ can be considered as the flux of the unbroken $U(1)\subset SO(3)$ on patch $l$. 
If we have strict inequality in \eqref{eq:semistabconds} for all $l$, we call the tuple of $p_l$ \textit{stable}. If there is at least one $l$ for which the inequality is saturated, $p_l + p_{l+1} = p_{l+2}$, we call it \textit{strictly semi-stable}. 

We define the vector $p$ of equivariant fluxes by 
\be
p=(p_1,p_2,p_3), 
\ee
and the total flux $P$,
\be
\label{eq:DefP}
P=p_1+p_2+p_3.
\ee
We also introduce the quadratic form $Q:\mathbb{Z}^3\to \mathbb{Z}$,
\be
\label{eq:Qp}
Q(p)=2p_1p_2+2p_2p_3+2p_1p_3-p_1^2-p_2^2-p_3^2,
\ee
which has indefinite signature $(1,2)$. The instanton number $k$ reads in terms of the triple $p$,
\be
k=\frac{1}{4}Q(p).
\label{eq:k_def}
\ee
The Hurwitz class number $H(4k)$ enumerates triples $p$ with fixed $Q(p)$ modulo automorphisms. The generating function of $H(4k)$ is a mock modular form \cite{Zagier:1975}. See also the discussion in Sec. \ref{sec:comp2*}.

We define the parameter $a^{(l)}$ in each patch as
\be
\label{eq:aell}
2a^{(l)}=2a+p_{l}\varepsilon^{(l)}_1+p_{l+1}\varepsilon^{(l)}_2.
\ee 
The shift of the $a^{(l)}$ by local fluxes on a toric variety is argued, among other places in \cite{BonTanMarYagi_2013}, or we can argue with the explicit formulae and derivations from \cite{Bershtein:2015xfa}.

Similarly for coupling the hypermultiplet to the background flavor symmetry bundle $\CL$, we define the equivariant fluxes 
\be
b=(b_{1}, b_{2}, b_{3}),
\ee
with $b_{l}$ the flux for patch $l$, and the total flux 
\be
B=2\bfk=\sum_{l=1}^3 b_{l}.
\ee
We define the equivariant mass $M^{(l)}$ in each patch as
\be
\label{eq:Ml_shift}
M^{(l)}=M+\frac{1}{2}\left( b_l\, \varepsilon_1^{(l)} + b_{l+1} \,\varepsilon_2^{(l)}\right).
\ee 
In case of multiple hypermultiplets, we add a further subscript $j$ to $M$, $B$ and $b_l$.
Analogously to Eq. \eqref{eq:apm}, we also introduce the notation
\begin{align}
\label{eq:apm2}
    a_\pm^{(l)}  &= \frac{p_l \epsilon_1^{(l)} \pm p_{l+1} \epsilon_2^{(l)}}{2}, \\
    \label{eq:bpm}
    b_\pm^{(l)}  &= \frac{b_l \epsilon_1^{(l)} \pm b_{l+1} \epsilon_2^{(l)}}{2},
\end{align}
such that
\begin{align}
    a^{(l)} &= a + a_+^{(l)}, \\
    M^{(l)} &= M + b_+^{(l)}.
\end{align}

We will now proceed with the evaluation of the integral in \eqref{eq:NekPartFctnGeneral}. We consider the perturbative and non-perturbative contributions separately.

\subsection{Perturbative part}
We denote by $b_j$, for $1 \leq j \leq N_f$, both the single $j$-th triple of background flavour fluxes, and the collection $\{ b_j \}_{j=1}^{N_f}$ of all $N_f$ such triples. The usage will be clear from the context. Analogously, we use the notation $M_j$ for both the single $j$-th mass parameter and the set of all hypermultiplet masses.\par
We consider the perturbative part of the integrand in Eq. (\ref{eq:NekPartFctnGeneral}),
\begin{equation}
Z_{{\rm pert}}^{(N_f)}(a,\Lambda, q,M_j,b_j,p,\varepsilon_i)=\prod_{l=1}^3 Z^{(N_f)}_{\mathbb{C}^2,{\rm pert}}(a^{(l)},\Lambda,q,M_j^{(l)},  \varepsilon_i^{(l)}),
\end{equation}
with $Z^{(N_f)}_{\BC^2, \rm pert}$ defined in Eq. (\ref{eq:Zpert_C2}). The dependence on the triples $b$ and $p$ is contained in $M^{(l)}$ and $a^{(l)}$ respectively. We determine $Z^{(N_f)}_{\rm pert}$ by separately evaluating the vector, fundamental and adjoint contributions on the full $\mathbb{CP}^2$.\par
To this end, we introduce the function $W$,
\be
\label{eq:Wxd}
W(x,d;\Lambda,\epsilon_i)=\prod_{l=1}^3 \exp[- \gamma_{\varepsilon_1^{(l)},\varepsilon_2^{(l)}}(x^{(l)};\Lambda)],
\ee 
with $d=(d_1,d_2,d_3)\in \mathbb{Z}^3$ a triple, and
\be
x^{(l)}=x+\frac{1}{2}\left(d_{l}\varepsilon_1^{(l)}+d_{l+1}\varepsilon_2^{(l)}\right).
\ee 
For the vector and hypermultiplets, we will find that $D=d_1+d_2+d_3\in 2\mathbb{Z}$. For such $D$, $W(x,d)$ can be expressed as a finite product using \eqref{eq:gammalog} and first analytic continuation. One finds for such cases \cite{Bershtein:2015xfa}
\be
\label{eq:def_W}
W(x,d; \Lambda, \epsilon_i)=\left\{\begin{array}{rr} \prod_{i=0}^{D/2}\prod_{j=0}^{{D/2}-i}\frac{(x+(d_1/2-j)\varepsilon_1+(d_2/2-i)\varepsilon_2)}{\Lambda},  & \,   D\geq 0, \\ 
1, &  \, D=-2,-4, \\
\prod_{i=0}^{|D/2|-3}\prod_{j=0}^{|D/2|-3-i}\frac{(x+(d_1/2+1+j)\varepsilon_1+(d_2/2+1+i)\varepsilon_2)}{\Lambda} , & \, D\leq -6.
\end{array}\right.
\ee
We combine the cases with $D<0$, by defining that the product over an empty set, such as $\prod_{j=0}^{-2}$, is equal to 1. We denote the number of factors in $W$ by $f(D)$. This evaluates for $D\in 2\mathbb{Z}$ to
\begin{equation}
    \label{eq:W_no_factors}
    f\left(D\right) = \frac{1}{2} \left(1+ \frac{D}{2} \right)\left(2+ \frac{D}{2}\right). 
\end{equation}
Notice, that $f (-2) = f(-4) = 0$, and further $f\left( D \right) = f\left(- D -6 \right)$. 

As mentioned above, the contribution from the vector multiplets reduces to a product with a finite number of terms \cite{Bershtein:2015xfa, Bonelli:2020xps}, namely 
\be
\label{eq:Zpertvec}
\begin{split}
Z_{{\rm pert, vec}}(a,\Lambda,p,\varepsilon_i)&=\prod_{l=1}^3 Z_{\BC^2, \rm pert, vec}(a^{(l)}, \Lambda,\varepsilon_i^{(l)})\\
& =W(2a,2p; \Lambda, \epsilon_i)\,W(-2a,-2p; \Lambda, \epsilon_i).
\end{split}
\ee 
Or fully expanded with $P\geq 3$ as in (\ref{eq:DefP}), 
\be 
\label{eq:Z1vecP2}
\begin{split}
Z_{{\rm pert, vec}}(a,\Lambda,p,\varepsilon_i)&=\prod_{i=0}^P\prod_{j=0}^{P-i}\frac{(2a+(p_1-j)\varepsilon_1+(p_2-i)\varepsilon_2)}{\Lambda}\\
&\quad \times \prod_{i=0}^{P-3}\prod_{j=0}^{P-3-i} -\frac{(2a+(p_1-1-j)\varepsilon_1+(p_2-1-i)\varepsilon_2)}{\Lambda}.
\end{split}
\ee 
Using \eqref{eq:W_no_factors}, we see that the overall power of $\Lambda$ in Eq. \eqref{eq:Z1vecP2} reads
\be
\label{eq:zpertvec_lambdas}
\Lambda^{-2-P^2}.
\ee

We next proceed with the contribution from the fundamental hypermultiplet.
Following the quantization condition \eqref{eq:quantcond}, we deduce that $P+B\in 2\mathbb{Z}+1$. Crucially with this condition, the perturbative contribution from a single hypermultiplet does reduce to a product with a finite number of factors. We directly read off the contribution from $j$-th fundamental hypermultiplet, $Z_{{\rm pert,fund}}$ as,
\be
\label{eq:Zpert_hyp_P2}
\begin{split}
&Z_{{\rm pert,fund}}(a,\Lambda,M_j,b_j,p,\varepsilon_i)=\prod_{l=1}^3 Z_{\BC^2, \rm pert, fund}(a^{(l)}, \Lambda,M_j^{(l)},\varepsilon_i^{(l)})\\
&\qquad = \frac{1}{W(a+M_j,p+b_j-(1,1,1); \Lambda, \epsilon_i)\,W(-a+M_j,-p+b_j-(1,1,1); \Lambda, \epsilon_i)}.
\end{split}
\ee
Since \eqref{eq:W_no_factors} holds for all cases of $\frac{P+B_j-3}{2}$ and $\frac{P-B_j+3}{2}$, the overall factor of $\Lambda$ of $Z_{{\rm pert,hyp}}$ is
\begin{equation}
\label{eq:Zperthyp_lambdas}\Lambda^{\frac{1}{4}(-1+B_j^2+P^2)}.
\end{equation}

Finally, for formulation of the $\CN=2^*$ theory on $\mathbb{CP}^2$, we have the condition that $B\in 2\mathbb{Z}+1$ by the quantization condition \eqref{eq:quantcond_2*}. The perturbative contribution of the adjoint hypermultiplet contains again a finite number of factors and reads
\be
\begin{split}
\label{eq:Zpert_adj_P2}
&Z_{{\rm pert,adj}}(a,\Lambda,M,b,p,\varepsilon_i)=\prod_{l=1}^3 Z_{\BC^2, \rm pert, adj}(a^{(l)}, \Lambda,M^{(l)},\varepsilon_i^{(l)})\\
&\qquad = \frac{1}{W(2a+M,2p+b-(1,1,1); \Lambda, \epsilon_i)\,W(-2a+M,-2p+b-(1,1,1); \Lambda, \epsilon_i)}\\
&\qquad \quad \times \frac{1}{W(M,b-(1,1,1);\Lambda,\epsilon_i)}.
\end{split}
\ee
The overall power of $\Lambda$ of $Z_{{\rm pert,adj}}$ is
\be
\Lambda^{P^2+\frac{3}{8}(B^2-1)}.
\ee

\subsection{Observables}
A key aspect of the theory are the $Q$-exact observables \cite{Witten:1988ze, Moore:1997pc, LoNeSha}. For a point $p\in X$, the point observable in the non-equivariant theory is
\be
u(p)=\frac{1}{8\pi^2}{\rm Tr}[\phi(p)^2],
\ee
while the surface observable is
\begin{equation}
    I(\Sigma)=\frac{1}{4\pi^2}\int_\beta \mathrm{Tr}\left[ \frac{1}{8}\psi\wedge \psi-\frac{1}{\sqrt{2}}\phi\, F\right],
\end{equation}
for a 2-cycle $\Sigma\in H_2(X,\mathbb{Z})$. Due to the topological invariance of the theory, correlation functions of such observables are independent of the position $p\in X$, and only dependent on the homology class of $\Sigma$ in $H_2(X,\mathbb{Z})$. There are ambiguities in the precise definition of $u$ in theories with hypermultiplets. See for example \cite{Manschot:2019pog, Manschot:2021qqe}.

We review in the following the equivariant observables \cite{Ne, NekOk, Gottsche:2006tn}. To this end, recall that the observables correspond to restrictions of the Pontryagin class, or better instanton density, of the universal bundle over $M_k\times X$ to $M_k$.
Let $\CF = F + \psi + \phi$ be the equivariant curvature on the universal bundle $\CE$ associated to the $U(2)$ bundle $\tilde E$ introduced in Section \ref{sec:classasp}. Denote $H_{U(1)^2}^\ast (\mathbb{CP}^2)$ the equivariant cohomology with respect to the toric $U(1)^2$ action on $\mathbb{CP}^2$ and let  $\Omega \in  H_{U(1)^2}^\ast (\mathbb{CP}^2)$. The observable is defined as 
\begin{equation}
    \label{eq:observable_def}
    \CO (\Omega, \CF) \coloneqq  -\frac{1}{8\pi^2}\int_X \Omega \wedge {\rm Tr} (\CF^2).
\end{equation}
We let $\Omega$ be a sum of surface observable and point observable, 
\be
\Omega=\alpha x+\beta z,
\ee
with $\alpha$ an equivariant 0-form, and $\beta$ an equivariant 2-form, Poincar\'e dual to $\Sigma$, and $x\in \mathbb{R}$, $z\in \mathbb{R}^{b_2}$ fugacities. Thus for $X=\mathbb{CP}^2$, both $x,z\in \mathbb{R}$. Moreover, $x$ has dimension $\Lambda^{-2}$, and $z$ has dimension $\Lambda^{-1}$.

The precise expression for $\CO$ can be derived using the lift of the $SO(3)$ bundle to a $U(2)$ bundle. In algebraic geometry, the analogue of the integral over $X$ is the slant product. For $X=\mathbb{C}^2$, application of Atiyah-Bott localization to the toric fixed point $0$ at the origin of $\mathbb{C}^2$, gives
\be
{\rm ch}(\CE)/[\mathbb{C}^2]=\frac{1}{\epsilon_1\epsilon_2} \iota^*_{0\times M_k} {\rm ch}(\CE).
\ee

The Chern character of the universal $U(r)$ instanton was calculated, at a fixed point of the toric action in $M_k$ labelled by $\vec{Y}$, to be \cite{Gottsche:2006tn}
\begin{equation}
    \label{eq:ch2_univ_atfp}
    \iota_{(0,\vec{Y})}^\ast {\rm ch} (\CE) = \sum_{\alpha=1}^r e^{a_\alpha} \left( 1 - (1 - e^{-\epsilon_1})(1 - e^{-\epsilon_2}) \sum_{s\in Y_\alpha} e^{- l'(s)\epsilon_1 - a'(s)\epsilon_2} \right),
\end{equation}
where $a'(s)=j-1$ and $l'(s)=i-1$ for a box $s=(i,j)$. From this equation, we read off the first and second Chern characters. We restrict to rank $r=2$,
\be
\begin{split}
& \iota_{(0,\vec{Y})}^\ast c_1(\CE)=a_1+a_2,\\
& \iota_{(0,\vec{Y})}^\ast {\rm Ch}_2(\CE)=\frac{1}{2}(a_1^2+a_2^2)-\epsilon_1\epsilon_2 |\vec Y|.
\end{split}
\ee
Since ${\rm ch}_2=\frac{1}{2}c_1^2-c_2$, we find therefore for the instanton density in terms of the Pontryagin class $p_1$ of the $SO(3)$ universal bundle $\CP$,
\be
-\iota_{(0,\vec{Y})}^\ast\frac{1}{4}p_1(\CP)=\iota_{(0,\vec{Y})}^\ast\left(c_2(\CE)-\frac{1}{4}c_1(\CE)^2\right)=\epsilon_1\epsilon_2 |\vec Y|-\frac{1}{4}(a_1-a_2)^2.
\ee

The exponentiated observables are then included through
\be
\exp\left( \left(|\vec Y|-\frac{a^2}{\epsilon_1\epsilon_2}\right) \iota_{0,\vec Y}^\ast\Omega \right).
\label{eq:obs_localized_factors}
\ee
The term with $|\vec Y|$ is easily included by making the replacement for 
\begin{equation}
\label{eq:obs_lambdashift_inst}
q \to q\exp\!\left( \iota_{(0)}^\ast \Omega \right),   
\end{equation}
or $\Lambda \to \Lambda \exp\!\left( \frac{\iota_{(0)}^\ast \Omega}{4-N_f}  \right)$ for $N_f\leq 3$,
in the instanton contributions $Z^{(N_f)}_{\BC^2, \rm inst}$, while the term with $a^2$ is added to the integrand as an additional factor. 

We choose the equivariant extension of $\alpha$ and $\beta$ such that at the fixed points of $\mathbb{CP}^2$,\footnote{For the explicit form of $\Omega$, see \citep[Eq. (3.6) - (3.8)]{Bershtein:2015xfa}, where we chose $h = h' = h'' = K = 0$. Note that in \eqref{eq:P2_obs_Omega_choice} $\epsilon_i^2$ denotes the square of $\epsilon_i$, and not $\epsilon_i^{(l)}$ as in Table  \ref{tab:epsilon_l}.} 
\be
\label{eq:P2_obs_Omega_choice}
\iota_{(P_l)}^\ast \Omega=\left\{\begin{array}{rr} 0, & \qquad l=0, \\  -\epsilon_1\,z + \epsilon_1^2\, x , & \qquad l=1, \\ -\epsilon_2\,z + \epsilon_2^2\, x   , & \qquad l=2.\end{array}\right.
\ee

\subsection{Equivariant Nekrasov partition function on \texorpdfstring{$\mathbb{CP}^2$}{P^2}}
\label{Chap:Partitionfctn_P2}

We can now combine the previous results, and provide the explicit expression for the $U(2)$ equivariant partition function as in \eqref{eq:NekPartFctnGeneral} on $\mathbb{CP}^2$. This step builds on and generalizes the results from \cite{Bershtein:2015xfa, Bonelli:2020xps}. \par
To this end, we evaluate the residue in $a=0$ for a given flux configuration $p$. We assume generic masses $M_j \neq 0$ for all $j = 1, \dots, N_f$. 
Then, the perturbative contributions due to hypermultiplets as in \eqref{eq:Zpert_hyp_P2}, \eqref{eq:Zpert_adj_P2} do not have poles in $a=0$. Hence, only the three instanton parts contribute poles. From \eqref{eq:InstRecRel_Nf1}, we see that $Z^{(N_f)}_{\BC^2, {\rm inst}} (a^{(l)})$ has a pole in $a=0$, only if $p^l, p^{l+1} \geq 1$. If $p^{l+2}$ were zero,  $Z^{(N_f)}_{\BC^2, {\rm inst}} (a^{(l)})$ exhibits a simple pole, while the factors $Z^{(N_f)}_{\BC^2, {\rm inst}} (a^{(l+1)}),  Z^{(N_f)}_{\BC^2, {\rm inst}} (a^{(l+2)})$ are regular. This pole cancels with a zero from $Z_{{\rm pert, vec}, \mathbb{CP}^2}$ as in \eqref{eq:Z1vecP2}, hence we can assume $p^l \geq 1, l = 1, 2, 3$. In this case, we get three poles in $a=0$ from the instanton factors, and a double zero from the vector multiplet perturbative factor \cite{Bershtein:2015xfa, Bonelli:2020xps}. The residue of the resulting overall simple pole in $a=0$ is the contribution to the partition function for the equivariant gauge flux configuration $p$ as discussed in the following. 

In the case of $w_2(P)$ odd, semi-stability implies \textit{ stability}. 
By the quantization condition \eqref{eq:quantcond}, this means that
the background flux $B_j$ must be even for $j=1,\dots,N_f$. By \eqref{eq:quantcond_2*}, we have  $B\in 2\mathbb{Z}+1$ on $\mathbb{CP}^2$ for $\CN=2^*$, independently from $w_2(P)$. Further, since $\sum_i p_i  +
w_2(P) \in 2\mathbb{Z}$, we are precisely in the situation that the perturbative part reduces to a finite product.  For the case of $w_2$ even, some instanton solutions can be reducible, which gives rise to singularities in the moduli space $M_{k,\mu}$. In terms of the triple $p$, these correspond to the \textit{strictly semi-stable} configurations. The contribution of such a triple to the partition function \eqref{eq:Zfinal_c1odd_new} is multiplied by a factor of $\frac{1}{2}$ \cite{Bershtein:2015xfa, Bonelli:2017, Bonelli:2020xps, Kim:2025fpz}. Further, for $w_2$ even  the background fluxes $B_j$ must be odd for $1\leq j\leq  4$ by \eqref{eq:quantcond}. Similarly to \cite{Kim:2025fpz}, we thus introduce the function
\begin{equation}
    \Theta_\mu (p) = \left\{ 
    \begin{array}{rl}
        1, & \quad p \ \text{  stable}, \\
        \frac{1}{2}, & \quad p \ \ \text{strictly semi-stable}. 
    \end{array}
    \right.
\end{equation}
Given a fixed $p$, we now collect all factors after taking the residue. From the perturbative part, we get a factor of 
\be
\begin{split}
\lim_{a\to 0} \frac{1}{a^2} Z^{(N_f)}_{\rm pert}(a,\Lambda,q, M_j,b_j, p, \epsilon_i)&=   \Tilde{Z}^{(N_f)}_{\rm pert}( M_j,b_j,p, \epsilon_i) \ \ q^{- \frac{P^2}{4}}\\
\qquad 
&\times \left\{\begin{array}{rr}\Lambda^{-\frac{8+N_f}{4}+\frac{1}{4}\sum_j B_j^2}, & \qquad N_f=0,\dots, 4, \\
\Lambda^{\frac{3}{8}(B^2-1)}, & \qquad \CN=2^*.
\end{array}\right. 
\end{split}
\ee
By the above discussion and \eqref{eq:Z1vecP2}, the limit exists for $p^{(l)}  \geq 1$. We determined the powers of $\Lambda$ from (\ref{eq:zpertvec_lambdas}) and (\ref{eq:Zperthyp_lambdas}). Above equation implicitly defines
\begin{equation}
    \label{eq:Zpert_p2_reduced}
    \begin{split}
   \Tilde{Z}^{(N_f)}_{\rm pert}(M_j,b_j, p, \epsilon_i) 
    & = \Tilde{W}(2p; 1, \epsilon_i)\, \Tilde{W}( -2p; 1, \epsilon_i)\\
      \times & \left\{ 
      \begin{array}{rr}
       \prod_{j=1}^{N_f} Z_{{\rm pert,fund}}(0,1,M_j,b_j,p,\varepsilon_i) ,  & \qquad N_f = 0,1,2,3,4,  \\
        Z_{{\rm pert,adj}}(0,1,M_j,b,p,\varepsilon_i), & \qquad \CN = 2^\ast,
    \end{array}
    \right .
    \end{split}
\end{equation}
where we have replaced the scale $\Lambda$ by 1. $\Tilde{W}$ is defined as
\begin{equation}
\tilde W(\pm 2p;1,\epsilon_i)= \lim_{a\to 0} \frac{1}{a} W(a,\pm 2p;1,\epsilon_i).
\end{equation}
The factor $a^{-1}$ removes the factor $(i, j ) = (p_1, p_2)$ for the argument $+2p$, while it removes the factor $(i,j)=(p_1 -1, p_2 -1)$ for $-2p$.\par

We next include the contribution from the instanton parts using the recursion formula (\ref{eq:InstRecRel_Nf1}) at the order 3 pole in $a=0$. In the residue, this evaluates $a^{(l)}$ \eqref{eq:aell} to $a_+^{(l)}$ in the $Z_{\BC^2, {\rm inst}}^{(N_f)}$-factor from the $l$-th open patch. 
$Z_{\BC^2, {\rm inst}}^{(N_f)}, H^{(N_f)}_{\BC^2}$ and ${C^{(N_f)}}$ are related as in \eqref{eq:defH}. Thus, taking the residue amounts to picking up a factor $C^{(N_f)}$ evaluated at $a_+^{(l)}$ and  
the summand in \eqref{eq:InstRecRel_Nf1} with $m=p_l$, $n=p_{l+1}$. 

First, we collect all $p$-dependent factors of $\Lambda$ and $q_\UV$. The three terms $X^{mn}$ in the numerator of the summand of \eqref{eq:InstRecRel_Nf1} and, for $N_f=4$, the factors $\left(\frac{q_\UV}{16\,q_\SW}\right)^{mn}$ in \eqref{eq:C4_factor_a_out}, combine with the exponent of $q$ from the vector perturbative part to the quadratic form $Q(p)$ \eqref{eq:Qp}. By (\ref{eq:K}), we rescale the partition function by a factor of $\Lambda^{-1}$, arriving at an overall power of $\Lambda$ and $q$,
\be
\label{eq:Lq}
\Lambda^{d(B_j)}\,q^{Q(p)/4},
\ee
where we defined $d(B_j)$ as the $k$-independent part of ${\rm vdim}_\BC(\CM^{Q,N_f}_{k})$,
\be
d(B_j)=\left\{\begin{array}{rr}
        -\frac{12 + N_f}{4}  + \frac{1}{4} \sum_{j = 1}^{N_f} B_j^2, & \qquad N_f=0,\dots,4, \\
        \frac{3}{8}(B^2-9),&\qquad \CN=2^*.
\end{array}\right.
\label{eq:def_D_exponent}
\ee
Note the power of $\Lambda$ in \eqref{eq:Lq} precisely equals ${\rm vdim}_\BC(\CM^{Q,N_f}_{k})$ and ${\rm vdim}_\BC(\CM^{Q,*}_{k})$ for $\mathbb{CP}^2$ as in \eqref{eq:vdim_def}. \par
The remaining factors from the instanton part third-order pole are given by
\begin{equation}
\begin{split}
    &\Tilde{Z}_{\rm inst}^{(N_f)}(q,M_j,b_j,p,x,z,\epsilon_i)  \coloneqq \\
    &\qquad - \left( \prod_{l=1}^3 \left(X^{(l)} \right)^{-p_l p_{l+1}} \right) \  \lim_{a\to 0} a^3 \prod_{l=1}^3 Z^{(N_f)}_{\BC^2,\rm inst}(a^{(l)},q^{(l)},M^{(l)}_j,\epsilon_i^{(l)}),
    \end{split}
\end{equation}
with $q$ as in \eqref{eq:defq}, $X$ as in  (\ref{eq:defX}), and 
\begin{equation}
    \label{eq:def_ql}
   q^{(l)} \coloneqq q\, e^{\iota^\ast_{P_l} (\alpha x + \beta z)}, \qquad X^{(l)} \coloneqq X\left(q^{(l)}\right).
\end{equation}
Then, using again (\ref{eq:defH}), the recursion relation (\ref{eq:InstRecRel_Nf1}) and for $N_f=4$, \eqref{eq:C4_factor_a_out}, we obtain
\begin{equation}
    \label{eq:Zinst_P2_reduced}
\begin{split}
 &\Tilde{Z}_{\rm inst}^{(N_f)}(q,M_j,b_j,p,x,z,\epsilon_i) = \\
&\qquad  \frac{1}{4^3}\prod_{l=1}^3 \frac{\CR^{(N_f)}_{p_l, p_{l+1}}\!\left(M^{(l)}, \epsilon_i^{(l)}\right)}{2 \ a_+^{(l)}} \  Z^{(N_f)}_{\BC^2,\rm inst}\left(a_-^{(l)},  q^{(l)},M^{(l)}_j,\epsilon_i^{(l)} \right),
\end{split}
\end{equation}
with $\CR^{(N_f)}_{m,n}$ defined in \eqref{eq:InstRecRel_Nf1_Rmn}.

Lastly, we include contributions from observables. After shifting to $a^{(l)}$ and evaluating at the residue, the $a$-dependent factor in \eqref{eq:obs_localized_factors} contributes $\exp(-(a_+^{(l)})^2/4\epsilon^l_1\epsilon^l_2)$. This combines with the observable factors from the instanton parts to $\exp(-(a_-^{(l)})^2/4\epsilon^l_1\epsilon^l_2)$.

We introduce
\begin{equation}
    \label{eq:ZP2_O_reduced}
    \Tilde{\CO}(p, x, z, \epsilon_i) = \exp\left(-\sum_{l=1}^3 \frac{(a_-^{(l)})^2}{\epsilon_1^l \epsilon_2^l} \iota^\ast_{P_l}
    (\alpha z + px) \right).
\end{equation}
The full $U(2)$ partition function as in \eqref{eq:NekPartFctnGeneral} with observables and for $w_2(P)$ odd thus reads 
\be
\label{eq:Zfinal_c1odd_new}
\begin{split}
Z^{(N_f)}_{\mu}(\Lambda, q, M_j,b_j, x, z,\epsilon_i)&=\quad -2\,\Lambda^{d(B_j)} \sum_{p \atop \rm semi-stable} \Theta_\mu (p) \ q^{Q(p)/4}\, \Tilde{\CO}(p, x,z, \epsilon_i) \\
&\quad \times  \Tilde{Z}_{\rm pert}^{(N_f)}(M_j,b_j,p, \epsilon_i)\,\Tilde{Z}_{\rm inst}^{(N_f)} \left(q,M_j,b_j,p, x, z ,\epsilon_i \right).
\end{split}
\ee

The first few terms of this partition function can be evaluated with Mathematica or other computer programs. Note that to evaluate the first $\ell$ terms of the partition function in the $q$ expansion, $Z_{\rm inst}$ (\ref{eq:Zinst_iterative}) has to be determined up to $O(q^\ell)$. Moreover, all tuples $p$ such that $Q(p)\leq 4\ell$ have to be included. \par 
Further, note that the factors of $Z_{\BC^2, \rm inst}^{(N_f)}$ in \eqref{eq:Zinst_P2_reduced} can be explicitly computed using the iterative formula \eqref{eq:Zinst_iterative}, although this formula exhibits spurious poles in $a$. Hence, computation using the recursion formula \eqref{eq:InstRecRel_Nf1} can be more efficient.\par
We note, that formula \eqref{eq:Zfinal_c1odd_new} for  $Z_{\mu}^{(0)}$ is consistent with the results in \cite{Bershtein:2015xfa, gottsche_instanton_2006}, up to a constant overall factor.

\subsubsection*{\texorpdfstring{$SO(3)$}{SO3} partition function $H^{(N_f)}_{\mu}$} 
We also considers the \textit{reduced} partition function $H^{(N_f)}_{\mu}$ on $\mathbb{CP}^2$. Similarly to the discussion below (\ref{eq:defH}), we interpret this as the partition function of the theory with gauge group $SO(3)$, obtained by ``dividing  out'' a suitable factor $D^{(N_f)}$ (\ref{eq:DefDNf}) below associated to the $U(1)\subset U(2)$ degrees of freedom. The comparison with $u$-plane results for $SO(3)$ in Section \ref{sec:compwuplane} provides further evidence for this intepretation. 

Analogously to \eqref{eq:Zinst_P2_reduced}, we define
\begin{equation}
\begin{split}
    \label{eq:HTilde_P2}
    &\Tilde{H}^{(N_f)} \left(q, M_j,b_j, p, x, z,  \epsilon_i \right) =  \\
    &\quad \frac{1}{4^3}\prod_{l=1}^3 \frac{\CR^{(N_f)}_{p_l, p_{l+1}}\!\left(M^{(l)}, \epsilon_i^{(l)}\right)}{2 \ a_+^{(l)}} \  H^{(N_f)}_{\BC^2}\left(a_-^{(l)},  q^{(l)},M^{(l)}_j,\epsilon_i^{(l)} \right),
\end{split}
\end{equation}
with $H^{(N_f)}_{\BC^2}$ as in Eq. \eqref{eq:defH}.
Then we define $H^{(N_f)}_{\mu}$ by replacing $\tilde Z^{(N_f)}_{\rm inst}$ by $\Tilde{H}^{(N_f)}$ in $Z^{(N_f)}_{\mu}$ \eqref{eq:Zfinal_c1odd_new} ,
\begin{equation}
\begin{split}
    \label{eq:Hfinal_c1odd}
    H^{(N_f)}_{\mu}(\Lambda, q, M_j, b_j, x, z,\epsilon_i)&=\quad -2\,\Lambda^{d(B_j)} \sum_{p \atop \rm semi-stable} \Theta_\mu (p) \ q^{Q(p)/4}\, \Tilde{\CO}(p, x,z, \epsilon_i) \\
&\quad \times  \Tilde{Z}_{\rm pert}^{(N_f)}(M_j,b_j,p, \epsilon_i)\,\Tilde{H}^{(N_f)} \left(q,M_j,b_j,p, x, z ,\epsilon_i \right).
\end{split}
\end{equation}
 This corresponds to substituting $Z_{\BC^2, {\rm inst}}^{(N_f)}$ in \eqref{eq:NekPartFctnGeneral} by 
\begin{equation}
    \begin{split}
        H_{\BC^2}^{(N_f)}, \qquad &N_f = 0,1,2,3,\ast, \\
        \left( \frac{q_{\rm UV}}{16\, q_{\rm SW}}\right)^{\frac{a^2}{\epsilon_1 \epsilon_2}}H_{\BC^2}^{(4)}, \qquad &N_f = 4,
    \end{split}
\end{equation}
with $H_{\BC^2}^{(N_f)}$ as in \eqref{eq:defH}. $Z_{\mu}^{(N_f)}$ and $H_{\mu}^{(N_f)}$ are then related by 

\begin{equation}
    \label{eq:Z_H_D_relation}
    H_{\mu}^{(N_f)} = \frac{Z_{\mu}^{(N_f)}}{D^{(N_f)}},
\end{equation}
where 
\begin{equation}
\label{eq:DefDNf}
    D^{(N_f)}(q,M_j,b_j,x,z, \epsilon_i)=\left\{ 
    \begin{array}{rr}
    \prod_{l=1}^3 C^{(N_f)}( \Lambda\,e^{\iota_{(P_l)}^\ast \Omega/(4-N_f)},M_j^{(l)}, \epsilon_i^{(l)}),   &N_f = 1,2,3, \\ \vspace{-.2cm} & \\
     \prod_{l=1}^3 C^{(4)}(a=0,q^{(l)},M_j^{(l)},\epsilon_i^{(l)}), &N_f = 4, \\ \vspace{-.2cm} & \\
     \prod_{l=1}^3 C^{(\ast)}(q^{(l)},M_j^{(l)},\epsilon_i^{(l)}), & \CN = 2^\ast. 
    \end{array}
    \right.
\end{equation}

Here, for $N_f=4$, $D^{(4)}$ factors out the $a$-independent part for each patch, or $C^{(4)}(0,q_\UV,M_j,\epsilon_i)$.
Explicitly, the above definitions yield
\begin{equation}
    \begin{split}
      &  D^{(0)} = D^{(1)} = 1,\\
    & D^{(2)} = \exp{\left[\frac{(-1 + e^{\eps_2 (\eps_2 x - z)})\eps_1 - (1 - e^{\eps_1 (\eps_1 x - z)})\eps_2)}{2 \eps_1 \eps_2 (\eps_1 - \eps_2)}\Lambda^2 \right]}.
    \end{split}
\end{equation}
For later discussion, observe that $D^{(2)}$ is $b_j$-independent and invariant under $(z, \eps_1, \eps_2) \to  (-z,  -\eps_1, -\eps_2)$. The equivariant expression for $D^{(3)}$ is rather lengthy, hence we do not give it explicitly. However, we note that it does \textit{not} exhibit invariance under $(z, b_j, \eps_1, \eps_2) \to  (-z, -b_j, -\eps_1, -\eps_2)$, which is due to the term $\sim \Lambda \frac{\eps_1 + \eps_2}{\eps_1 \eps_2}$ in $C^{(3)}$ in \eqref{eq:instrecrel_def_C}. 

Further, we observe that for vanishing observables $x = z = 0$, we have $D^{(N_f)}  = 1$ for $N_f = 0,1,2,3$. For $N_f = 2,3$, the non-equivariant limits $\lim_{\eps_i \to 0} D^{(N_f)} (\eps_i)$ with non-vanishing observables $x, z \neq 0$ are given by
\be
\label{eq:factorNf}
\begin{split}
&D^{(N_f)}(\Lambda,M_j,B_j, x,z)\\
&\quad =\left\{ \begin{array}{rr} e^{-\Lambda^2 (2x+z^2)/4}, & \quad N_f=2,\\
\exp\left[ \frac{1}{64} \Lambda (-16(6+\sum_j B_j)z - \Lambda z^2 - (\Lambda + 16 \sum_j M_j)(2x + z^2) )\right], & \quad N_f=3.
\end{array}
\right.
\end{split}
\ee
For $N_f=4$, we have equivariantly and with vanishing observables
\be
\begin{split}
D^{(4)}(q_\UV, B_j, \eps_i)= (1-q_\UV)^{\frac{1}{16}(6+\sum_j B_j)^2}\,\vartheta_3(2\tau_\SW)^{-9+\sum_j B_j^2/2}.
\end{split}
\ee
and
\be
\begin{split}
\label{eq:Dstar_explicit}
&D^{(*)}(q_\UV, B, \eps_i)= \left(\prod_{n=1}^\infty (1-q_{\UV}^n)\right)^{-\frac{3+B^2}{2}}.
\end{split}
\ee
Note that both $D^{(4)}$ and $D^{(\ast)}$ are independent of $\epsilon_i$ without taking the non-equivariant limit, and the expressions are independent of $M_j$. We can extrapolate these functions to a generic four-manifold $X$ with Euler number $\chi$ and signature $\sigma$,
\be
\begin{split}
&D^{(4)}_X(q_\UV, B_j)=(1-q_\UV)^{\frac{1}{16}(2c_1(T_X)+\sum_j B_j)^2}\,\vartheta_3(2\tau_\SW)^{-(2\chi+3\sigma)+\sum_j B_j^2/2},\\
&D^{(*)}_X(q_\UV, B_j)=\left(\prod_{n=1}^\infty (1-q_{\UV}^n)\right)^{-2\chi- (B^2-2\chi-3\sigma)/2}.
\end{split}
\ee
For $\CN=2^*$, this agrees with the power of the Dedekind $\eta$-function for a four-manifold  \cite{Manschot:2021qqe}. The action of $\CS$ and $\CT$ \eqref{eq:STtriality} does not leave invariant $D^{(4)}_X(q_\UV, B_j)$. One finds
\be
\begin{split}
&\CS:\qquad  D^{(4)}_X(\CS q_\UV, \CS B_j)=(q_\UV)^{\frac{1}{16}(2c_1(T_X)+2M_1)^2}\,\vartheta_3(2\tau_\SW)^{-(2\chi+3\sigma)+\sum_j B_j^2/2}\\
&\CT:\qquad  D^{(4)}_X(\CT q_\UV, \CT B_j)= (1-q_\UV)^{-\frac{1}{16}(2c_1(T_X)-2M_4+\sum_j B_j)^2}\\
&\hspace{4cm}\qquad \times \,\vartheta_4(2\tau_\SW)^{-(2\chi+3\sigma)+\sum_j B_j^2/2}.
\end{split}
\ee

\subsubsection*{Mass decoupling limit for \texorpdfstring{$N_f = 1$}{Nf 1}}
As a first consistency check of our partition function $Z^{(N_f)}_{\mu}$ \eqref{eq:Zfinal_c1odd_new} and $H^{(N_f)}_{\mu}$ \eqref{eq:Hfinal_c1odd}, we check that the large mass decoupling limit of the $N_f=1$ theory yields the $N_f = 0$ pure theory partition function. In the following, we derive this consistency explicitly for $Z^{(N_f)}_{\mu}$, and expect analogous derivations to hold for the $N_f = 3 \to N_f = 2$ and $N_f = 2 \to N_f = 1$ decoupling limits.\par
First note in \eqref{eq:Zfinal_c1odd_new}, the factors $ \Tilde{W}(2p; 1, \epsilon_i) \Tilde{W}(-2p; 1, \epsilon_i)$ in $\Tilde{Z}_{\rm pert}^{(1)}$ \eqref{eq:Zpert_p2_reduced}, the $2a_+^{(l)}$ in $ \Tilde{Z}_{\rm inst}^{(1)}$ \eqref{eq:Zinst_P2_reduced} and the factors from the $ \prod''$ products in $\CR^{(1)}_{p_l, p_{l+1}}$ \eqref{eq:InstRecRel_Nf1_Rmn} already appear in the pure theory partition function $Z_{\mu}^{(0)}$. 

To recover the pure theory from the $N_f = 1$ theory, we take the mass of the hypermultiplet to infinity and consider the double scaling limit \cite{Seiberg:1994aj}
\begin{equation}
\label{eq:doublescalelimit}
  M \to \infty, \quad \Lambda \to 0, \quad \text{while} \  \Lambda_0^4 = M \Lambda^3 \ \text{is fixed}.
\end{equation}
Here $\Lambda_0$ denotes the scale of the pure theory. Now, according to \eqref{eq:doublescalelimit}, we substitute $M= \Lambda_0^4 / \Lambda^3$ in $Z_{\mu}^{(1)}$
\eqref{eq:Zfinal_c1odd_new}. 
From the $M$-dependent part of $\Tilde{Z}_{\rm pert}^{(1)}$ and the numerators of $\CR^{(1)}_{p_l, p_{l+1}}$ \eqref{eq:InstRecRel_Nf1_Rmn}  appearing in $\Tilde{Z}_{\rm inst}^{(1)}$, we factor out an overall factor of 
\begin{equation}
    \label{eq:Massdecouple_lambdas1}
    \left( \frac{\Lambda_0^4}{\Lambda^3} \right)^{\frac{Q(p)}{4}-\frac{1}{4}(-1 + B^2)}
\end{equation}
after the substitution.
The remaining factors from the numerator of $\CR^{(1)}_{p_l, p_{l+1}} $ then have the limit
\begin{equation} 
    {\prod_{r,s}}' \frac{1}{2} \left(2 - \frac{\Lambda^3}{\Lambda_0^4} (r\epsilon_1 + s \epsilon_2) \right) \ \overset{\Lambda \to 0}{\longrightarrow} 1,
\end{equation}
and similarly for the remaining factors from the functions $W(M)$ in $\Tilde{Z}_{\rm pert}^{(1)}$. 
\eqref{eq:Massdecouple_lambdas1} combines with the  $\Lambda^{d(B)} \Lambda^{3 \frac{Q(p)}{4}}$ in \eqref{eq:Zfinal_c1odd_new} to an overall
\begin{equation}
    \Lambda_0^{Q(p) - 3} \times \left( \frac{\Lambda_0}{\Lambda}\right)^{-B^2 + 4}.
\end{equation}
Here the first factor corresponds to the scale prefactor of a summand of the pure theory full Nekrasov partition function on $\mathbb{CP}^2$, as expected from \eqref{eq:vdim_def}.
The second factor, independent of the gauge fluxes $p$, suggests to normalize the function in the limit $M \to \infty$ by an overall factor of \begin{equation}
    \left( \frac{\Lambda_0}{\Lambda}\right)^{B^2 - 4}.
\end{equation} 
This is in accordance with the results from \cite{Aspman:2022sfj}, equation (3.21). 
In the double scaling limit the three  instanton functions $Z^{(1)}_{\BC^2, \rm inst}$ in \eqref{eq:Zinst_P2_reduced}  become the instanton partition functions $Z^{(0)}_{\rm inst, \BC^2}$ from the pure theory. This can be seen from the sum over partitions:  For a fixed set of Young diagrams $\Vec{Y}$, consider a summand in \eqref{eq:Zinst_iterative}. Combining the overall factor $\left(\Lambda^3\right)^{|\Vec{Y}|}$ with the numerator of the fundamental factor \eqref{eq:Zinst_iterative_fund}, where we substitute $M$ as in  \eqref{eq:doublescalelimit}, gives a product containing a number of $|\Vec{Y}|$ factors of the form
\begin{equation}
    \left(\Lambda^3 \ \chi_{\alpha, s_\alpha} + \Lambda^3 \frac{\Lambda_0^4}{\Lambda^3}\right)  \ \overset{\Lambda \to 0}{\longrightarrow} \ \Lambda_0^4,
\end{equation}
for $\alpha=1,2, \ s_\alpha \in Y_\alpha$.
Potential mass-shifts by the background fluxes $b_i$ vanish in the limit the same way as $\chi$.
Hence we are left with only the vector part and $Z^{(1)}_{\rm inst, \BC^2}$ goes to $Z^{(0)}_{\rm inst, \BC^2}$. Lastly, the observable factor $\Tilde{\CO}$ \eqref{eq:ZP2_O_reduced} is not affected by the decoupling limit.\\
Combining all these findings, we see that in the double scaling limit,  
\begin{equation}
    \left( \frac{\Lambda_0}{\Lambda}\right)^{B^2 - 4}  Z^{(1)}_{\mu} \ \longrightarrow \  Z^{(0)}_{\mu}.
    \label{eq:Z1_final_decouplinglimit}
\end{equation}

\subsubsection*{$\epsilon_i$-reflection symmetry for \texorpdfstring{$H_{\mu}$}{H} on \texorpdfstring{$\mathbb{CP}^2$}{P2}}
The analysis of the $u$-plane integral \cite[Eq. (A.14)]{Aspman:2022sfj}
demonstrates the $u$-plane integral changes by the sign $(-1)^{2\mu+1}$ under the reflection $(z,B_j)\to -(z,B_j)$.
Using the $\epsilon_i$-reflection symmetry on $\BC^2$ \eqref{eq:HC2_epsilon_flip}, we can extend this symmetry for $N_f= 0,1,2,3,4,\ast$ on $\mathbb{CP}^2$ to 
\begin{equation}
    \label{eq:HP2_c1odd_eps_flip}
    H_{\mu}^{(N_f)} (\Lambda, q,  M_j, -b_j, x, -z, -\epsilon_i) = (-1)^{2\mu+1}  H_{\mu}^{(N_f)} ( \Lambda, q, M_j, b_j, x,z,\epsilon_i ).
\end{equation}
A proof using explicit formulae is given in appendix \ref{App:eps_refl_sym_proof}. 

By \eqref{eq:Z_H_D_relation} and the subsequent discussion of $D^{(N_f)}$, we can deduce from \eqref{eq:HP2_c1odd_eps_flip}, that the above symmetry also holds equivariantly for $Z_{\mu}^{(N_f)}$ for $N_f = 0,1,2, \ast$, while it does not hold for $N_f = 3,4$ for general $z, B_j$.

\section{Explicit results}
\label{sec:ExResults}
We use our formula \eqref{eq:Zfinal_c1odd_new} to compute explicit results for correlation functions. For each $1\leq N_f\leq 4$ and $\CN=2^*$, we provide tables for $U(2)$, $SO(3)$ and $SU(2)$ bundles, where we let ``{\bf $SO(3)$ bundles}'' refer to those bundles with $w_2(P)$ odd, and ``{\bf $SU(2)$ bundles}'' to those bundles with $w_2(P)$ even, ie the $SO(3)$ bundles which can be lifted to an $SU(2)$ bundle. We concentrate in this section on presenting the explicit results, while the following Section \ref{sec:compwuplane}, will provide a comparison of our results with the literature.\par

While we can easily determine equivariant correlation functions with distinct masses $M_j$, the expressions are rather lengthy. We will therefore restrict to list the non-equivariant limits of correlators with equal masses $M_j=M$. The limits are given for either purely point correlators ($z=0$) and purely surface correlators ($x=0$). The results pass the non-trivial consistency check that the non-equivariant limit is finite, except for $\CN=2^*$ with $B=\pm 3$. Moreover, the non-equivariant limit only depends on the total non-equivariant background flux $B=\sum_{l=1}^3 b_l$, and is independent of the distribution of equivariant fluxes $b_l$ for fixed $B$. 

As a further consistency check, we have examined whether the  correlators for $N_f = 1,2,3$ hypermultiplets with generic masses $M_j$ reduce to the corresponding correlators for $N_f-1$ hypermultiplets in the double scaling limit $M_{N_f} \to \infty, \ \Lambda = \Lambda_{N_f} \to 0$, where \cite{Seiberg:1994aj}
\begin{equation}
    M_{N_f}\Lambda^{4-N_f}_{N_f} = \Lambda_{N_f-1}^{4 - (N_f - 1)}.
\end{equation}
Following \cite{Marino:1998tb, Aspman:2022sfj}, we multiply the $N_f$ correlation function by an overall factor before taking the limit, namely
\begin{equation}
\label{eq:MassDecoup}
    \left( \frac{\Lambda_{N_f}}{\Lambda_{N_f -1}} \right)^{-\alpha}\left<\CO_1\dots \CO_p \right>_{N_f}\to \left<\CO_1\dots \CO_p \right>_{N_f-1},
\end{equation}
where, for $\mathbb{CP}^2$,
\begin{equation}
    \alpha = \frac{1}{4} \left( (5 - N_f)c_1(\CL_{N_f})^2 + \sum_{j=1}^{N_f - 1} c_1 (\CL_j)^2 - 16 \right).
\end{equation}

\subsection{\texorpdfstring{$N_f = 1$}{Nf = 1} hypermultiplet}
\paragraph{$SO(3)$ bundles} \mbox{}\\
For the case of a (proper) $SO(3)$ gauge bundle, all semistable equivariant gauge flux configurations are strictly stable. In this case, we directly compute the correlation function $Z^{(1)}_{1/2}$ by \eqref{eq:Zfinal_c1odd_new}, using the explicit choice of observables as in \eqref{eq:P2_obs_Omega_choice}.
The non-equivariant limits of the point correlators for $w_2(P)$ odd with $N_f = 1$ fundamental hypermultiplet are listed in Table \ref{tab:oddptcorrelator}. We observe that they are identical for $\pm B$, in agreement with \eqref{eq:HP2_c1odd_eps_flip}.

Explicitly taking the decoupling limit \eqref{eq:Z1_final_decouplinglimit} of the polynomials in Table \ref{tab:oddptcorrelator} according to \eqref{eq:doublescalelimit}, \eqref{eq:Z1_final_decouplinglimit}, recovers precisely the corresponding terms in known expansions for Donaldson invariants, for example in \cite{ellingsrud1995wall, Gottsche:1996aoa} and \citep[(3.71)]{Bershtein:2015xfa} for $z=0$. In particular, the coefficients $\frac{19}{32}x^2$, $\frac{85}{768}x^4$ appear for each of the values of $B$ at instanton numbers $k=11/4$ and $k=15/4$ respectively. These correspond to the  Donaldson invariants for $\mathbb{CP}^2$ with the insertions of two and four point observables respectively.  

The surface correlators are listed in Table \ref{tab:oddsfcorrelator} for positive values of $B$. The partition function for negative values of $B$ follows from  $Z_{1/2}(M,-B,0,-z,)=Z_{1/2}(M,B,0,z)$. The coefficients $\frac{1}{128}z^4, \frac{29}{1290240}z^8$ survive the mass decoupling limit to $N_f=0$, and precisely reproduce the 4- and 8-fold surface observables \citep[(3.71)]{Bershtein:2015xfa} for $x=0$. 

\setlength\tabcolsep{6pt}
\begin{center}
\renewcommand{\arraystretch}{1.5}
\begin{table}[h!]
\begin{tabular*}{\linewidth}{@{\extracolsep{\fill}}|c|r|}
\hline
$\pm B$ & $Z^{(1)}_{1/2}(M,B,x,0)$  \\
\hline 
0 & $\frac{M}{\Lambda}+ \frac{7}{16}x\Lambda^2 + \frac{19}{32} x^2 M^2 \Lambda^2+ \left( \frac{7}{32} x^3 M+\frac{85}{768}x^4M^3\right)\Lambda^5+\left( \frac{1093}{98304}x^4+\dots\right)\Lambda^8 + \dots
$   \\
2 & $1+\frac{19}{32} x^2 M \Lambda^3+\left( \frac{11}{192} x^3 +\frac{85}{768}x^4M^2\right)\Lambda^6+\left(\frac{7613}{491520}x^5M+\dots \right)\,\Lambda^9+\dots$  \\
4 & $\frac{\Lambda^3}{M^3}+\left(\frac{15}{16M^6}-\frac{21x}{16M^4}+\frac{19x^2}{32M^2}\right)\Lambda^6+(\frac{33}{32M^9}-\frac{3x}{2M^7}+\frac{67x^2}{64M^5}-\frac{41x^3}{96M^3} + \frac{85x^4}{768 M})\Lambda^9+\dots$ \\
6 & $\frac{\Lambda^8}{M^8}+\left(\frac{45}{8M^{11}}-\frac{7x}{2M^9}+\frac{19x^2}{32M^7} \right)\Lambda^{11}$ \\
  & $+\left(\frac{591}{32M^{14}}-\frac{263x}{16M^{12}}+\frac{201x^2}{32M^{10}} -\frac{79x^3}{64M^8}+\frac{85x^4}{768M^6}\right) \Lambda^{14}+\dots$\\
\hline
\end{tabular*}
\caption{$N_f = 1$ correlation functions $Z^{(1)}_{1/2}(M,B,x,0)$ for point observables and for different choices of $B$. 
}
\label{tab:oddptcorrelator}
\end{table}
\end{center}

\begin{center}
\renewcommand{\arraystretch}{1.5}
\begin{table}[h!]
\begin{tabular*}{\linewidth}{@{\extracolsep{\fill}}|c|r|}
\hline
$B$ & $Z^{(1)}_{1/2}(M,B,0,z)$  \\
\hline 
0 & $\frac{M}{\Lambda}+ \left(
    \frac{z^2}{32}+\frac{M^2z^4}{128}\right)\Lambda^2+\left(
    \frac{Mz^6}{3840}+\frac{29M^3z^8}{1290240}\right)\Lambda^5+ \left( \frac{29z^8}{18350080}+\dots\right)\Lambda^8+\dots$   \\
2 & $1+\left(\frac{z^3}{32}+\frac{Mz^4}{128}\right)\Lambda^3+\left(\frac{11z^6}{23040}+\dots \right)\Lambda^6+\left(\frac{6787}{1486356480} z^9+\dots\right)\,\Lambda^9+\dots$  \\
4 & $\frac{\Lambda^3}{M^3}+\left(\frac{15}{16M^6}-\frac{3z}{4M^5}-\frac{3z^2}{32M^4}+\frac{z^3}{16M^3}+\frac{z^4}{128M^2}\right)\Lambda^6+\left(\frac{33}{32M^9}-\frac{9z}{8M^8} +\dots \right)\Lambda^9+\dots$ \\
6 & $\frac{\Lambda^8}{M^8}+\left( \frac{45}{8M^{11}}+\frac{3z}{M^{10}}-\frac{z^2}{4M^9}-\frac{3z^3}{32M^8}+\frac{z^4}{128M^7}\right)\Lambda^{11}+\left(\frac{591}{32M^{14}}+\frac{33z}{2M^{13}}+\dots\right)\Lambda^{14}+\dots$\\
\hline
\end{tabular*}
\caption{$N_f = 1$ correlation functions $Z^{(1)}_{1/2}(M,B,0,z)$ for surface observables  and  for different choices of $B$. 
}
\label{tab:oddsfcorrelator}
\end{table}
\end{center}

\paragraph{$SU(2)$ bundles} \mbox{} \\
We list the partition function for point observables for $N_f = 1$ in Table \ref{tab:evenptcorrelator} for $B<0$. In agreement with the non-equivariant limit of \eqref{eq:HP2_c1odd_eps_flip}, we observe also computationally that $Z^{(1)}_{0}(M,-B, x, 0) = - Z^{(1)}_{0}(M,B, x, 0)$. All point correlators vanish in the large mass decoupling limit, as expected from \citep[(3.72)]{Bershtein:2015xfa}.

Also in Table \ref{tab:evenptcorrelator}, we recognize coefficients $\frac{3}{2}, \frac{13}{16}, \frac{293}{2048}$ times $|B|/2$ at $k=1,2,3$.
These coincide with Donaldson invariants for the $N_f=0$ theory as in equation \citep[(3.72)]{Bershtein:2015xfa} at $\CO (z)$. This is not a coincidence but suggests that the insertion of a surface observable and the characteristic class of the index bundle give rise to the same intersection number.

Surface correlators for $N_f = 1$ are listed in Table \ref{tab:evensfcorrelator}. The mass decoupling limit (\ref{eq:MassDecoup}) yields for each choice of $B$,
\begin{equation}
    - \frac{3 z}{2} \Lambda_0 + \frac{z^5}{120}\Lambda_0^5 + \frac{z^9}{120960} \Lambda_0^9+\dots.
\end{equation}
This precisely coincides with the Donaldson invariants for the surface correlators of the pure theory. 

\begin{center}
\renewcommand{\arraystretch}{1.5}
\begin{table}[h!]
\centering
\begin{tabular*}{\linewidth}{@{\extracolsep{\fill}}|c|r|}
\hline
$B$ & $Z^{(1)}_0(M,B,x,0)$  \\
\hline 
$-1$ & $\frac{3}{4}+\frac{13Mx^2}{32}\Lambda^3+\left(\frac{113x^3}{3072}+\frac{293M^2x^4}{4096}\right) \Lambda^6+ 
\left( \frac{1879 M x^5}{196608} + \frac{815 M^3 x^6}{131072} \right)\Lambda^9 $\\
& $+ \left(\frac{33 x^6}{131072} +  \frac{7241 M^2 x^7}{7340032} + \frac{170053 M^4 x^8}{528482304}  \right)\Lambda^{12} +\dots$   \\
$-3$ & $\frac{9}{4M^2}\Lambda^2+\left(\frac{15}{16M^5}-\frac{31x}{16M^3}+\frac{39x^2}{32M}\right)\Lambda^5$ \\
  & $+\left( \frac{175}{256M^8}-\frac{315x}{256M^6}+\frac{567x^2}{512M^4}-\frac{289x^3}{512M^2}+\frac{879x^4}{4096}\right)\Lambda^8+\dots$  \\
$-5$ & $\frac{15}{4 M^6}\Lambda^6+\left(\frac{91}{8M^9} -\frac{155x}{16M^7}+\frac{65x^2}{32M^5}\right)\Lambda^9$ \\
  & $+\left(\frac{12155}{512M^{12}}-\frac{1719x}{64M^{10}}+\frac{13391x^2}{1024M^8}-\frac{1225x^3}{384M^6}+\frac{1465x^4}{4096M^4}\right)\Lambda^{12}+\dots$
\\
\hline
\end{tabular*}
\caption{$N_f =1$ correlation functions $Z^{(1)}_0(M,B,x,0)$ for different choices of $B$.
}
\label{tab:evenptcorrelator}
\end{table}
\end{center}

\begin{center}
\renewcommand{\arraystretch}{1.5}
\begin{table}[!h]
\centering
\begin{tabular*}{\linewidth}{@{\extracolsep{\fill}}|c|r|}
\hline
$B$ & $Z^{(1)}_0(M,B,0,z)$  \\
\hline 
$-1$ & $\frac{3}{4}-\frac{3Mz}{2}+\left(\frac{z^3}{24}-\frac{Mz^4}{48}+\frac{M^2z^5}{120} \right)\Lambda^3+\left(-\frac{z^6}{5760}+\frac{Mz^7}{6720} -\frac{M^2z^8}{26880} + \frac{M^3 z^9}{120960}\right)\Lambda^6$ \\
$-3$ & $\left(\frac{9}{4M^2}-\frac{3z}{2M} \right)\Lambda^2+\left(\frac{15}{16M^5}+\frac{3z}{4M^4}-\frac{z^2}{M^3}+\frac{3z^3}{8M^2}-\frac{z^4}{16M}+\frac{z^5}{120} \right)\Lambda^5$ \\
  & $+\left(\frac{175}{256M^8} +\dots \right)\Lambda^8+\dots $ \\
$-5$ & $\left(\frac{15}{4M^6}-\frac{3z}{2M^5}\right)\Lambda^6+\left(\frac{91}{8M^9}+\frac{35z}{8M^8}-\frac{5z^2}{M^7}+\frac{25z^3}{24M^6} -\frac{5z^4}{48M^5}+\frac{z^5}{120M^4}\right)\Lambda^9+\dots$ \\
\hline
\end{tabular*}
\caption{$N_f =1$ correlation functions $Z^{(1)}_0(M,B,0,z)$ for different choices of $B$.}
\label{tab:evensfcorrelator}
\end{table}
\end{center}

\subsection{$N_f = 2$ hypermultiplets}

\paragraph{$SO(3)$ bundles} \mbox{} \\
The point correlators for $w_2(P)$ odd with $N_f = 2$ fundamental hypermultiplets and equal masses are listed in Table \ref{tab:c1odd_Nf2_results_x}. 
Taking the $M_2$ mass decoupling limit for the $B_1 =0,2, 4$ point correlators,
we recover the $B=0,2, 4, \ N_f=1$ correlators respectively, as listed for equal masses in Table \ref{tab:oddptcorrelator}. 
Consistently, we again observe the $SO(3)$ pure theory Donaldson invariants $\frac{19}{32}, \frac{85}{768}, \frac{29557}{2949120}$ appearing for $x^2, x^4, x^6$ and instanton numbers $k=\frac{11}{4}, \frac{15}{4}, \frac{19}{4}$ respectively. 

While the decoupling limit reproduces the pure Donaldson invariants, for comparison with the $u$-plane results on gauge groups $SO(3)$ and $SU(2)$ in the next section, we need to consider the modified partition function $H_\mu^{(2)}$ \eqref{eq:Z_H_D_relation} as discussed in Section \ref{Chap:Partitionfctn_P2}. Results for this function are presented in Table \ref{tab:c1odd_Nf2_results_x}. 

\begin{center}
\renewcommand{\arraystretch}{1.5}
\begin{table}[h!]\centering
\begin{tabular*}{\linewidth}{@{\extracolsep{\fill}}|c|r|}
    \hline
    $(B_1, B_2 )$ & $Z^{(2)}_{1/2} (M,B_j,x, 0)$ \\
    \hline
       $(0,0)$  &  $\frac{M^2}{\Lambda^2} + \frac{3}{16} + \frac{3}{8} M^2 x + \frac{19}{32} M^4 x^2 +  \left( -\frac{3}{32}x + \frac{3}{64} M^2 x^2 + \frac{9}{64} M^4 x^3 + \frac{85}{768} M^6 x^4 \right) \Lambda^2$ \\ & $+  \left(\frac{245}{8192}x^2  - \frac{139}{6144}M^2 x^3 + \frac{53}{16384} M^4 x^4 + \frac{2357}{122880} M^6 x^5 + \frac{29557}{2949120} M^8 x^6 \right) \Lambda^4 + \dots$ \\
    $(2,0)$ & $\frac{M}{\Lambda} + \frac{1}{32} x (-2M + 19 M^3x)\Lambda + \left( \frac{M x^2}{64} - \frac{M^3 x^3}{48} + \frac{85 M^5 x^4}{768} \right) \Lambda^3
       $ \\
       & $+\left( -\frac{83 M x^3}{24576} + \frac{383 M^3 x^4}{98304} - \frac{1273 M^5 x^5}{491520} + \frac{29557 M^7 x^6}{2949120} \right) \Lambda^5 + \dots$ \\
       $(4, 0)$ & $\frac{\Lambda^2}{M^2} + \left( \frac{3}{8M^4}  - \frac{11x}{8M^{2}}  + \frac{19 x^2}{32} \right) \Lambda^4$ \\ & $+ \left( \frac{3}{16 M^6} - \frac{9x}{16M^4} + \frac{31x^2}{32M^2} - \frac{97 x^3}{192}+ \frac{85 M^2 x^4}{768}\right) \Lambda^6 + \dots$\\
        \hline  
    \end{tabular*}
    \caption{$N_f = 2$ correlation functions $Z^{(2)}_{1/2}(M,B_j,x,0)$ for point observables for different choices of $B_j$ in the equal mass $M_1 = M_2 = M$, non-equivariant limit.}
    \label{tab:c1odd_Nf2_results_x}
\end{table}
\end{center}

\begin{center}
\renewcommand{\arraystretch}{1.5}
\begin{table}[h!]
\begin{tabular*}{\linewidth}{@{\extracolsep{\fill}}|c|r|}
    \hline
    $(B_1, B_2 )$ & $H^{(2)}_{1/2} (M,B_j, x, 0)$ \\
    \hline
       $(0,0)$  &  $\frac{M^2}{\Lambda^2} + \frac{3}{16}  + \frac{7}{8} M^2 x + \frac{19}{32} M^4 x^2 +  \left(\frac{23}{64} M^2x^2 + \frac{7}{16} M^4 x^3 + \frac{85}{768} M^6 x^4 \right)  \Lambda^2$ \\ & $+   \left(\frac{53}{8192}x^2 + \frac{421}{6144}M^2 x^3 + \frac{2421}{16384} M^4 x^4 + \frac{9157}{122880} M^6 x^5 + \frac{29557}{2949120} M^8 x^6 \right) \Lambda^4 + \dots$ \\
          $(2,0)$ & 
        $\frac{M}{\Lambda} + \left( \frac{7}{16} Mx + \frac{19}{32} M^3 x^2 \right) \Lambda + \left( \frac{7}{64} M x^2 + \frac{53}{192} M^3 x^3 + \frac{85}{768} M^5 x^4 \right) \Lambda^3 $\\
        & $+ \left( \frac{143}{8192} M x^3 + \frac{6655}{98304} M^3 x^4 + \frac{25927}{491520} M^5 x^5 + \frac{29557}{2949120} M^7 x^6 \right) \Lambda^5 + \dots$ \\
       $(4, 0)$ & $\frac{\Lambda^2}{M^2} + \left( \frac{3}{8} \frac{1}{M^4} - \frac{7}{8} \frac{x}{M^2} + \frac{19}{32} x^2 \right) \Lambda^4$\\
       & $+ \left( \frac{3}{16} \frac{1}{M^6} - \frac{3}{8} \frac{x}{M^4} + \frac{13}{32} \frac{x^2}{M^2} - \frac{5}{24} x^3 + \frac{85}{768} M^2 x^4 \right) \Lambda^6 + \dots$ \\
       \hline  
    \end{tabular*}
    \caption{$N_f = 2$ correlation functions $H^{(2)}_{1/2}(M,B_j, x,0)$ for point observables for different choices of $B_j$ in the equal mass $M_1 = M_2 = M$, non-equivariant limit.}
    \label{tab:H_c1odd_Nf2_results_x}
\end{table}
\end{center}

\paragraph{$SU(2)$ bundles} \mbox{} \\
In Table \ref{tab:evenptcorrelator_Nf2} we list our nonequivariant results for the point correlators for $w_2 (P)=0$, $N_f = 2$ with equal masses $M_1 = M_2 = M$. 
We find some coefficients  $\frac{3}{2}, \frac{13}{16}, \frac{293}{2048}$ appearing (up to some overall constant) as in the pure theory partition function (cf. equation (3.72), \cite{Bershtein:2015xfa}). For generic masses $M_1, M_2$, we find that the mass decoupling limits $M_2\to \infty$ of the $B=(1,1), (3,1)$ correlators reproduce the respective $B=1,3, \ N_f=1$ point correlators as in Table \ref{tab:evenptcorrelator}, up to the respective order in $\Lambda$. Consistently, the $M_1 \to \infty$ decoupling limits of the $B=(1,1), (3,1)$ correlators reproduce the $B=1$ point correlator.

\begin{center}
\renewcommand{\arraystretch}{1.5}
\begin{table}[h!]
\centering
\begin{tabular*}{\linewidth}{@{\extracolsep{\fill}}|c|r|}
\hline
$(B_1, B_2)$ & $Z^{(2)}_0(M,B_j, x,0)$ \\
\hline 
$(1,1)$ & $- \frac{3 M}{2 \Lambda} - \frac{1}{16}  M x (-2 + 13 M^2 x)\Lambda $\\ & $- M x^2 (\frac{15}{512} - \frac{29}{768} M^2 x + \frac{293}{2048} M^4 x^2) \Lambda^3 + \dots$\\
$(3,1)$ & $ - \frac{3}{M} \Lambda + \left( \frac{1}{2M^3} + \frac{5 x}{2M} - \frac{13 M x^2}{8} \right)  \Lambda^3 $\\ 
&$+ \left( \frac{1}{4 M^5} - \frac{3x}{4M^3} - \frac{7 x^2}{8M} + \frac{655 M x^3}{768} - \frac{293 M^3 x^4}{1024} \right) \Lambda^5 + \dots$ \\
\hline
\end{tabular*}
\caption{Correlation functions $Z^{(2)}_0(M,B_j,x,0)$ for different choices of $B_1, B_2$, for $w_2$ even, $N_f =2$, equal masses $M$. 
}
\label{tab:evenptcorrelator_Nf2}
\end{table}
\end{center}

\begin{center}
\renewcommand{\arraystretch}{1.5}
\begin{table}[h!]
\centering
\begin{tabular*}{\linewidth}{@{\extracolsep{\fill}}|c|r|}
\hline
$(B_1, B_2)$ & $H^{(2)}_0(M,B_j, x,0)$ \\
\hline 
$(1,1)$ & $- \frac{3}{2} \frac{M}{\Lambda} - \left( \frac{5}{8} Mx + \frac{13}{16} M^3 x^2\right) \Lambda $\\ 
& $- \left(  \frac{79}{512} M x^2 + \frac{283}{768} M^3 x^3 + \frac{293}{2048} M^5 x^4 \right) \Lambda^3 + \dots$\\
$(3,1)$ & $- 3 \frac{\Lambda}{M} + \left( \frac{1}{2} \frac{1}{M^3} + \frac{x}{M} - \frac{13}{8} M x^2 \right) \Lambda^3 $ \\
&$+ \left( \frac{1}{4} \frac{1}{M^5} - \frac{1}{2} \frac{x}{M^3}  + \frac{31}{768} M x^3 - \frac{293}{1024} M^3 x^4 \right) \Lambda^5+ \dots$\\
\hline
\end{tabular*}
\caption{Correlation functions $H^{(2)}_0(M,B_j, x,0)$ for different choices of $B_1, B_2$, for $w_2$ even, $N_f =2$, equal masses $M$. 
}
\label{tab:H_evenptcorrelator_Nf2}
\end{table}
\end{center}

\subsection{\texorpdfstring{$N_f = 3$}{Nf = 3} hypermultiplets}

\paragraph{$SO(3)$ bundles} \mbox{} \\
The generating function $Z^{(3)}_{1/2}$ for point observables for $N_f = 3$ fundamental hypermultiplets with equal mass $M$ is listed in Table \ref{tab:c1odd_Nf3_results_x}, and $H^{(3)}_{1/2}$ in Table \ref{tab:c1odd_Nf3_results_x}. We have checked that decoupling the mass $M_3$ in the $B =(0,0,0)$, point correlator yields precisely the $(M_1, M_2)$-dependent $N_f =2$ point correlators.

\paragraph{$SU(2)$ bundles} \mbox{} \\
We list the generating function $Z^{(3)}_0$ for $N_f = 3, B= (1,1,1)$ and equal masses in Table \ref{tab:evenptcorrelator_Nf3}, and $H^{(3)}_0$ in Table \ref{tab:evenptcorrelator_Nf3_H}. The $M_3$ decoupling limit yields again the expected $N_f=2$ result.

\begin{table}[h!]
    \centering
\renewcommand{\arraystretch}{1.5}
\begin{tabular*}{\linewidth}{@{\extracolsep{\fill}}|c|r|}
    \hline
    $(B_1, B_2, B_3)$ & $Z^{(3)}_{1/2} (M,B_j,  x, 0)$ \\
    \hline
       $(0,0,0)$  & $\frac{M^3}{\Lambda^3} + \frac{1}{32}\frac{M^2}{\Lambda^2} \left(18 - 6 M^2 x + 19 M^4 x^2\right) $ \\
       & $ + \frac{5}{768} \frac{M^3}{\Lambda} x \left( -72 + 36 M^2 x - 36 M^4 x^2 + 17 M^6 x^3 \right) $ \\
       & $+ \frac{5}{4096} + \frac{15 M^2 x}{2048} + \frac{2235 M^4 x^2}{8192} - \frac{1015 M^6 x^3}{6144} + \frac{3065 M^8 x^4}{32768}$ \\
       & $- \frac{4443 M^{10} x^5}{81920} + \frac{29557 M^{12} x^6}{2949120}  + \dots$ \\
         \hline
    \end{tabular*}
    \caption{$N_f = 3$ correlation functions $Z^{(3)}_{1/2}(M,B_j, x,0)$ for point observables for different choices of $B$ in the equal mass $M_1 = M_2 = M_3 = M$, non-equivariant limit.}
    \label{tab:c1odd_Nf3_results_x}
\end{table}

\begin{table}[h!]
    \centering
\renewcommand{\arraystretch}{1.5}
  \begin{tabular*}{\linewidth}{@{\extracolsep{\fill}}|c|r|}
    \hline
    $(B_1, B_2, B_3)$ & $H^{(3)}_{1/2}(M,B_j,  x, 0)$ \\
    \hline
       $(0,0,0)$  & $ \frac{M^3}{\Lambda^3} + \left( \frac{9}{16} M^2 + \frac{21}{16} M^4 x + \frac{19}{32} M^6 x^2  \right) \frac{1}{\Lambda^2}  $ \\
       & $+ \left( \frac{13}{32} M^3 x + \frac{69}{64} M^5 x^2 + \frac{21}{32} M^7 x^3 + \frac{85}{768} M^9 x^4 \right) \frac{1}{\Lambda}$ \\
       & $+\frac{5}{4096} + \frac{51}{2048} M^2 x + \frac{1995}{8192} M^4 x^2 + \frac{3419}{6144} M^6 x^3 + \frac{13433}{32768} M^8 x^4 $ \\
       & $ + \frac{9157}{81920} M^{10} x^5 + \frac{29557}{2949120} M^{12} x^6 + \dots$\\
         \hline
    \end{tabular*}
    \caption{$N_f = 3$ correlation functions $H^{(3)}_{1/2}(M,B_j, x,0)$ for point observables for different choices of $B$ in the equal mass $M_1 = M_2 = M_3 = M$, non-equivariant limit.}
    \label{tab:c1odd_Nf3_results_H}
\end{table}

\begin{center}
\renewcommand{\arraystretch}{1.5}
\begin{table}[h!]
\centering
\begin{tabular*}{\linewidth}{@{\extracolsep{\fill}}|c|r|}
\hline
$(B_1, B_2, B_3)$ & $Z^{(3)}_0(M,B_j, x,0)$ \\
\hline 
$(1,1,1)$ & $- \frac{9 M^2 }{4} \Lambda^{-2} + \left( - \frac{3 M }{16} + \frac{11 M^3 x }{8} - \frac{39 M^5 x^2}{32} \right) \Lambda^{-1} $ \\
& $+  \frac{1}{512} + \frac{39 M^2 x   }{256} - \frac{189 M^4 x^2}{256} + \frac{403 M^6 x^3}{512} - \frac{879 M^8 x^4}{4096} + \dots $ \\
\hline
\end{tabular*}
\caption{Correlation function $Z^{(3)}_0(M,B_j, x,0)$ for $(B_1, B_2, B_3) = (1,1,1)$, for $w_2$ even, $N_f =3$, equal masses $M$. 
}
\label{tab:evenptcorrelator_Nf3}
\end{table}
\end{center}

\begin{center}
\renewcommand{\arraystretch}{1.5}
\begin{table}[h!]
\centering
\begin{tabular*}{\linewidth}{@{\extracolsep{\fill}}|c|r|}
\hline
$(B_1, B_2, B_3)$ & $H^{(3)}_0(M,B_j, x,0)$ \\
\hline 
$(1,1,1)$ 
& $  - \frac{9}{4} M^2 \Lambda^{-2} - \left( \frac{3}{16} M + 2 M^3 x + \frac{39}{32} M^5 x^2 \right) \Lambda^{-1} $ \\ 
&$ \frac{1}{512} - \frac{51}{256} M^2 x - \frac{309}{256} M^4 x^2 - \frac{533}{512} M^6 x^3 - \frac{879}{4096} M^8 x^4 + \dots $ \\
\hline
\end{tabular*}
\caption{Correlation function $H^{(3)}_0(M,B_j, x,0)$ for $(B_1, B_2, B_3) = (1,1,1)$, for $w_2$ even, $N_f =3$, equal masses $M$. 
}
\label{tab:evenptcorrelator_Nf3_H}
\end{table}
\end{center}

\subsection{{\texorpdfstring{$N_f = 4$}{Nf = 4} hypermultiplets}}
\label{sec:resultsNf4}
Our next theory is the superconformal $N_f=4$ theory, which is of particular interest due to its triality symmetry discussed in Section \ref{sec:Coulbranch}. The appropriate coupling constant for modular transformations is $\tau_\SW$, or better $2\tau_\SW$, rather than $\tau_\UV$. Indeed, we will observe in the tables below  that the coefficients of the $q_\UV$-series are rational, while upon substitution of (\ref{eq:qUVqSW}) for $q_\UV$, we find that the coefficients of the $q_\SW$-series are integers.

It is expected that the transformations (\ref{eq:STtriality}) leave the partition function invariant, such that the masses multiply (vector-valued) modular forms or mock modular form with the appropriate transformation properties.
From our results, it is straightforward to verify invariance of the partition functions under $\CT$. However, since equivariant localization only provides us with a finite number of terms, we do not have enough information for the verification of $\CS$. We find that the triality group \eqref{eq:STtriality} together with the action on the background fields $B_j$ as in \eqref{eq:actionTBj} does leave the partition functions invariant up to a non-trivial weight and multiplier system. Based on the examples below, we propose for the $\CT$-transformation,
\be
\label{eq:H(4)trafo}
H^{(4)}_{\mu}(\CT q_\UV,\CT M_j,  \CT B_j)=\left\{ \begin{array}{rr} e^{-\pi i/4}H^{(4)}_{1/2}(q_\UV,M_j,  B_j), & \qquad \mu=1/2,\\
\vspace{-.2cm} & \\
H^{(4)}_{0}(q_\UV,M_j,  B_j), & \qquad \mu=0. \end{array} \right.
\ee

This clearly indicates an intriguing structure of the partition functions in terms of polynomial functions of the masses, ie the Seiberg-Witten mass combinations \cite{Seiberg:1994aj}, and (mock) modular forms for congruence subgroups which together are triality invariant. Automorphic forms for $D_4$ triality were recently studied in \cite{Sakai:2023fnp, Sakai:2025ybk}. We leave the precise determination of partition functions in terms of mock modular forms for future work.

\paragraph{$SO(3)$ bundles} \mbox{} \\
Before listing tables, we first list the first two coefficients for generic masses for $B_j=0$,
\be
\begin{split}
Z^{(4)}_{1/2}(q_\UV,M_j,  B_j=0)&=\frac{M_1M_2M_3M_4}{\Lambda^4} \,q_{\UV}^{3/4}\\
&\quad +\frac{3(-16M_1M_2M_3M_4+\sum_{i<j}M_i^2M_j^2)}{16\,\Lambda^4}\,q_\UV^{7/4}+\dots,
\end{split}
\ee
while the function $H^{(4)}_{1/2}$ is
\be
\begin{split}
H^{(4)}_{1/2}(q_\UV,M_j,  B_j=0)&=\frac{M_1M_2M_3M_4}{\Lambda^4} \,q_{\SW}^{3/4} +\frac{3 \sum_{i<j}M_i^2M_j^2}{\Lambda^4}\,q_\SW^{7/4}+\dots.
\end{split}
\ee
Considering the action of $\CT$, we find that $Z^{(4)}_{1/2}$ is {\it not} invariant under $\CT$, due to odd and even powers of $M_4$ in the coefficient of $q_\UV^{7/4}$. However, we deduce based on the first four coefficients of the $q_\SW$-series for $H^{(4)}_{1/2}$ do match with \eqref{eq:H(4)trafo}. As mentioned above, for non-vanishing background fluxes, one needs to include the action on $B_j$ \eqref{eq:actionTBj}.

We list partition functions for the $SO(3)$, $N_f=4$ theory with equal masses  and without observables in Table \ref{tab:oddptcorrelator_Nf4}. We again observe that the $q_\UV$-series of $Z_\mu^{(4)}$ have fractional coefficients, while the $q_\SW$-series have integer coefficients with alternating signs. Moreover, the examples for $H^{(4)}_\mu$ demonstrate coefficients with a smaller growth and non-alternating sign. As mentioned above, we will not attempt here to connect these series for equal masses to modular forms.

Instead we consider a mass configuration for which a result is available in terms of mock modular forms \cite{Malmendier:2008db}. It is derived from the $u$-plane and expressed in terms of mock modular forms. This corresponds to the configuration $M_1=M_2=M$ and $M_3=M_4=0$, which is invariant under triality. 
Table \ref{tab:oddptcorrelator_Nf4MM00} lists the $\epsilon_i\to 0$ limit of the equivariant partition functions $Z^{(4)}_{1/2}$ for two choices of $B_j$ expressed in terms of both $q_\UV$ and $q_\SW$. For the comparison, we also present the first coefficients of $H_{1/2}^{(4)}$ \eqref{eq:Z_H_D_relation} in Table \ref{tab:H12Nf4}. The discussion in Section \ref{eq:compNf4} below indicates that $Z^{(4)}_{1/2}$ has modular weight 0, while $H^{(4)}_{1/2}$ has weight $9/2$ for $B_j=0$.
The latter is consistent with the derivation based on the $u$-plane integral \cite{moore_talk2018}.

A consequence of the triality symmetry is that the coefficients of the $q_\SW$-series of $H^{(4)}_{1/2}$ vanishes for exponents $3/4 \mod 2$. This is indeed borne out by the series in the Tables. The precise result from the $u$-plane integral and comparison is further discussed in Section \ref{sec:compwuplane} below.

\begin{center}
\renewcommand{\arraystretch}{1.5}
\begin{table}[h!]
\centering
\begin{tabular*}{\linewidth}{@{\extracolsep{\fill}}|c|r|}
\hline
$(B_1, B_2, B_3,B_4)$ & $Z^{(4)}_{1/2}(q_\UV,M,B_j)$ \\
\hline 
(0,0,0,0) & $\frac{M^4}{\Lambda^4}q_\UV^{3/4}(1-\frac{15}{8}q_\UV+\frac{15}{32}q_\UV^2+\frac{595}{1024}q_\UV^3-\frac{1095}{8192}q_\UV^4+\dots)    $ \\
& $= 8\frac{M^4}{\Lambda^4}q_\SW^{3/4}(1-36\,q_\SW+567\,q_\SW^2-3928\,q_\SW^3-23751\,q_\SW^4+\dots)$\\
(2,0,0,0) & $\frac{M^3}{\Lambda^3}q_\UV^{3/4}(1-\frac{63}{16}q_\UV+\frac{183}{32}q_\UV^2-\frac{14413}{4096}q_\UV^3+\frac{5115}{8192}q_\UV^4+\dots)$ \\
& $=8\frac{M^3}{\Lambda^3}q_\SW^{3/4}(1-69\,q_\SW+2372\,q_\SW^2-54216\,q_\SW^3+924729\,q_\SW^4+\dots)$\\
\hline
\end{tabular*}
\caption{Partition functions $Z^{(4)}_{1/2}(q_\UV,M, B_j)$ for $w_2$ odd, equal masses $M_j=M$. 
}
\label{tab:oddptcorrelator_Nf4}
\end{table}
\end{center}

\begin{center}
\renewcommand{\arraystretch}{1.5}
\begin{table}[h!]
\centering
\begin{tabular*}{\linewidth}{@{\extracolsep{\fill}}|c|r|}
\hline
$(B_1, B_2, B_3,B_4)$ & $H^{(4)}_{1/2}(q_\UV,M,B_j)$ \\
\hline 
(0,0,0,0) & $8\frac{M^4}{\Lambda^4}q_\SW^{3/4}(1+18\,q_\SW+63\,q_\SW^2+194\,q_\SW^3+315\,q_\SW^4+\dots)$\\
(2,0,0,0) & $8\frac{M^3}{\Lambda^3}q_\SW^{3/4}(1+9\,q_\SW+19\,q_\SW^2+50\,q_\SW^3+51\,q_\SW^4+\dots)$\\
\hline
\end{tabular*}
\caption{Partition functions $H^{(4)}_{1/2}(q_\UV,M, B_j)$ expressed as a $q_\SW$-series for $w_2$ odd, equal masses $M_j=M$. 
}
\label{tab:oddptcorrelator_Nf4H12}
\end{table}
\end{center}

\begin{center}
\renewcommand{\arraystretch}{1.5}
\begin{table}[h!]
\centering
\begin{tabular*}{\linewidth}{@{\extracolsep{\fill}}|c|r|}
\hline
$(B_1, B_2, B_3,B_4)$ & $Z^{(4)}_{1/2}(q_\UV,M_j,B_j)$ \\
\hline 
(0,0,0,0) & $\frac{M^4}{\Lambda^4} q_\UV^{7/4}\left(\frac{3}{16}-\frac{15}{32}\,q_\UV+\frac{735}{2048}\,q_\UV^2-\frac{285}{4096}\,q_\UV^3+\dots\right)$ \\
& $=24 \frac{M^4}{\Lambda^4}\, q_\SW^{7/4} \left( 1-54\,q_\SW+1489\,q_\SW^2-27790\,q_\SW^3+\dots\right)$\\
(2,0,0,0) & $\frac{M^3}{\Lambda^3} q_\UV^{7/4}\left(\frac{3}{16}-\frac{3}{4}\,q_\UV+\frac{4607}{4096}\,q_\UV^2-\frac{3069}{4096}\,q_\UV^3+\dots\right)$ \\
& $=24 \frac{M^3}{\Lambda^3}\, q_\SW^{7/4} \left( 1-78\,q_\SW+9188\,q_\SW^2-80808\,q_\SW^3+\dots\right)$\\
\hline
\end{tabular*}
\caption{Partition functions $Z^{(4)}_{1/2}(q_\UV,M_j, B_j)$ for $w_2$ odd, with masses $M_1=M_2=M$ and $M_3=M_4=0$. 
}
\label{tab:oddptcorrelator_Nf4MM00}
\end{table}
\end{center}

\begin{center}
\renewcommand{\arraystretch}{1.5}
\begin{table}[h!]
\centering
\begin{tabular*}{\linewidth}{@{\extracolsep{\fill}}|c|r|}
\hline
$(B_1, B_2, B_3,B_4)$ & $H^{(4)}_{1/2}(q_\UV,M_j,B_j)$ \\
\hline 
(0,0,0,0) &  $24 \frac{M^4}{\Lambda^4}\, q_\SW^{7/4} \left(1+13\,q_\SW^2+42\,q_\SW^4+\dots\right) $\\
(2,0,0,0) &  $8 \frac{M^3}{\Lambda^3}\, q_\SW^{7/4} \left(3+20\, q_\SW^2+47\,q_\SW^4+\dots \right)$\\
\hline
\end{tabular*}
\caption{Partition functions $H^{(4)}_{1/2}(q_\UV,M_j, B_j)$ expressed as a $q_\SW$-series for $w_2$ odd, with masses $M_1=M_2=M$ and $M_3=M_4=0$. 
}
\label{tab:H12Nf4}
\end{table}
\end{center}

\paragraph{$SU(2)$ bundles} \mbox{} \\
We first list a result for generic masses and $B_j=1$, $j=1,\dots, 4$. We find for the first two coefficients
\be
\begin{split}
Z^{(4)}_{0}(q_\UV,M_j, B_j)&=-\frac{3\sum_{i<j<k}M_iM_jM_k}{4\,\Lambda^3}\,q_\UV\\
& \qquad - \frac{\sum_{j\neq k }M_j^2 M_k-80\sum_{i<j<k}M_iM_jM_k}{16\, \Lambda^3}\,q_\UV^2+\dots
\end{split}
\ee
and
\be
\begin{split}
H^{(4)}_{0}(q_\UV, M_j,B_j)&=-\frac{12\sum_{i<j<k}M_iM_jM_k}{\Lambda^3}\,q_\SW\\
&\qquad  - \frac{16\sum_{j\neq k }M_j^2 M_k-8\sum_{i<j<k}M_iM_jM_k}{ \Lambda^3}\,q_\SW^2+\dots
\end{split}
\ee
Based on the first four coefficients of these series, we find that $H_0^{(4)}$ satisfies \eqref{eq:H(4)trafo}. We list results for partition functions $Z^{(4)}_0$ and $H^{(4)}_0$ for equal masses and a few choices of $B_j$ in Table \ref{tab:evenptcorrelator_Nf4} and \ref{tab:evenptcorrelator_Nf4H12}.

\begin{center}
\renewcommand{\arraystretch}{1.5}
\begin{table}[h!]
\centering
\begin{tabular*}{\linewidth}{@{\extracolsep{\fill}}|c|r|}
\hline
$(B_1, B_2, B_3,B4)$ & $Z^{(4)}_0(q_\UV,M,B_j)$ \\
\hline 
(1,1,1,1) & $\frac{M^3}{\Lambda^3}(-3q_\UV+\frac{77}{4}q_\UV^2-\frac{6675}{128}q_\UV^2+\frac{157335}{2048}q_\UV^4+\dots)$ \\
 & $=16\frac{M^3}{\Lambda^3}(-3\,q_\SW+332\,q_\SW^2-18410\,q_\SW^3+682462\,q_\SW^4+\dots)$\\
(3,1,1,1) & $\frac{M}{\Lambda}(-\frac{9}{2}\,q_\UV+\frac{375}{8}\,q_\UV^2-\frac{56621}{256}\,q_\UV^3+\frac{2555241}{4096}\,q_\UV^4+\dots)$ \\
& $=8\frac{M}{\Lambda}(-9\,q_\SW+1572\,q_\SW^2-137638\,q_\SW^3+8058018\,q_\SW^4+\dots)$\\
\hline
\end{tabular*}
\caption{Correlation function $Z^{(4)}_0(q_\UV,M,B_j)$ for $w_2$ even, $N_f =4$, equal masses $M$. 
}
\label{tab:evenptcorrelator_Nf4}
\end{table}
\end{center}

\begin{center}
\renewcommand{\arraystretch}{1.5}
\begin{table}[h!]
\centering
\begin{tabular*}{\linewidth}{@{\extracolsep{\fill}}|c|r|}
\hline
$(B_1, B_2, B_3,B4)$ & $H^{(4)}_0(q_\UV,M,B_j)$ \\
\hline 
(1,1,1,1) & $-16\frac{M^3}{\Lambda^3}(3\,q_\SW+10\,q_\SW^2+14\,q_\SW^3+30\,q_\SW^4+\dots)$\\
(3,1,1,1) &  $-8\frac{M}{\Lambda}(9\,q_\SW-222\,q_\SW^2+3034\,q_\SW^3-31686\,q_\SW^4+\dots)$\\
\hline
\end{tabular*}
\caption{Correlation function $H^{(4)}_0(q_\UV,M,B_j)$ for $w_2$ even, $N_f =4$, equal masses $M$. 
}
\label{tab:evenptcorrelator_Nf4H12}
\end{table}
\end{center}

\subsection{\texorpdfstring{$\CN=2^*$}{N} theory}
\label{sec:N=2*}

We collect in this section our results on the partition functions of the Donaldson-Witten twist of the $\CN=2^*$ theory with background flux for the $U(1)$ flavor symmetry.

\subsubsection*{Results for $SO(3)$}

We find a significant dependence of the partition functions on the background flux. First for background fluxes $b$ such that $B=\pm 3$, or $\vdim_{\BC}(\CM^{Q,*}_{k})=0$, the equivariant partition function does not depend on $\epsilon_i$.
For a background fluxes such that $B=\pm 1$, the dependence on $\epsilon_i$ is contained in an overall prefactor $K_b$. The factorization of the partition function suggests that the weights of the $U(1)^2$ action on the tangent space at the fixed points is the same for all fixed points. For all cases checked, we find
\be
\label{eq:Kb}
\begin{split}
& B=1: \qquad \quad K_{1/2,b}=\frac{1}{8\Lambda^3}(2M+(b_1-1)\epsilon_1+(b_2-1)\epsilon_2)\\
&\qquad \qquad \qquad \hspace{1cm} \times (2M+(b_1+1)\epsilon_1+(b_2-1)\epsilon_2)\\
&\qquad \qquad \qquad  \hspace{1cm}  \times (2M+(b_1-1)\epsilon_1+(b_2+1)\epsilon_2),\\
& B=-1: \qquad \, K_{1/2,b}=\frac{1}{8\Lambda^3}(2M+(b_1+1)\epsilon_1+(b_2+1)\epsilon_2)\\
&\qquad \qquad \qquad \hspace{1cm} \times (2M+(b_1+1)\epsilon_1+(b_2-1)\epsilon_2)\\
&\qquad \qquad \qquad  \hspace{1cm}  \times (2M+(b_1-1)\epsilon_1+(b_2+1)\epsilon_2),\\
&B=\pm 3:\qquad \quad K_{1/2,b}=1.
\end{split}
\ee

The first coefficients of the $q$-series for $\mu=1/2$ are given in Table \ref{tab:ZN=2*12}. We will discuss in more detail in Section \ref{sec:comp2*} that
 the coefficients of the $q$-series for $B=\pm 1$ agree with Eq. \eqref{eq:ZN=2*} derived in \cite[Eq. (6.35)]{Manschot:2021qqe}. 

\begin{center}
\renewcommand{\arraystretch}{1.5}
\begin{table}[h!]
\centering
\begin{tabular*}{\linewidth}{@{\extracolsep{\fill}}|c|r|}
\hline
$B$ & $Z^{(*)}_{1/2}(q,M,  b)$  \\
\hline 
$\pm 1$ & $K_{1/2,b}\,(q^{3/4}+11\,q^{7/4}+42\,q^{11/4}+\dots)$ \\
$\pm 3$ & $K_{1/2,b}\,(q^{3/4}+9\,q^{7/4}+48\,q^{11/4}+203\,q^{15/4}+\dots)$ \\
\hline
\end{tabular*}
\caption{Partition functions $Z^{(*)}_{1/2}(q,M,  b)$ of the $\CN=2^*$ theory for different choices of $b$.}
\label{tab:ZN=2*12}
\end{table}
\end{center}

The factorization with all $\epsilon_i$-dependence in the pre-factor does not occur for $|B|\geq  5$ for which $\vdim_{\BC}(\CM^{Q,*}_{k})<0$. Then the coefficients are rational functions of $\epsilon_i$ and $M$.  On the other hand, we do determine a finite $q$-series in non-equivariant limit, $\varepsilon_i\to 0$, independent of distribution of the equivariant flux $b$ for fixed non-equivariant flux $B$. For example,
\be
B=\pm 5:\qquad \lim_{\eps_i\to 0} Z^{(*)}_{1/2}(q,M,b,\eps_i)= \frac{\Lambda^6}{M^6}\left(q^{3/4}-10\,q^{7/4}+250\,q^{11/4}+\dots\right).
\ee
It would be interesting to determine if these coefficients are those of a mock or quasi-modular form since the the partition functions for $B=\pm 1, \pm 3$ are. Moreover, it would be interesting to derive these from the $u$-plane integral as for $B=\pm 1,\pm 3$ in \cite{Manschot:2021qqe}. See Section \ref{sec:comp2*} for more discussion.

\subsubsection*{Results for $SU(2)$}
We next gives results for $\mu=0$. This case is more subtle due to the contribution to the partition function of strictly semi-stable bundles. As before, we take into account the strictly semi-stable fluxes by an additional factor $\frac{1}{2}$. For $B=\pm 1$, we find a similar structure as above for $\mu=1/2$, that is to say we find that 
\be
\label{eq:Z*0B1}
Z^{(*)}_0(q,M,b)=K_{0,b}\,(3\,q+20\,q^2+73\,q^3+214\,q^4+\dots)
\ee
with the $\epsilon_i$-dependence contained in the overall factor $K_{0,b}$, times the non-equivariant partition function \eqref{eq:ZN=2*}. With $K_{0,b}=K_{1/2,b}$ for $B=-1$, and $K_{0,b}=-K_{1/2,b}$ for $B=1$.

On the other hand for $B=\pm 3$, we do not observe this structure. We summarize some of our findings:
\begin{itemize}
\item 
For generic non-zero $M$, the equivariant partition function $Z^{(*)}_0(q,M,b,\eps_i)$ is {\it not} independent of $\epsilon_i$ as is the case for $B=\pm 3$, $\mu=1/2$ above. Rather, the $\eps_i$ dependence of $Z^{(*)}_0(q,M,b,\eps_i)$ does depend on the distribution of the equivariant flux $b$ given the total flux $B$ over the three patches, and does not factor out as for $B=\pm 1$, $\mu=0$ above. 

The non-equivariant limit $\epsilon_i\to 0$ is finite, independent of the equivariant flux $b$ and independent of the mass $M$. The first few terms of the $q$-series are
\be
\label{eq:SU2nonequi}
\lim_{\eps_i\to 0} Z^{(*)}_0(q,M,b,\eps_i)= -9\,q +168\,q^2 - 1945\,q^3+18594\,q^4 + \dots.
\ee

\item
For generic non-vanishing $\eps_i$, the limit $M\to 0$ does depend on $\eps_i$ and on the equivariant flux $b$. For $b=(1,1,1)$, the limit is independent of $\eps_i$,
\be
\label{eq:SU2massless}
\lim_{M\to 0} Z^{(*)}_0(q,M,b,\eps_i)= \frac{3}{2}\,q+12\,q^2+\frac{125}{2}\,q^3+\frac{513}{2}\,q^4+\dots.
\ee
However for other choices of $b_l$, the result is in general not independent of the $\epsilon_i$ and the $\epsilon_i\to 0$ limit leads to different $q$-series. To illustrate this, we give the limits for three choices of $b$,
\be
\begin{split}
&\lim_{\eps_i \to 0} \lim_{M\to 0}Z^{(*)}_0(q,M,b,\eps_i)\\
&\qquad =\left\{\begin{array}{rr} 
2\,q+15\,q^2+76\,q^3+\dots, & b=(2,2,-1),\\
\frac{13}{3}\,q+\frac{6584}{143}\,q^2+\frac{30039575}{94809}\,q^3+\dots, & b=(3,0,0),\\
\sim O(\eps_i)/M\,\,{\rm divergent}, & \quad b=(3,-3,3).
\end{array} \right.
\end{split}
\ee
Thus the non-equivariant limit and the massless limit do not commute for this partition function. This is not unexpected, however we will find in Section \ref{sec:comp2*} that (\ref{eq:SU2massless}) comes closed to the expected result rather than (\ref{eq:SU2nonequi}).
\end{itemize}

These observations clearly have some tension with the expected
properties. Our impression is that the sum over triples $p$ misses one
or multiple contributions, which leads to this unexpected
behaviour. The connection of the
non-equivariant partition function to indefinite theta series suggests that the missing triples
could possibly be of the form $p=(\ell,\ell,0)$, $\ell\in \mathbb{N}$
and permutations, which all correspond to a vanishing instanton
number.\footnote{A similar resolution was suggested in \cite{Khlaif:2026rnb}.}

We discuss in Section \ref{sec:comp2*} how the above results fit with results in the literature on non-equivariant partition functions.
As even that case requires sophisticated techniques from Donaldson-Thomas theory and intersection theory \cite{Yoshioka1995, Manschot:2016gsx}, we leave a further exploration to future work.

\section{Comparison with $u$-plane results}
\label{sec:compwuplane}
As mentioned in the introduction, a different approach to topological
correlation functions is using low energy field theory. This results
in the $u$-plane integral \cite{Moore:1997pc}. This method is also
applied to theories with hypermultiplets, for massless $N_f=2,3,4$ theories in
\cite{Malmendier:2008db} and including the mass dependence in
\cite{Aspman:2023ate} and $\CN=2^*$ \cite{Labastida:1998sk,  Manschot:2021qqe}. Before comparing the
results of Section \ref{sec:ExResults} in with those obtained with the $u$-plane approach, we provide a brief review. More details can found in the references. The $u$-plane integral for an observable $\CO$ takes the form
\be
\Phi_\mu^J[\CO]=\int_{\CF} d\tau\wedge d\bar \tau\,\nu(\tau,m_j)\,\CO\,\Psi^J_\mu(\tau,\bar \tau, \sum_j \bfk_j v_j, \sum_j \bfk_j\bar v_j),
\ee
where $\nu$ is a measure arising from topological couplings of the theory and $\Psi^J_\mu$ a theta series of Siegel-Narain type. The integration domain $\CF$ is the domain of the effective coupling on the Coulomb branch of the theory. It is a modular curve for $N_f=0$ \cite{Seiberg:1994aj, matone1996, Nahm:1996di}, for generic masses it is typically a branched cover of a modular curve \cite{Aspman:2021vhs}.
The integral  can be evaluated using mock modular forms \cite{Moore:1997pc, Malmendier:2008db, Korpas:2019cwg, Manschot:2021qqe, Aspman:2022sfj}.

The resulting expressions obtained are rather different from Eq. (\ref{eq:Zfinal_c1odd_new}) and its non-equivariant limit. In this section, we will make a rather detailed comparison between results based on equivariant localization (\ref{eq:Zfinal_c1odd_new}) and $u$-plane integrals \cite{Malmendier:2008db,  Manschot:2021qqe, Aspman:2022sfj} at low instanton order. We take away the following lessons from the comparison:
\begin{itemize}
\item For $N_f=1,2,3$: While there is good agreement between the two methods for the partition function without insertion of observables and with vanishing background fluxes, there are minor discrepancies beyond this case. We think these are mainly due to difficulties with the evaluation of $u$-plane integral with observables and background fluxes. We expect that the discrepancies can be clarified using the better understanding of the couplings obtained in \cite{Furrer:2026byd}.
\item For $N_f=4$: this theory displays a rich structure of $S$-duality and triality, which motivates further study of both methods.
\item For $\CN=2^*$: There is a discrepancy for $\mu=0$ and $B=\pm 3$, which is the canonical Spin$^c$ structure equivalent with the VW twist. This demonstrates that the integration contour for $a$ still remains to be better understood. 
\end{itemize}

 We have ordered the section from $N_f=1$ to 4, followed by $\CN=2^*$. Readers interested in the superconformal theories can easily skip the earlier subsections.

\subsection{\texorpdfstring{$N_f = 1$}{Nf = 1} hypermultiplet}
\noindent \paragraph{The $SO(3)$ case} \mbox{} \\
We can compare the results in Tables \ref{tab:oddptcorrelator} and \ref{tab:evenptcorrelator} with the results of Ref. \cite{Aspman:2023ate}, where correlation functions of point observables are determined using integration over the $u$-plane. In \cite{Aspman:2023ate}, the background flux is denoted by $k_1$, the mass by $m$ and the fugacity for point observables by $p$. We furthermore use $\Lambda^{AFM}$ for the scale $\Lambda_1$ in \cite[Eq. (11.5)]{Aspman:2023ate}.  Explicit relations between the variables both papers can be derived by comparison of the first three terms of the partition functions, for example for $k_1=0$, $c_1(\tilde E)=1$. These are given in Table \ref{tab:variaident}. The overall factor for $w_2(P)$ odd is $r= 2^{2/3}$. Note the fugacity for the point observable $p$ is dimensionless in \cite{Aspman:2023ate}, while $x$ has dimension $\Lambda^{-2}$.
\begin{center}
\renewcommand{\arraystretch}{1.5}
\begin{table}[h!]
\centering
\begin{tabular}{|l|r|}
\hline
 Background flux & $B=2k_1$ \\
 Mass/scale ratio & $\frac{M}{\Lambda}=2^{2/3}\frac{m}{\Lambda^{AFM}}$ \\
Fugacity & $x\Lambda^2=-2^{-1/3}p$ \\
Partition function & $Z_{\mu}(\Lambda,M,x)= r \,  Z^{AFM}_{\mu}(\Lambda^{AFM},m,p)$ \\
\hline
\end{tabular}
\caption{Relation between the variables in Ref. \cite{Aspman:2023ate} and this paper, for $N_f = 1$. See the main text for the values of $r$.}
\label{tab:variaident}
\end{table}
\end{center}
For $w_2(P)\neq 0$ and $B=2k_1=0$, we find precise agreement with \cite[Eq. (11.5) and Table 4]{Aspman:2023ate}. 
Similarly for $B=2k_1=2$, we find precise agreement with \cite[Table 11]{Aspman:2023ate}. For $B=2k_1=4,6$, we also find agreement of most terms with the following exceptions:
\begin{itemize}
\item For $B=4$: \cite{Aspman:2023ate} does not list terms corresponding to $O(\Lambda^9)$ for $x^\ell$ with $\ell=0,1,2,3$, while these are present in Table \ref{tab:oddptcorrelator}. While the caption in \cite[Table 11]{Aspman:2023ate} claims accuracy to this order, it may be that not enough terms in \cite[Table 11]{Aspman:2023ate} the inverse mass expansion are determined. 
\item For $B=6$: \begin{itemize}

\item the terms 
corresponding to $O(\Lambda^{14})$ for $x^\ell$, $\ell=0,1$, are not present in \cite[Table 11]{Aspman:2023ate}. 
\item  Table \ref{tab:oddptcorrelator} does not reproduce the term of $O(\Lambda^{9})$ for $\ell=4$ in \cite[Table 11]{Aspman:2023ate}. Non-vanishing of this term in \cite{Aspman:2023ate} is somewhat questionable, since the power of $\Lambda$ does not agree with $2 \mod 3$ as expected for this choice of background flux. 
\end{itemize}
\end{itemize}

Ref. \cite{Furrer:2026byd} derived improved expressions for the couplings to background fluxes. Application to the $u$-plane approach may clarify the observed differences.

\paragraph{The $SU(2)$ case} \mbox{} \\
The parameter conversions to match our results with the literature are again given by Table \ref{tab:variaident}, where we get different values for $r$, depending on the choice of background fluxes $B_i$.

We compare our nonequivariant results for the point correlators for $w_2(P)$ even, $N_f = 1$ in Table \ref{tab:evenptcorrelator} with \citep[Table 10, p.44]{Aspman:2023ate}.
\begin{itemize}
\item For $B = -1$ we have $r = \sqrt{2}$. Then the terms up to $x^4$ in Table \ref{tab:evenptcorrelator} agree with \cite[Table 10]{Aspman:2023ate}, except for the term $\sim \frac{\Lambda^9}{m^3}$ for $\ell=3$ in \cite{Aspman:2023ate}, which we do not reproduce with the equivariant calculation. 
\item For $B = -3$  we have $r = 2^{-5/6}$. The equivariant results match all terms from \cite{Aspman:2023ate} up to order $\Lambda^8$, except that the terms $\sim \frac{\Lambda^5}{M^5}x^0, \frac{\Lambda^8}{M^8}x^0$ and $\sim \frac{\Lambda^8}{M^6}x$ in Table \ref{tab:evenptcorrelator} are absent in \cite{Aspman:2023ate}. 

\item For $B=-5$, we find $r=2^{-3/2}$, and then agreement with the terms in \cite[Table 10]{Aspman:2023ate}. The equivariant calculation provides further subleading terms, which are not given in \cite[Table 10]{Aspman:2023ate}. The calculation in \cite{Aspman:2023ate} possibly was not carried out to sufficiently high order in the inverse mass expansion.
\end{itemize}

\subsection{\texorpdfstring{$N_f = 2$}{Nf = 2} hypermultiplets}

We compare our results for $N_f=2$ with the literature for  $\mu=1/2$ with $B_1=B_2 = 0$ \cite{Malmendier:2008db, Aspman:2023ate}.
More precisely, we consider the equal mass, non-equivariant limit of the partition function $H_{1/2}^{(2)}$ presented in Table \ref{tab:H_c1odd_Nf2_results_x} and compare with Eq. (11.6) in \citep[p. 37]{Aspman:2023ate} determined in the large mass regime. The latter results match with the results for vanishing masses case obtained in \cite[4.6.2]{Malmendier:2008db}.
Our calculation ran up to $O(x^4,\Lambda^4)$. For all terms up to this order, we find agreement for $\mu=1/2$, $B_j=0$ with the identification between the parameters as in Table \ref{tab:variaidentNf2}.\\
\begin{center}
\renewcommand{\arraystretch}{1.5}
\begin{table}[h!]
\centering
\begin{tabular}{|l|r|}
\hline
 Background flux & $B_j=2k_j$ \\
 Mass/scale ratio & $\frac{M}{\Lambda}=\frac{2m}{\Lambda^{AFM}}$ \\
Fugacity & $x\Lambda^2=- p/2$ \\
Partition function & $Z_{1/2}(\Lambda,M,x)= 4\, \  Z^{AFM}_{1/2}(\Lambda^{AFM},m,p)$ \\
\hline
\end{tabular}
\caption{Relation between the variables in Ref. \cite{Aspman:2023ate} and this paper, for the case $w_2(P)$ odd, $N_f = 2$. }
\label{tab:variaidentNf2}
\end{table}
\end{center}

\subsection{\texorpdfstring{$N_f = 3$}{Nf = 3} hypermultiplets}
Similarly to $N_f=2$, we compare the correlation functions based on the $u$-plane method \cite{Malmendier:2008db, Aspman:2022sfj} to $H^{(3)}_{1/2}$, ie $Z^{(3)}_{1/2}$ divided by the factor in Eq. \eqref{eq:factorNf}.
We thus compare our results in Table \ref{tab:c1odd_Nf3_results_H} to results in \cite{Aspman:2022sfj}. Eq. (11.7) and Table 6 of \cite{Aspman:2022sfj} are determined in the large mass regime while \cite[Table 9]{Aspman:2022sfj} is determined in the small mass regime. Those results match with the results for vanishing masses case obtained in \cite[4.6.3]{Malmendier:2008db}.

By comparing the partition functions for $B_j=2k_j=0$, with $x=p=0$, we determine the overall factor and relation between the masses as in Table \ref{tab:variaidentNf3}. With these two identifications, we find that the three terms of the partition function match. With further relating the fugacities $x$ and $p$, we find agreement with 6 out of the 10 terms with positive powers of $p$ in \cite[Eq. (11.7)]{Aspman:2022sfj}. For each power of $p$ the most subleading term in the inverse mass expansion does not appear to match.

\begin{center}
\renewcommand{\arraystretch}{1.5}
\begin{table}[h!]
\centering
\begin{tabular}{|l|r|}
\hline
 Background flux & $B_j=2k_j$ \\
 Mass/scale ratio & $\frac{M}{\Lambda}=\frac{4m}{\Lambda^{AFM}}$ \\
Fugacity & $x\Lambda^2=- p/2^3$ \\
Partition function & $Z_{1/2}(\Lambda,M,x)= 2^6\, \  Z^{AFM}_{1/2}(\Lambda^{AFM},m,p)$ \\
\hline
\end{tabular}
\caption{Relation between the variables in Ref. \cite{Aspman:2023ate} and this paper, for the case $w_2(P)$ odd, $N_f = 3$. }
\label{tab:variaidentNf3}
\end{table}
\end{center}

\subsection{\texorpdfstring{$N_f = 4$}{Nf = 4} hypermultiplets}\label{eq:compNf4}

Our results for $N_f=4$ can be compared to explicit results by Malmendier and Ono \cite{Malmendier:2008db} for $\mathbb{CP}^2$, while the structure of the $u$-plane integral is also discussed for more general manifolds in \cite{moore_talk2018}.

Ref. \cite{Malmendier:2008db} evaluates $u$-plane integral $\Phi^{(4)}_{1/2}$ for the massless $SO(3)$ theory on $\mathbb{CP}^2$ with $w_2(P)$ odd and vanishing background fluxes. Their result that the integral vanishes is clearly in agreement with our results in Tables \ref{tab:oddptcorrelator_Nf4} and \ref{tab:oddptcorrelator_Nf4MM00}. Ref. \cite{Malmendier:2008db} also determine the integral to first non-vanishing order in an expansion around $M_j=0$, which allows us to make a more non-trivial comparison. Note that such a result does crucially depend on the details of the expansion, which can be seen for example from our results: the difference of $Z^{(4)}_{1/2}(q,M,b)$ in Tables \ref{tab:oddptcorrelator_Nf4} and \ref{tab:oddptcorrelator_Nf4MM00}. 

Ref. \cite[Def. 4.22]{Malmendier:2008db} starts with the theory corresponding to Tables \ref{tab:oddptcorrelator_Nf4MM00} and \ref{tab:H12Nf4}, ie with  $M_1=M_2=M$, $M_3=M_4=0$. This theory is invariant under the triality group, such that the partition function is expected to transform as a modular form under the full $SL_2(\mathbb{Z})$ modular group. We let $\Phi^{(4)}_{1/2}(\tau,M)$ be the $u$-plane integral defined in \cite{Malmendier:2008db}. Then \cite{Malmendier:2008db} evaluates the function $\tilde \Phi^{(4),\rm MO}_{1/2}(\tau)$ defined by
\be
\tilde \Phi^{(4),\rm MO}_{1/2}(\tau)=\lim_{M\to 0}\frac{1}{M^4\,\eta(\tau)^{12}}\,\Phi^{(4),\rm MO}_{1/2}(\tau,M),
\ee
with the Dedekind $\eta$-function defined as
\be
\eta(\tau)=q^{\frac{1}{24}}\prod_{n=1}^\infty (1-q^n).
\ee
Intriguingly, $\tilde \Phi^{(4),\rm MO}_{1/2}(\tau)$ is a mock modular form of weight 0 \cite[Theorem 1.2]{Malmendier:2008db}. To state this function, we introduce a number of modular and mock modular forms. First, let $Q$ be the weight $1/2$ mock modular form for the congruence subgroup $\Gamma^0(4)$, defined as
\be
\label{eq:Qtau}
\begin{split}
Q(\tau)&=\frac{1}{6}h^{(2)}(\tau)+\frac{7}{6}\frac{\vartheta_2(\tau)^4+\vartheta_3(\tau)^4}{\eta(\tau)^3}\\
&=q^{-1/8}\left( 1+28\,q^{1/2}+39\,q+196\,q^{3/2}+161\,q^2+\dots\right),
\end{split}
\ee
where $h^{(2)}(\tau)$ is the weight $1/2$ mock modular form for $SL_2(\mathbb{Z})$, famous for its relation to the Mathieu Moonshine phenomenon \cite{Eguchi:2010ej, Dabholkar:2012nd},
\be
\begin{split}
h^{(2)}(\tau)&=\frac{\vartheta_2(\tau)^4-\vartheta_4(\tau)^4}{\eta(\tau)^3}-\frac{24}{\vartheta_3(\tau)}\sum_{n=1}^\infty \frac{q^{n^2/2-1/8}}{1+q^{n-1/2}}\\
&=q^{-1/8}\left(-1+45\,q+231\,q^2+770\,q^3+\dots \right).
\end{split}
\ee
As $Q$ is a mock modular form, it is the non-holomorphic modular completion of $Q$, $\widehat Q$,
\be
\widehat Q(\tau,\bar \tau)=Q(\tau)-\int^{i\infty}_{-\bar \tau}\frac{\eta(w)^3}{\sqrt{-i(w+\tau)}}dw,
\ee
which transforms as a modular form of weight $1/2$ for $\Gamma^0(4)$. By \eqref{eq:Qtau}, this also defines the modular completion $\widehat h^{(2)}$ of $h^{(2)}$. We introduce furthermore the modular form
\be
g(\tau)=-\frac{1}{36}\left( \left(\frac{\vartheta_2(\tau)}{\eta(\tau)} \right)^8-\left(\frac{\vartheta_2(\tau)}{\eta(\tau)} \right)^4\left(\frac{\vartheta_3(\tau)}{\eta(\tau)} \right)^4 +\left(\frac{\vartheta_3(\tau)}{\eta(\tau)} \right)^8\right).
\ee
$\tilde \Phi^{(4), \rm MO}_{1/2}(\tau)$ is then given as \cite[Lemma 4.24]{Malmendier:2008db}
\be
\label{eq:Z4MO}
\begin{split}
\tilde \Phi^{(4),\rm MO}_{1/2}(\tau)&=\left( g(\tau)+\frac{1}{2} \frac{q}{\eta(\tau)^4} \frac{d}{dq} \left( \frac{q}{\eta(\tau)^4} \frac{d}{dq}\right)\right) \frac{Q(\tau)}{\eta(\tau)}\\
&=3\,q^{1/2}(1+22\,q+213\,q^2+1460\,q^3+7938\, q^4+\dots).
\end{split}
\ee

Since $Q(\tau)$ is a mock modular form, $\tilde \Phi^{(4),\rm MO}_{1/2}(\tau)$ does not transform as a modular form under $S$-duality. We therefore conjecture that the full path integral of this theory is complemented by a non-holomorphic term arising from replacing $Q$ by $\widehat Q$ in \eqref{eq:Z4MO} such that $\tilde \Phi^{(4),\rm MO}_{1/2}$  transforms as a modular form. We expect that this can be derived from the $u$-plane integral as in the case of $\CN=2^*$ \cite{Manschot:2021qqe}, or from equivariant localization as for $\CN=4$ in \cite{Bonelli:2020xps}.

We next aim to relate this function to our non-equivariant result $H^{(4)}_{1/2}(q_\UV,M_j,B_j)$ in Table \ref{tab:H12Nf4} for the theory with $M_{1,2}=M$, $M_{3,4}=0$ and $B_j=0$. First taking into account \cite[Def (2.12)]{Malmendier:2008db}, we identify $\tau$ by $2\tau_\SW$ to match our conventions. Then based on the first few coefficients of $H_{1/2}^{(4)}$ as in Table \ref{tab:H12Nf4}, we find that $H_{1/2}^{(4)}$ agrees with,
\be
8\,\frac{M^4}{\Lambda^4}\,\eta(2\tau_\SW)^9\,\tilde \Phi^{(4),\rm MO}_{1/2}(2\tau_\SW). 
\ee
This is an non-trivial confirmation of both methods. We note that since $Z^{(4),\rm MO}_{1/2}$ has modular weight 0, $H^{(4)}_{1/2}$ has modular weight $9/2$. This is in agreement with the proposed weight $w$ of the $u$-plane integral, $w=(2\chi+3\sigma)/2$ \cite{moore_talk2018}.

\subsection{\texorpdfstring{$\CN=2^*$}{N=2*} theory}
\label{sec:comp2*}
Let us first recall the explicit results of the non-equivariant $u$-plane integral $\Phi^{(*)}_{\mu}$ evaluated in \cite{Manschot:2021qqe}. It is convenient to parametrize the partition function 
as
\be
\label{eq:ZN=2*}
\Phi^{(*)}_{\mu}(\tau,M;B)=\left(\frac{\Lambda}{M} \right)^\ell \frac{g_{\mu}(\tau;B)}{\eta(\tau)^{2\chi+4\ell}},
\ee
with $\ell=(B^2-2\chi-3\sigma)/8$ half the real dimension of the moduli space of abelian monopole equations. Ref. \cite{Manschot:2021qqe} evaluated  $g_\mu$ using the $u$-plane integral for background fluxes $B=\pm 1$ and $B=\pm 3$. For $B=\pm 3$, $g_{\mu}(\tau;B)$, $\mu=0,1/2$ is a vector-valued mock modular form of weight $3/2$, whose coefficients are a multiple of the Hurwitz class numbers,
\be
\label{eq:gmu3}
g_{\mu}(\tau;3)=3\sum_{n\in \mathbb{Z}} H(4n-2\mu)\,q^{n-\mu/2}.
\ee
The first few terms are
\be
\label{eq:gmutau3}
g_{\mu}(\tau;3)=\left\{ \begin{array}{rr} -\frac{1}{4}+\frac{3}{2}\,q+3\,q^2+4\,q^3+\dots, & \quad \mu=0, \\ q^{3/4}+3\,q^{7/4}+3\,q^{11/4}+\dots, & \quad \mu=1/2. \end{array}\right.
\ee
The coefficients of the $q$-series for $\pm 3$ agree with the
generating function of class numbers \cite{Klyachko1991}, which are a
count of the number of toric fixed points in the moduli space. These
partition functions are well-known to be the partition functions of
the Vafa-Witten twist applied to $\CN=4$, $SU(2)$ Yang-Mills
\cite{Vafa:1994tf, Dabholkar:2020fde}.  

For $B=\pm 1$, $g_{\mu}(\tau;B)$ is a vector-valued quasi-mock modular forms of weight $9/2$ whose closed form is given in \cite[Eq. (6.35)]{Manschot:2021qqe}. The first coefficients are \cite{Manschot:2021qqe}
\be
\label{eq:gmutau1}
g_{\mu}(\tau;1)=\left\{ \begin{array}{rr} 3\,q+14\,q^2+30\,q^3+\dots, & \quad \mu=0, \\ q^{3/4}+9\,q^{7/4}+19\,q^{11/4}+\dots, & \quad \mu=1/2. \end{array}\right.
\ee
Such functions require an additional non-holomorphic term to transform
as a modular form, derived for example in \cite{Vafa:1994tf}. Other path integral
derivations are in \cite{Bershtein:2015xfa, Dabholkar:2020fde, Manschot:2021qqe}.

Now comparing with our results in Section \ref{sec:N=2*}, we find agreement of Eq. (\ref{eq:gmutau1}) for $B=\pm 1$ with Table \ref{tab:ZN=2*12} for $\mu=1/2$, and Eq. \eqref{eq:Z*0B1} for $\mu=0$. Moreover, for $B=\pm 3$ and $\mu=1/2$, we find agreement of Eq. \eqref{eq:gmutau3} with Table \ref{tab:ZN=2*12}. The case $B=\pm 3$ and $\mu=0$ is more subtle. Here we find that the partition function in the massless limit \eqref{eq:SU2massless} agrees with the expansion of
\be
\frac{g_0(\tau;3)+\frac{1}{4}}{\eta(\tau)^6},
\ee
with $g_0(\tau;3)$ as in Eq. \eqref{eq:gmu3}. As mentioned earlier, this agreement up to the constant term seems to suggest that the contribution $-\frac{1}{4}$ from the trivial instanton with $k=0$ is missing in the equivariant approach. Deriving the proper form of these contributions will hopefully also resolve the issues with the different non-equivariant limits.
  
 \appendix 
 
 \section{The function \texorpdfstring{$\gamma_{\epsilon_1, \epsilon_2}$}{gamma_e1,e2}}
 \label{App:gamma}

We recall the special function $\gamma_{\varepsilon_1,\varepsilon_2}(x;\Lambda)$ \cite[App. A]{NekOk}\cite[App. E]{Nakajima:2003uh}. The specialization $\eps_1=-\eps_2$ appeared as the free energy of the A-model topological string on the resolved conifold \cite{Vafa:2000wi}, and appears through a Schwinger type computation of BPS particles  \cite{Gopakumar:1998ii}.

For ${\rm Re}(x)>0$, it is defined as the integral
\be
\label{eq:Defgammaeps}
\gamma_{\varepsilon_1,\varepsilon_2}(x;\Lambda)=\left.\frac{d}{ds}\right|_{s=0} \frac{\Lambda^s}{\Gamma(s)}\int_0^\infty dt\,t^{s-1}\,\frac{e^{-xt}}{(1-e^{\varepsilon_1 t})(1-e^{\varepsilon_2 t})},
\ee   
with
\be 
\Gamma(s)=\int_0^\infty dt\,t^{s-1}\,e^{-t}.
\ee 
We introduce the functions $c_n(\epsilon_1,\epsilon_2)$ as the coefficients in the small $\epsilon_i$-expansion
\be
\frac{1}{(1-e^{\epsilon_1 t})(1-e^{\epsilon_2 t})}=\sum_{n=0}^\infty \frac{c_n(\epsilon_1,\epsilon_2)}{n!}\,t^{2-n}
\ee

This functions has an expansion for large $x$, or small $\epsilon_i$
\cite{Nakajima:2003uh},
\be
\label{eq:gammaepsExp}
\begin{split}
\gamma_{\varepsilon_1,\varepsilon_2}(x;\Lambda)&=\frac{1}{\varepsilon_1\varepsilon_2}\left(-\frac{1}{2}x^2\log\left(\frac{x}{\Lambda}\right)+\frac{3}{4}x^2 \right)+\frac{\varepsilon_1+\varepsilon_2}{2\varepsilon_1\varepsilon_2}\left( -x\log\left(\frac{x}{\Lambda}\right)+x\right)\\
&\quad -\frac{\varepsilon_1^2+\varepsilon_2^2+3\varepsilon_1\varepsilon_2}{12\eps_1\eps_2}\log\left( \frac{x}{\Lambda}\right)+\sum_{n=3}^\infty \frac{c_n}{n(n-1)(n-2)}\,x^{2-n}
\end{split}
\ee
where the $c_n$ are homogenous functions of degree $n-2$ in $\epsilon_i$. This identity implies 
\be
\label{eq:gammaepsLambdashift}
\gamma_{\varepsilon_1,\varepsilon_2}(x;\Lambda\,e^u)=\gamma_{\varepsilon_1,\varepsilon_2}(x;\Lambda)+\frac{u}{2\eps_1\eps_2}\left( x^2+(\eps_1+\eps_2)x+\tfrac{1}{6}(\varepsilon_1^2+\varepsilon_2^2+3\varepsilon_1\varepsilon_2) \right).
\ee
In terms of the Barnes' double Gamma function $\Gamma_2$, we have the relation
\be
\label{eq:BarnesDGamma}
\gamma_{\varepsilon_1,\varepsilon_2}(x;1)=\log\left( \Gamma_2(x+\varepsilon_1+\varepsilon_2|\varepsilon_1,\varepsilon_2) \right),
\ee 
which is related to the Barnes $G(z;\tau)$ function as,
\be
\Gamma_2(z|\omega_1,\omega_2)=(2\pi)^{\frac{z}{2\omega_1}}\omega_2^{-\frac{z^2-z(\omega_1+\omega_2)}{2\omega_1\omega_2}-1}\,G\!\left(\frac{z}{\omega_1};\frac{\omega_2}{\omega_1}\right).
\ee
Another identity is
\be
\gamma_{-\epsilon_1,-\epsilon_2}(x;\Lambda)=\gamma_{\epsilon_1,\epsilon_2}(x-\epsilon_1-\epsilon_2;\Lambda).
\ee
For any $x\in \mathbb{C}$, we use the formal identity 
\begin{equation}
    \label{eq:char_prod_id}
    \prod_i \left( \frac{x_i}{\Lambda} \right) = \exp \left( - \frac{d}{ds} \left[ \frac{\Lambda^s}{\Gamma(s)} \int_0^\infty \frac{dt}{t}t^s \sum_i e^{-x_i t} \right]_{s=0} \right),
\end{equation}
to switch between the character 
\be
\chi(\{x_i\})=\sum_i e^{-x_i},
\ee
and product representation of a function.
Expanding the denominators in \eqref{eq:Defgammaeps}, $\gamma_{\epsilon_1, \epsilon_2}$ is then evaluated as
\be
\label{eq:gammalog}
\gamma_{\varepsilon_1,\varepsilon_2}(x;\Lambda)= \left\{\begin{array}{rr}  \sum_{m,n=1}^\infty \log\left( \frac{\Lambda}{x+m\varepsilon_1+n\varepsilon_2} \right), & \qquad \varepsilon_1,\varepsilon_2>0,\vspace{.1cm}\\
-\sum_{m=1,n=0}^\infty \log\left( \frac{\Lambda}{x+m\varepsilon_1-n\varepsilon_2} \right), & \qquad \varepsilon_1>0,\varepsilon_2<0,\vspace{.1cm}\\
-\sum_{m=0,n=1}^\infty \log\left( \frac{\Lambda}{x-m\varepsilon_1+n\varepsilon_2} \right), & \qquad \varepsilon_1<0,\varepsilon_2>0,\vspace{.1cm}\\
\sum_{m,n=0}^\infty \log\left( \frac{\Lambda}{x-m\varepsilon_1-n\varepsilon_2} \right), & \qquad \varepsilon_1,\varepsilon_2<0.\end{array}\right. 
\ee 
Note the sum on the right hand side diverges, and is better expressed in terms of the product formula for the Barnes function $G(z;\tau)$. On the other hand, the linear combinations of $\gamma_{\epsilon_1, \epsilon_2}$ appearing for compact toric manifolds as in Eq. \eqref{eq:Wxd} are finite.

There is an alternative analytic continuation of $\gamma_{\varepsilon_1,\varepsilon_2}$ \cite{Manschot:2019pog}. For ${\rm Re}(x)<0$, $\gamma_{\varepsilon_1,\varepsilon_2}$ is analytically continued by 
\be
\label{fn7}
\gamma_{\varepsilon_1,\varepsilon_2}(x;\Lambda)=\gamma_{\varepsilon_1,\varepsilon_2}(-x-\varepsilon_1-\varepsilon_2;\Lambda)
\ee 
While this analytic continuation is well-adapted to perturbative calculations in $\varepsilon_{1,2}$ (symmetry of the perturbative prepotential under change of sign of $\varepsilon_{1,2}$), we will use the analytic continuation of Eq. (\ref{eq:gammalog}), which is better suited in combination with the non-perturbative partition functions. 

\section{Proof of \texorpdfstring{$\epsilon_i$}{e}-reflection symmetry}
\label{App:eps_refl_sym_proof}
To prove the  $\eps_i$-reflection symmetry of  $H_{\mu}^{(N_f)}$ in formula \eqref{eq:HP2_c1odd_eps_flip}, we fix $b$, consider a summand of $H_{\mu}^{(N_f)}(\Lambda, q, M_j,b_j, x, z,\epsilon_i)$ in \eqref{eq:Hfinal_c1odd} with equivariant gauge flux $p = (p_1, p_2, p_3)$, and apply the transformation $(\epsilon_1, \epsilon_2) \to -(\epsilon_1, \epsilon_2)$. \par
The exponent $d(B_j)$ of $\Lambda$ in \eqref{eq:def_D_exponent} is independent of $\epsilon_i$ and invariant under $b_j \to -b_j$. By our above choice \eqref{eq:P2_obs_Omega_choice}, the observable factor $\Tilde{\CO}$ behaves as $\Tilde{\CO}(z, -\epsilon_1, -\epsilon_2) = \Tilde{\CO}(-z, \epsilon_1, \epsilon_2)$. \par 
Consider now $\Tilde{H}^{(N_f)}$ as in \eqref{eq:HTilde_P2}. Under the $\epsilon_i$ transformation, the three denominators $a_+^{(l)}$ give an overall factor of $(-1)$. As discussed below \eqref{eq:recrel_fs}, the factors $\CR^{(N_f)} (M_j, \epsilon_i)$ are invariant  under changing the sign of $(\epsilon_1, \epsilon_2)$. By \eqref{eq:HC2_a_flip} and \eqref{eq:HC2_epsilon_flip}, the factor $H^{(N_f)}_{\BC^2}$ in \eqref{eq:HTilde_P2} is invariant under both a sign flip in $a$ and in $\eps_i$. Again, by our choice of observables, the argument $q^{(l)}$ as in \eqref{eq:def_ql} is mapped under the $\epsilon_i$ transformation to the expression for $(-z, \epsilon_1, \epsilon_2)$. 
Note that $M_j^{(l)} (b_j, -\epsilon_i) = M_j^{(l)} (-b_j, \epsilon_i)$. Hence, we have $\Tilde{H}^{(N_f)} \left(M_j, b_j, p, x, z,  -\epsilon_1, -\epsilon_2 \right) = \Tilde{H}^{(N_f)} \left(M_j, -b_j, p, x, -z,  \epsilon_1, \epsilon_2 \right)$. This means, under $\epsilon_i \to -\epsilon_i$, this factor gets mapped to the factor for the same $p$ in the partition function for the theory with $-z$ and $-b_j$.\par
Lastly, we have to discuss the perturbative part as in \eqref{eq:Zpert_p2_reduced}. Similarly to \eqref{eq:zpertvec_lambdas}, from the two factors $\Tilde{W}$ in \eqref{eq:Zpert_p2_reduced} under $(\epsilon_1, \epsilon_2) \to -(\epsilon_1, \epsilon_2)$ we obtain a factor of $(-1)^{-P^2} = (-1)^{2\mu} $. 
Now for some $1 \leq j \leq 4 = N_f$, consider a factor of 
\begin{equation}
    \label{eq:App_epssymm_W}
    \frac{1}{W(M_j, p+b_j - (1,1,1); 1, \epsilon_i)\, W(M_j, -p+b_j - (1,1,1); 1, \epsilon_i)}
\end{equation}
from \eqref{eq:Zpert_p2_reduced}, \eqref{eq:Zpert_hyp_P2}.
Consider the case $P+B_j-3 \geq 0$. This is equivalent to $-P + (-B_j) -3 \leq -6$. By \eqref{eq:def_W}, the first factor in the denominator in \eqref{eq:Zpert_p2_reduced} is, after the $\epsilon_i$ transformation,
\begin{equation}
\label{eq:epsflip_Widentity}
\begin{split}
    &W(M, p+b_j - (1,1,1); 1, -\epsilon_i) \\
    &= \prod_{i=0}^{\frac{P+B_j-3}{2}}\prod_{j=0}^{\frac{P+B_j-3}{2}-i} \left( M + \left( \frac{p_1 + (b_j)_1 -1}{2} -j \right)(-\epsilon_1) + \left( \frac{p_2 + (b_j)_2 -1}{2} - i\right) (-\epsilon_2)\right) \\
    &= \prod_{i=0}^{\left| -\frac{P+B_j+3}{2}\right|  - 3} \ \prod_{j=0}^{\left| -\frac{P+B_j+3}{2}\right|-3-i} \\ & \qquad \times \left( M + \left( \frac{-p_1 + (-(b_j)_1) -1}{2} + 1+j \right)\epsilon_1 + \left( \frac{-p_2 + (-(b_j)_2) -1}{2} +1+ i\right) \epsilon_2 \right)\\
    &= W(M, -p+ (-b_j) - (1,1,1); 1, \epsilon_i). 
\end{split}
\end{equation} 
But this precisely corresponds the second $W$-factor of the $j$-th hypermultiplet product of the form \eqref{eq:App_epssymm_W}  of the summand with flux $p$ from the $H_{\mu}^{(N_f)}$  with $-b_j$. More generally, this identity can easily be deformed to see that the cases $P-B_j-3\geq 0$ ($\leq -6$ resp.) always transform to the case $-P+ (-B_j)-3\leq -6$ ($\geq 0$ resp.), and the term $W(M, p+b_j - (1,1,1); 1, -\epsilon_i)$ equals $W(M, -p+ (-b_j) - (1,1,1); 1, \epsilon_i)$. We have an analogue one-to-one correspondence for the other factor $W$ in the perturbative part. Finally, we have  $P + B_j-3 = -2 \iff -P + (-B_j) - 3 = -4$ and similar relations, for which $W=1$, and the two cases transform into each other. These assignments work for all $1\leq j \leq 4$. For $\CN=2^\ast$, similar arguments follow with \eqref{eq:Zpert_p2_reduced} and \eqref{eq:Zpert_adj_P2}, by replacing $p$ by $2p$ and $0$ in \eqref{eq:epsflip_Widentity} and its variations. \par
In total, we see that the summand in $H_{\mu}^{(N_f)} (\Lambda, q, M_j, B_j, x, z,  -\epsilon_1, -\epsilon_2)$ with equivariant gauge flux $p$ is equal to the summand for $p$ in $(-1)^{2\mu + 1}H_{\mu}^{(N_f)} (\Lambda, q, M_j, -b_j, x, -z,  \epsilon_1, \epsilon_2)$, proving \eqref{eq:HP2_c1odd_eps_flip}.

\clearpage

%\bibliography{references} 
%\bibliographystyle{JHEP} 

\providecommand{\href}[2]{#2}\begingroup\raggedright\endgroup

\end{document}